\documentclass[11pt]{article}
\input{preamble}
\hypersetup{pdftitle={Linear Single-Source Shortest Path and Minimum Cut, Almost}}
\title{Linear Single-Source Shortest Path and Minimum Cut, Almost}
\author{Iman Gholami\thanks{University of Maryland. Email: \texttt{igholami@umd.edu}.}
\and MohammadTaghi Hajiaghayi\thanks{University of Maryland. Email: \texttt{hajiagha@umd.edu}.}}
\date{}
\begin{document}
\maketitle
\begin{abstract}
We give a comparison--addition algorithm for single-source shortest paths in undirected graphs with nonnegative real edge lengths. For each input with $n$ vertices and $m$ edges, its expected running time is $O(m+n\log\log\log n)$. It terminates almost surely. Every terminating run returns exact distances and a shortest-path tree on the reachable vertices. The bound is independent of the ratio of edge lengths and is linear whenever $m=\Omega(n\log\log\log n)$.

We also give randomized comparison--addition algorithms for exact global minimum cut in undirected graphs with arbitrary nonnegative real capacities. For every fixed positive integer $h$, one algorithm runs in worst-case time $O_h(m+n\log^{(h)}n)$, where $\log^{(h)}n$ is the lower-truncated $h$-fold logarithm; a uniform algorithm runs in worst-case time $O((m+n)2^{O(\log^* n)})$. Both bounds charge individually for every fair random bit, succeed with probability at least two thirds, and always return a nontrivial cut with its exact original capacity.

\end{abstract}
\clearpage
\tableofcontents
\clearpage
\section{Introduction}\label{sec:intro}
The single-source shortest paths problem asks for the distance from a vertex $s$ to every vertex of a weighted graph. Even on a star, sorting the distances orders an arbitrary list of real numbers. Yet computing the distances on that star requires only reading the edge weights. This distinction between computing a collection of distances and ordering it is central to the complexity of the problem.

Dijkstra's algorithm settles the vertices in nondecreasing distance order~\cite{DBLP:journals/nm/Dijkstra59}. Fredman and Tarjan's Fibonacci heaps give it a running time of $O(m+n\log n)$ on nonnegative real lengths~\cite{DBLP:journals/jacm/FredmanT87}. For integer weights on undirected graphs, Thorup obtained a linear worst-case bound by using a hierarchical ordering of groups of vertices~\cite{DBLP:journals/jacm/Thorup99}. In the comparison--addition model, edge lengths occupy opaque real registers. Only combinatorial indices can be used as addresses. Pettie and Ramachandran showed how to operate a hierarchy using comparisons and additions. Their preprocessing bound includes the smaller of $n\log n$ and $n\log\log r$, where $r$ bounds the positive weight spread~\cite{DBLP:journals/siamcomp/PettieR05}.

A second line of work broke the sorting bound for arbitrary nonnegative real lengths: first by a randomized algorithm on sparse undirected graphs~\cite{DBLP:conf/focs/DuanMSY23}, and subsequently by a deterministic algorithm on sparse directed graphs~\cite{DBLP:conf/stoc/DuanMMSY25}. In an undirected graph, a short route to an already known vertex can be traversed in either direction. We use such routes to organize the computation without fixing a global distance order or enumerating all weight scales.

\begin{theorem}\label{thm:main}
Given an adjacency-list undirected multigraph with $n$ vertices, $m$ edges of arbitrary finite nonnegative real length, and a source $s$, \tref{Algorithm}{alg:sssp} returns all exact source distances and a shortest-path tree on the reachable vertices. It terminates almost surely, and its expected running time is
\begin{align*}
 O(m+n\log\log\log n).
\end{align*}
Every terminating outcome is correct.
\end{theorem}

We use the comparison--addition model. Each edge length occupies an opaque real register, and copying, adding, or comparing two register values takes constant time. Vertex identifiers, array indices, and counters use words of $\Theta(\log(n+m+2))$ bits with unit-cost indexing, integer addition and subtraction, comparisons, shifts, and bit operations. For each fixed input, the expectation in \tref{Theorem}{thm:main} is over the independent uniform random words used for sampling and distance estimation. \tref{Section}{sec:prelim} gives the full model.

We also study the global minimum cut problem, which asks for a partition of the vertices of an undirected graph into two nonempty sets minimizing the total capacity of the edges between them. Karger's spanning-tree approach led to a randomized near-linear-time algorithm: construct trees that are likely to cross a minimum cut only a few times, then search the cuts described by those tree crossings~\cite{Karger2000}. Subsequent work reduced the time for the weighted problem to $O(m\log^2 n)$ with high probability~\cite{GMW2024}, raising the question of how close exact weighted minimum cut can come to linear time.

For an undirected multigraph $G$ with $n\ge 2$ vertices, $m$ edges, and nonnegative real capacities, let $\log^{(h)}n$ denote the $h$-fold iterated base-two logarithm (truncated below at a constant), and let $\log^*n$ denote the number of applications of the base-two logarithm needed to reduce $n$ to at most two.

\begin{theorem}[Fixed-depth minimum-cut bounds]\label{thm:main-fixed}
For every fixed positive integer $h$, exact global minimum cut in $G$ can be computed with success probability at least two thirds in worst-case time and space
\[
 O_h\bigl(m+n\log^{(h)}n\bigr).
\]
The algorithm always returns a nontrivial cut with its exact capacity. It uses only comparison, addition, and subtraction on input-derived real quantities, ordinary word operations on combinatorial data, and individually charged fair random bits.
\end{theorem}

\begin{theorem}[Uniform minimum-cut bound]\label{thm:main-uniform}
There is one algorithm in the same model with success probability at least two thirds and worst-case time and space
\[
 O\bigl((m+n)2^{O(\log^* n)}\bigr).
\]
Every output is again an actual cut with its exact capacity. All constants in this bound are independent of the input size.
\end{theorem}

Theorems~\ref{thm:main-fixed} and~\ref{thm:main-uniform} are Monte Carlo guarantees: a failed run may return a cut larger than the minimum, and for any fixed $d>0$, independent repetition reduces the failure probability to at most $n^{-d}$ with an $O_d(\log(2n))$ factor in time. On sparse graphs the uniform bound is asymptotically smaller than each fixed-depth vertex bound, whereas on denser graphs the fixed-depth family retains a linear edge term (and is linear whenever $m=\Omega(n\log^{(h)}n)$ for some fixed $h$). The main idea is to stop the usual forest construction before its forests span, complete one chosen partial forest recursively on its quotient while maintaining three guarantees for both genuine cuts and arbitrary edge sets, and then apply a linear-time two-respecting cut algorithm.

\subsection{Related Work}\label{sec:related}
\paragraph{Distance-ordered search and component hierarchies.}
Dijkstra's algorithm extracts vertices in nondecreasing distance order, running in $O((m+n)\log n)$ time with a binary heap and $O(m+n\log n)$ with a Fibonacci heap~\cite{DBLP:journals/jacm/FredmanT87} on directed or undirected graphs. For undirected graphs with positive integer weights, Thorup achieved linear $O(m+n)$ time using word operations~\cite[Theorem~37]{DBLP:journals/jacm/Thorup99}. In the comparison--addition model, Pettie and Ramachandran preprocess an undirected graph in $O(\mathrm{MST}(m,n)+\min\{n\log n,n\log\log r\})$ time (where $r$ bounds the positive weight spread) and answer source queries via a hierarchy~\cite[Theorem~1.1]{DBLP:journals/siamcomp/PettieR05}; their lower bound~\cite[Property~6.1 and Theorem~6.1]{DBLP:journals/siamcomp/PettieR05} applies to algorithms that also output a separation-respecting vertex order. In their adapted lower-bound tree~\cite[Figure~4]{DBLP:journals/siamcomp/PettieR05}, outward states on distinct arms have no transitions between them, so condensation can place a longer arm first; \tref{Section}{sec:overview} explains which transitions do impose an order.

\paragraph{Breaking the sorting bound for SSSP.}
For arbitrary nonnegative real lengths, a recent line of work broke the $O(m+n\log n)$ sorting barrier (on connected undirected inputs or reachable directed subgraphs, up to $O(m+n)$ reachability work): Duan, Mao, Shu, and Yin gave a randomized $O(m\sqrt{\log n\log\log n})$ undirected algorithm~\cite[Theorem~1]{DBLP:conf/focs/DuanMSY23}, later derandomized by Yan~\cite{Yan2026}; Duan, Mao, Mao, Shu, and Yin gave a deterministic $O(m\log^{2/3}n)$ directed algorithm~\cite[Theorem~1.1]{DBLP:conf/stoc/DuanMMSY25} and a subsequent $O(m\sqrt{\log n}+\sqrt{mn\log n\log\log n})$ directed bound~\cite[Theorem~1]{DBLP:conf/icalp/DuanMSY26}; Grishin and Zabokritskiy report a deterministic $O(n\sqrt{\log n}(\log\log n)^{1/4})$ bound for sparse directed graphs ($m=O(n)$)~\cite{GrishinZabokritskiy2026}; and Kadria and Roditty report $O(m\sqrt{\log n}(\log\log n\log\log\log n)^{1/4})$ high-probability undirected time using distances to sampled vertices~\cite[Theorem~1]{DBLP:journals/corr/abs-2609-15247}. By contrast, our route estimator samples edges in a smaller pivot graph using dyadic random partitions~\cite{DBLP:journals/jcss/FakcharoenpholRT04} only to certify detour bounds, while computing exact distances in the original physical metric. Table~\ref{tab:sssp-bounds} summarizes these bounds.

\begin{table}[ht]
\centering
\begin{tabular}{@{}p{4.4cm}p{5.5cm}p{4.0cm}@{}}
\toprule
Algorithm & Time & Setting / Guarantee \\
\midrule
Fredman and Tarjan \cite{DBLP:journals/jacm/FredmanT87}
 & $O(m+n\log n)$ & directed, deterministic \\
Thorup \cite{DBLP:journals/jacm/Thorup99}
 & $O(m+n)$ & undirected, integer weights \\
Pettie and Ramachandran \cite{DBLP:journals/siamcomp/PettieR05}
 & $O(\mathrm{MST}(m,n)+\min\{n\log n,n\log\log r\})$ & undirected, deterministic \\
Duan et al. \cite{DBLP:conf/focs/DuanMSY23}, Yan \cite{Yan2026}
 & $O(m\sqrt{\log n\log\log n})$ & undirected, deterministic \\
Duan et al. \cite{DBLP:conf/stoc/DuanMMSY25}
 & $O(m\log^{2/3}n)$ & directed, deterministic \\
Duan et al. \cite{DBLP:conf/icalp/DuanMSY26}
 & $O(m\sqrt{\log n}+\sqrt{mn\log n\log\log n})$ & directed, deterministic \\
Grishin and Zabokritskiy \cite{GrishinZabokritskiy2026}
 & $O(n\sqrt{\log n}(\log\log n)^{1/4})$ & directed ($m=O(n)$), det. \\
Kadria and Roditty \cite{DBLP:journals/corr/abs-2609-15247}
 & $O(m\sqrt{\log n}(\log\log n\log\log\log n)^{1/4})$ & undirected, high probability \\
This paper (\tref{Theorem}{thm:main})
 & $O(m+n\log\log\log n)$ & undirected, expected (Las Vegas) \\
\bottomrule
\end{tabular}
\caption{Selected single-source shortest paths bounds for nonnegative edge lengths. Except for Thorup's integer-weight word-RAM algorithm, the listed bounds apply to arbitrary real edge lengths in the comparison--addition model ($r$ denotes the positive weight spread in Pettie and Ramachandran's bound).}
\label{tab:sssp-bounds}
\end{table}

\paragraph{Tree packings, certificates, and minimum cut.}
Karger's near-linear minimum-cut algorithm~\cite{Karger2000} samples trees from a spanning-tree packing so that a minimum cut is crossed at most twice, then searches for the least two-respecting cut. Gawrychowski, Mozes, and Weimann gave a deterministic $O(m\log n)$ fixed-tree algorithm and an $O(m\log^2 n)$ global algorithm~\cite{GMW2024}, and subsequently simplified the approach of Mukhopadhyay and Nanongkai to obtain density-sensitive bounds including $O(m\log n+n^{1+\varepsilon})$ for any fixed $\varepsilon>0$~\cite{GMW2021}. Table~\ref{tab:bounds} compares these high-probability bounds with ours. Our base distribution builds on sparse certificates~\cite{NI1992}, cut sparsification~\cite{BK2015,KargerThesis,ForsterDeVos2023}, and a linear-time two-edge-cut cactus construction~\cite{MNS2017}, while our recursive completion replaces full tree packings with three weaker guarantees that survive contraction.

\begin{table}[ht]
\centering
\begin{tabular}{@{}p{4.6cm}p{5.9cm}p{3.2cm}@{}}
\toprule
Algorithm & Time & Success \\
\midrule
Karger \cite{Karger2000}
 & $O(m\log^3 n)$ & high probability \\
Gawrychowski et al. \cite{GMW2024}
 & $O(m\log^2 n)$ & high probability \\
Gawrychowski et al. \cite{GMW2021}
 & $O(m\log n+n^{1+\varepsilon})$ & high probability \\
This paper, each fixed $h$
 & $O_h((m+n\log^{(h)}n)\log n)$ & high probability \\
This paper, uniform
 & $O((m+n)\log n\cdot 2^{O(\log^*n)})$ & high probability \\
\bottomrule
\end{tabular}
\caption{Selected global weighted minimum-cut bounds. In the third row $\varepsilon$ is a fixed positive constant. Our bounds explicitly charge fair random bits and permit only comparison and addition on opaque capacities; the table does not identify the numerical sampling assumptions of the earlier algorithms with this model. Our high-probability bounds are obtained from the $2/3$-success bounds $O_h(m+n\log^{(h)}n)$ and $O((m+n)2^{O(\log^*n)})$ by $O(\log n)$ independent repetitions.}
\label{tab:bounds}
\end{table}

\paragraph{Comparison-based primitives and closeness to SSSP.}
Our minimum-cut algorithm uses linear-time comparison-based minimum spanning trees~\cite{KKT1995,KPS2026} and linear-time path-extremum queries~\cite{DRT1992,BuchsbaumEtAl2008}. Both our single-source shortest paths result (\tref{Theorem}{thm:main}) and our minimum-cut results (\tref{Theorem}{thm:main-fixed} and \tref{Theorem}{thm:main-uniform}) address the same central question in comparison--addition graph algorithms: how to keep edge scans linear while reducing vertex-level overhead to an iterated-logarithmic factor. In both problems, we avoid a global sort or full-graph scans across all weight scales by organizing the computation around a tree or partial forest hierarchy and compressing vertex-level work on quotients; although minimum cut does not call SSSP as a subroutine, both constructions share this structural paradigm.

\subsection{Technical Overview}\label{sec:overview}
Consider a shortest path that crosses from a region near one selected vertex, called a \emph{pivot}, to a region near another. Although a route through the two pivots may be longer than the crossing itself, its length can still tell us which part of the shortest path to process first. Our algorithm builds enough such comparisons to compute all distances without listing the vertices in distance order.

\paragraph{Which crossings need individual attention?}
Cut a minimum spanning tree into connected regions, each containing one pivot. For a vertex $v$ let $p(v)$ be its region's pivot, $R_v$ the length of the tree path to $p(v)$, and $r_v$ the largest edge on that path. The cuts ensure two properties: each path has at most $h$ edges, for $h$ polylogarithmic in $n$, and every edge $xy$ of weight $w$ between regions satisfies $r_x,r_y\le w$. Consequently $R_x+w+R_y$ is the length of a \emph{physical} (original-graph) walk from $p(x)$ to $p(y)$ and is at most $(2h+1)w$ (\tref{Lemma}{lem:owners}).

Form a smaller graph on the pivots by keeping one lightest projected edge $R_x+w+R_y$ (a \emph{request}) for each pivot pair that occurs. Let $t$ be the number of pivots. While more than $32t$ requests remain, sample edges and estimate the distance in the sample between the ends of every outstanding request. For connected endpoints the estimator returns a table value $Z$ at least their distance in the sample. Every sampled edge represents a physical walk, so $Z$ also bounds their distance in the original graph. Fix a polylogarithmic factor $P$. Among requests with the same table index, choose the smallest projected weight $\lambda$ satisfying $P\lambda\ge Z$, if one exists. Every request in that group whose projected weight is at least $\lambda$ then \emph{qualifies} and receives the certificate $\lambda$. \tref{Algorithm}{alg:name} accepts a trial only if at least half of the outstanding requests qualify, records their certificates, and continues with the rest. Once at most $32t$ requests remain, each supplies its own projected weight as its certificate. Every original interregion edge inherits the certificate of its pivot-pair representative. The estimator may overestimate, and the proof needs only a constant chance of acceptance per trial (\tref{Lemmas}{lem:estimate} and~\ref{lem:names}). This halving rule pays for all scans even when the original graph is dense.

\paragraph{A coordinate along a shortest path.}
For an interregion edge $e$ with certificate $\lambda_e$, put $d_e=P\lambda_e$, a bound on the length of a physical walk between its pivots, and set $d_e=0$ within one region. Call $e$ \emph{radial} when $d_e\le w(e)/2$ and \emph{macro} otherwise. Identify all pivots into a single root and give each radial edge the auxiliary length $w(e)+d_e$; denote the distance from $v$ to this root in the auxiliary graph by $J_v$. A route counted by $J_v$ unfolds into a physical walk from $v$ to its \emph{own} pivot $p(v)$: follow its radial edges to whichever pivot is reached, then traverse the paid-for pivot detours in reverse back to $p(v)$. Moreover, any radial edge changes $J$ by at most $3w(e)/2$ (\tref{Lemma}{lem:coordinate}), and any radial edge heavier than $J_x+d_e+J_y$ can be deleted since the route through the pivots is strictly shorter. \tref{Figure}{fig:detour} distinguishes the two walks.

\begin{figure}[ht]
\centering
\begin{subfigure}[b]{.47\linewidth}\centering
\begin{tikzpicture}[x=1cm,y=1cm]
\node[paper graph vertex] (p) at (0,1.2) {$p$};
\node[paper graph vertex] (x) at (.75,0) {$x$};
\node[paper graph vertex] (y) at (2.75,0) {$y$};
\node[paper graph vertex] (q) at (3.5,1.2) {$q$};
\draw[paper forest edge] (p)--node[left]{$R_x$}(x);
\draw[paper graph edge] (x)--node[below]{$w$}(y);
\draw[paper forest edge] (y)--node[right]{$R_y$}(q);
\end{tikzpicture}
\caption{The projected edge through $x$ and $y$.}\label{fig:projection}
\end{subfigure}\hfill
\begin{subfigure}[b]{.47\linewidth}\centering
\begin{tikzpicture}[x=1cm,y=1cm]
\node[paper graph vertex] (p) at (0,1.2) {$p$};
\node[paper graph vertex] (x) at (.75,0) {$x$};
\node[paper graph vertex] (y) at (2.75,0) {$y$};
\node[paper graph vertex] (q) at (3.5,1.2) {$q$};
\draw[paper forest edge] (x)--node[left]{$J_x$}(p);
\draw[paper pending edge] (p)--node[above]{$d_e$}(q);
\draw[paper forest edge] (q)--node[right]{$J_y$}(y);
\draw[paper graph edge] (x)--node[below]{$w$}(y);
\end{tikzpicture}
\caption{A certified alternative to the edge $xy$.}\label{fig:shortroute}
\end{subfigure}
\caption{The pivot pair is $p=p(x)$ and $q=p(y)$. In \tref{Figure}{fig:projection}, the tree routes give the input weight $R_x+w+R_y$ of a pivot edge. In \tref{Figure}{fig:shortroute}, the upper route represents physical walks with lengths bounded by the displayed quantities. The radial test compares $2d_e$ with $w$. A later deletion compares the entire upper bound $J_x+d_e+J_y$ with $w$.}\label{fig:detour}
\ifdefined\Description
\Description{Two diagrams have $p$ and $q$ above the endpoints $x$ and $y$ of an edge of weight $w$. In the first, the tree paths from the endpoints to their respective pivots have lengths $R_x$ and $R_y$. In the second, the two access bounds are $J_x$ and $J_y$, with a pivot detour bounded by $d_e$.}
\fi
\end{figure}
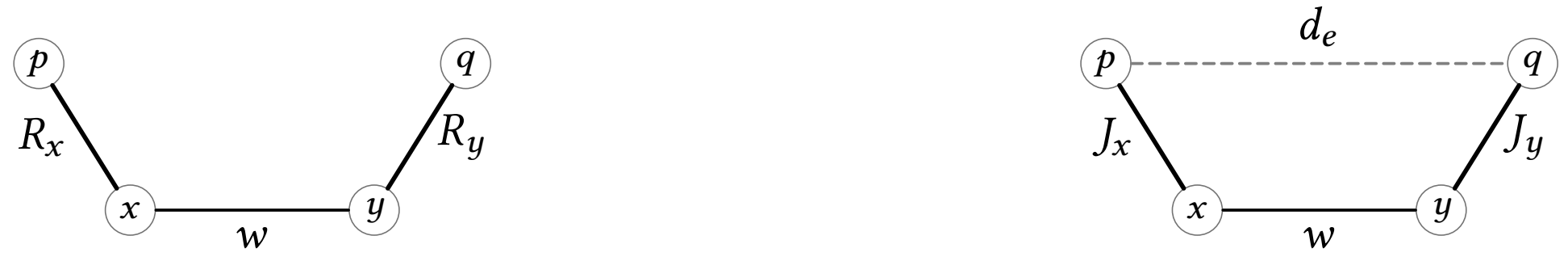

\paragraph{Why a few copies per vertex suffice.}
Suppose a radial subpath (using only radial edges) of a shortest path rises to a peak coordinate $M>0$ while both endpoints have coordinate at most $M/16$. Reaching the peak costs at least $5M/8$ by the bound on how fast $J$ can change. Bypassing the subpath via the pivots costs at most $M/8$ for its two endpoint walks ($J_x+J_y$) plus at most half the subpath's length for the pivot detours, which would strictly shorten the path. Thus, before a minimum of $J$ along the subpath, the current coordinate stays within a factor of $16$ of the smallest value seen; the reverse holds after the minimum (\tref{Lemma}{lem:peak}).

A radial edge between positive coordinates is near if their ratio is at most $16^2$, and far otherwise. Within each connected component of near edges, partition the positive coordinates into shells with successive boundaries in ratio $16$. Each positive-coordinate vertex has one native shell. Within a component the inward phase tracks the lowest shell reached and permits the shell immediately above it in the same window. The outward phase tracks the highest shell reached and permits the shell immediately below it. A far transfer resets the record to the destination's native shell. A vertex belongs to at most two windows per phase, so it needs only a constant number of copies, called \emph{states}. Traversing a macro edge restarts this rule at its head. The resulting directed state graph contains an equal-length realization of every shortest path from the source. An individual copy can have a larger directed distance, even infinity. The evaluator emits feasible upper bounds on physical distance and attains equality at the last copy of each realized shortest path. Taking the least emitted value at each reachable vertex therefore gives its exact distance.

\paragraph{Which state can be processed next?}
Organize the states into a hierarchy of strongly connected components that merge as an arc-length threshold increases. Where merge weights are grouped on a doubling grid, events in $[a,2a)$ are processed together, with each resulting merge assigned height $a$. Arcs shorter than $a$ determine the child order, and connections within the merge can use arcs shorter than $2a$. Process each group in distance intervals, called \emph{slots}, of width at most its height. If a state transition departs and arrives within the same slot, its length is strictly below the group height, so any such transition between immediate subgroups must move forward in a known topological order---a backward cycle of such transitions would already have merged those subgroups at a smaller threshold. An \emph{offer} is a tentative upper bound at a state. It is due when it lies in the current slot and satisfies the applicable deadlines. In child order the evaluator visits each registered live subgroup and checks contiguous runs of unregistered subgroups. A run with a due minimum splits at the subgroup containing that minimum. A run without one can be passed in a single check.

The cost of skipping subgroups without a due offer matters as much as finding one that is due. Once a group opens at some first offer $D_0$, the triangle inequality guarantees that every physically tight arrival into it occurs at or before $D_0$ plus the group's physical diameter bound. After processing this inclusive deadline, the group \emph{expires}. Likewise, a short physical walk from a newly active subgroup bounds how long a neighboring unvisited subgroup can be scanned unsuccessfully while the enclosing group remains active. \tref{Section}{sec:eval} builds a topological child order with short total connecting length and charges all unsuccessful checks to those walks; \tref{Algorithm}{alg:visit} implements the traversal.

\paragraph{Constructing the groups without already knowing $J$.}
Computing the coordinates $J_v$ is itself a shortest-path problem. \tref{Algorithm}{alg:aux} avoids circularity by partitioning the auxiliary graph into small connected sets and applying a separate local construction to each set. The local construction chooses pivots in its own minimum spanning tree and uses the maximum edge weight on the tree path to a vertex's local pivot as a coordinate available before any auxiliary distance is known (\tref{Section}{sec:base}). The original quantities $r_v$ and $R_v$ instead define a coarse time schedule that orders offers between sets. A local slot can straddle at most one boundary of that coarse schedule, and \tref{Lemma}{lem:clip} shows how to split and resume the search cleanly across it.

\paragraph{Where the third logarithm enters.}
Beyond its linear edge cost, the local construction spends $O(1+\log\log(k+4))$ per vertex on a set of size $k$. The auxiliary atoms are disjoint sets of auxiliary vertices. The final hierarchy uses disjoint sets of outer states, and each physical vertex has only a constant number of these states. In both applications $k$ is bounded by a fixed power of $\log n$. For the remaining hierarchy merge events (where subgroups unite at higher thresholds), the sampled certificates supply scale names for finite attachments and aligned blocks. Blocks on separate grids use their own radius bounds. In either case the search range spans only $O(\log\log n)$ consecutive powers of two, so binary search locates each event in $O(\log\log\log n)$ time. The initial minimum spanning tree, auxiliary width classification, component unions, and interval minima under splits and decreases also fit the $O(m+n\log\log\log n)$ budget. Sampling costs $O(m+n)$ in expectation (\tref{Section}{sec:full}).

\section{Preliminaries}\label{sec:prelim}

\paragraph{Graphs and distances.}
The input is an undirected multigraph $G=(V,E)$ given by adjacency lists,
a source $s\in V$, and a finite nonnegative real length $w(e)$ on every
edge. Write $n=|V|$ and $m=|E|$. A walk may repeat edges. Its length is
the sum of the traversed lengths, with repetitions. A path has no repeated
vertices. The distance $d_G(x,y)$ is the least length of an $x$--$y$ walk
and is $+\infty$ if no such walk exists. For reachable vertices a minimum
exists since deleting cycles of nonnegative length reduces an arbitrary
walk to a path of no greater length and the graph has finitely many paths.
A \emph{geodesic} attains this minimum. An arc $xy$ is \emph{tight} for
source $s$ if both endpoint distances are finite and $d_G(s,x)+w(xy)=d_G(s,y)$.
In an undirected graph either orientation may be used in this definition.

\paragraph{Notation.}
For a tree $T$, write $T[x,y]$ for its unique $x$--$y$ path, $D_T(x,y)$
for its length, and $b_T(x,y)$ for its greatest edge length, with
$b_T(x,x)=0$. A rooted-tree depth counts edges, and its root has depth zero.
We use $N$ for the vertex count in a construction and $M$ for the arc or
demand count where specified. All logarithms have base two. For integers
$k\ge1$ put
\begin{align*}
L_k=\lceil\log_2(k+2)\rceil,\qquad
\tau(k)=1+\log\log\log(k+16).
\end{align*}
The outer and local constructions assume $N>L_N^{64}$ and $N>2L_N^4$, respectively. The excluded sizes are absolutely bounded. A scan retaining a lightest edge per vertex pair then leaves a constant-size graph for direct source search or MST construction. For a local hierarchy, keep the original bidirected nonloop edges on one state per vertex and merge equal-weight MST edges simultaneously. The tree edges connect the merge children, and \tref{Lemma}{lem:mst} excludes lighter edges between them.

\paragraph{Operations on lengths and indices.}
Each input length occupies an opaque real register. Copying, comparison
and addition each take one operation. Infinity is a separate flag. To
compare $a-b$ with $c-d$, compare $a+d$ with $c+b$. More generally a
signed value is stored as two nonnegative sums. Addition is componentwise
and negation interchanges the components. This convention costs a constant
number of register operations per signed addition or comparison.

Multiplication of a real by $2^j$ takes $j$ doublings for $j\ge0$.
Whenever a construction uses a fixed power of two as a denominator,
one may represent every compared value on the corresponding common scale.
Values from outside the construction are rescaled on entry. Variable
numbers of doublings are charged in the relevant bounds. No real
register supplies an address, an integer part, a logarithm, or a bit.

Graph identifiers, array indices, and counters use a constant number of
words of $\Theta(\log(n+m+2))$ bits. Array access, shifts, bit operations,
integer addition, subtraction, and comparison take constant time per word.
An integer logarithm on a range of $k$ indices can be tabulated in $O(k)$
time by the recurrence for $\lfloor\log_2 i\rfloor$ under halving, starting
with $i=1$. Products used solely in integer size tests may be capped once
they reach the tested limit. For fixed-size blocks of any array, a scan
stores the block index and offset for every member. No integer division is
needed for these lookups.

Randomness consists of independent uniformly random words. To draw a
uniform integer from $0$ to $h-1$, take $\lceil\log_2 h\rceil$ unbiased
bits, rejecting outcomes at least $h$. For $h=1$ return zero. For
larger index ranges used here a constant number of words supplies the
bits and each attempt succeeds with probability greater than $1/2$.
Expectations are over these draws on a fixed input. \tref{Section}{sec:primitives}
proves the bounds used for graph searches, spanning trees, and structural
queries in this model.

\section{Graph and Data-Structure Primitives}\label{sec:primitives}
Several constructions need a linear edge term even when the number of
edges exceeds the number of vertices. Here every primitive is proved
with its dependence on both counts explicit.

\subsection{Searches and Selection}
A depth-first or breadth-first search marks a vertex at discovery and
scans its adjacency list once. To sort a directed acyclic graph
topologically, repeatedly remove a vertex of zero current indegree,
decrementing the indegrees of its successors. To sort $p$ records
stably on an integer coordinate in $[0,k)$, count each coordinate,
take prefix sums as starting positions, and write each record to its
coordinate's next position in input order. For a constant number of
coordinates apply these stable passes from last coordinate to first.

For nonnegative arc lengths we use Dijkstra's search~\cite{DBLP:journals/nm/Dijkstra59}.
Start a source at zero and all other tentative values at infinity. Extract
a vertex of smallest finite unsettled value, settle it, and offer that
value plus each outgoing arc length to the arc's head. Stop when the
queue is empty, or when the smallest finite value exceeds a specified
radius. Unvisited entries in a reused array carry a search identifier,
so each new search need only initialize the entries it touches.

For the rank-$r$ key among $p$ comparable keys, form groups of five and
at most one smaller group. Take the median of the group medians
recursively. Partition the original keys into the three sets below,
equal to, and above that pivot. Return the pivot if its equal block
contains rank $r$. In the remaining cases recurse into the appropriate strict
side with the adjusted rank. Use direct comparison for bounded $p$.
This is the selection method of Blum, Floyd, Pratt, Rivest, and
Tarjan~\cite{DBLP:journals/jcss/BlumFPRT73}.

\begin{lemma}\label{lem:basic-primitives}
Depth-first search, breadth-first search, and topological sorting cost $O(k+l)$ on $k$ vertices
and $l$ arcs. Stable integer sorting costs $O(p+k)$ per pass. The
selection algorithm returns its requested order statistic in $O(p)$
comparisons and elementary operations.
\end{lemma}
\begin{proof}
Each adjacency entry is examined once and each discovered or removed
vertex is queued once. A nonempty finite acyclic graph has a vertex of
indegree zero, since following incoming edges indefinitely would repeat
a vertex. Removal therefore yields an order in which every arc points
forward. Counting sort uses one pass per count, prefix sum, or output
write. Stability preserves the order of less significant coordinates.

For selection, at least half the group medians lie on each weak side
of the pivot. Except for at most one incomplete group, every such group
supplies three keys on that side. The strict side of the second recursive
call thus contains at most $7p/10+O(1)$ keys, including when keys tie.
The two recursive sizes sum to at most $9p/10+O(1)$. Choosing a fixed
bounded-input cutoff to absorb the additive constant proves a linear
bound by induction. Partition ranks show correctness in every branch.
\end{proof}

\begin{lemma}\label{lem:searches}
Dijkstra's search settles each reached vertex at its true distance. With
a radius cutoff $R$, it settles exactly the reachable vertices at
distance at most $R$. A binary heap on at most $k$ vertices costs
$O(1+\log(k+1))$ per insertion, decrease, or extraction. A scan for
each minimum gives a full search in $O(k^2+l)$ time.
\end{lemma}
\begin{proof}
A tentative value is the length of a discovered walk. If the first
erroneous extraction has value $z$ at $v$, follow a shortest source--$v$
path to its first unsettled vertex $u$. Its settled predecessor has
already offered its exact prefix length, or $u$ is the seeded source.
Thus the pending value at $u$ is at most $d_G(s,v)<z$, contrary to
minimal extraction. The same first-unsettled-vertex argument rules
out termination above radius $R$ while an unsettled path of length
at most $R$ exists. A binary heap restores heap order along one
root--leaf path after an operation, recording each moved vertex's
array address. A complete binary tree has logarithmic height. The
scan version spends $O(k)$ on each of at most $k$ extractions and
examines each outgoing incidence once.
\end{proof}

\subsection{Minimum Spanning Trees}
Assign each edge its unique input index and compare edges by the pair
(length, index). Maintain quotient vertices as connected sets of
original vertices. Each remaining quotient edge retains its original
index. \tref{Algorithm}{alg:bounded-prim} specializes the bounded tree growth of
Fredman and Tarjan~\cite{DBLP:journals/jacm/FredmanT87} to three passes
with the explicit heap limits below.
The \emph{current tree} in that algorithm comprises the vertices scanned
by the present growth. The vertices of completed growths are already
marked. A heap entry for an outside vertex names its lightest cut edge
from the current tree. Use a fresh heap and membership identifiers for
each growth so that old boundary entries are never reused.

\begin{algorithm}[ht]
\caption{One bounded spanning-tree growth pass}\label{alg:bounded-prim}
\Fn{\textsc{GrowthPass}($H,B$)}{
\KwIn{A connected quotient graph $H$ and an integer limit $B\ge1$ or $B=+\infty$.}
\KwOut{Retained original edge indices and the quotient after their contraction.}
$F\gets\varnothing$\;
Mark all vertices unvisited\;
\ForEach{still-unvisited start vertex $u$}{
 initialize an empty boundary heap (\tref{Lemma}{lem:heap-bound}) and a new growth identifier\;
 \While{true}{
  mark $u$ visited in this growth and scan its incident edges\;
  for each neighbor outside this growth, insert or improve its boundary edge\;
  \If{the heap is empty or its size exceeds $B$}{leave this growth\;}
  extract the minimum edge to its outside vertex $v$ and add its index to $F$\;
  \If{$v$ was visited by an earlier growth}{leave this growth\;}
  $u\gets v$\;
 }
}
\Return{$F$ and the loop-free quotient by the components of $F$}\;
}
\end{algorithm}

For a connected input with $k$ vertices take $B_1=\max(2,\lceil\log_2 L_k\rceil)$
in the first pass, $B_2=L_k$ in the second, and use $B=+\infty$ in the
last pass. Use the original $k$ to choose both finite limits. A singleton
requires no pass. Retain all selected indices across contractions.

We implement the queue by the Fibonacci heap. The operation bounds are
the insertion, decrease, and extraction cases of Fredman and Tarjan
\cite[Theorem~1]{DBLP:journals/jacm/FredmanT87}. We prove these cases
in \tref{Lemma}{lem:heap-bound}. Its roots and child lists
are doubly linked and store parent pointers, degrees, and one mark bit.
New singleton roots and decreased minima update a minimum pointer.
On extraction, remove the minimum, promote its children to roots,
and repeatedly link roots of equal degree by making the larger-keyed
root a child of the smaller. A degree-indexed array locates matches.
Rebuild the minimum pointer while collecting the remaining roots.
A decrease violating heap order cuts its node to the root list. When
a nonroot loses a child, mark it on its first loss since becoming
a child, cutting it and continuing upwards on its second loss.
Every newly linked child is unmarked, and roots are always unmarked.

\begin{lemma}\label{lem:heap-bound}
Heap insertion and decrease have constant amortized cost. Extraction from
a heap of size $s$ has amortized cost $O(1+\log(s+1))$.
\end{lemma}
\begin{proof}
Order the surviving children of a node by attachment time. When child
$j$ was linked, the parent still had at least the first $j-1$ children,
and the link rule required equal degrees. A surviving child loses at
most one child after that link, so its present degree is at least
$\max(0,j-2)$. Let $F_0=0,F_1=1$ and $F_{i+2}=F_{i+1}+F_i$.
Induction on the tree height gives a subtree size of at least $F_{d+2}$
for every node of degree $d$. Indeed, for $d=0$ or $1$ this follows
by direct counting, and for $d\ge2$ the lower bound is
\begin{align*}
 1+F_2+\sum_{j=2}^d F_j=F_{d+2}.
\end{align*}
The bound grows exponentially in $d$ since $F_{i+2}\ge2F_i$ for $i\ge1$.

Use a constant multiple of the number of roots plus twice the number
of marked nonroots as potential. Each cascading cut after the first
removes one mark and creates one root. The decrease can create at
most one final mark. The potential change pays for all but a constant
number of its steps. In an extraction, at most $O(1+\log(s+1))$
children become roots. Every link reduces the number of roots by one.
The roots scanned are bounded by the number of links plus the final
number of roots, which is at most one per possible degree. The degree
array need only have an explicit logarithmic size, e.g.\
$3+2\lceil\log_2(s+1)\rceil$, by the Fibonacci bound. This proves
the extraction cost. Discarding a heap releases at most one node per
earlier insertion and cannot increase the potential.
\end{proof}

\begin{lemma}\label{lem:growth-pass}
Every retained edge is in one common minimum spanning tree. A pass on
$k'$ vertices and $l'$ edges costs
$O(l'+k'(1+\log(\min(k',B)+1)))$ and, for finite $B$, leaves at most
$1+2l'/B$ quotient vertices.
\end{lemma}
\begin{proof}
Fix the tree $T$ obtained by starting from any weight-minimum spanning
tree and selecting, among all such trees, the lexicographically least
sorted list of edge pairs. For any cut, its least crossing edge $e$
belongs to $T$. Otherwise $T$ together with $e$ contains a cycle
crossing the cut at another edge $f$. Replacing $f$ by $e$ either
reduces total weight or, at equal weight, reduces the sorted list.
Both contradict the choice of $T$. The heap minimum is the least
edge across the current growth's vertex cut. In subsequent quotients
this cut lifts to a cut of original vertices because each quotient
vertex is an entire contracted set. The same $T$ therefore contains
all chosen edges. A chosen edge either reaches a new vertex or ends
the current growth upon reaching an earlier growth, so no retained
edge is selected twice.

The only growths that create new terminal components in this pass
are those stopping with a large heap or an empty heap. A growth attaching
to an earlier one adds no component. A growth with empty boundary
covers a whole connected quotient and consequently accounts for at
most one component. Each large-heap growth has more than $B$ distinct
boundary edges. Charge each to its scanned endpoint. Scanned vertex
sets of different growths are disjoint, so no edge is charged more
than twice. This proves the vertex bound.

Finally each vertex is
scanned once and every heap extraction uses a heap of size at most
$\min(k',B)$.

There is at most one extraction per scanned vertex.
\tref{Lemma}{lem:heap-bound} gives the claimed pass cost. Graph searches
in $F$ and a scan of all edge endpoints construct the quotient.
\end{proof}

\begin{lemma}\label{lem:mst}
On a connected undirected multigraph with $k\ge1$ vertices and $l$ edges,
a minimum spanning tree can be constructed in $O(l+k\tau(k))$ time.
In any minimum spanning tree $T$, $b_T(x,y)$ is the least possible
maximum edge length of an $x$--$y$ walk for distinct $x,y$. In
particular $b_T(x,y)\le w(xy)$ for every nonloop edge $xy$.
\end{lemma}
\begin{proof}
At infinite capacity a growth from one vertex can stop only after
scanning its whole connected graph. Thus the three passes connect
all input vertices using only edges of the tree in
\tref{Lemma}{lem:growth-pass}, and return that tree. The successive
vertex counts are at most $k$, $1+2l/B_1$, and $1+2l/B_2$. The sum of
the extraction terms is
\begin{align*}
 O\bigl(k(1+\log(B_1+1))
 +(1+l/B_1)(1+\log(B_2+1))
 +(1+l/B_2)(1+\log(k+1))\bigr)=O(k\tau(k)+l).
\end{align*}
The three scans cost $O(k+l)$. For the bottleneck assertion, deleting a
heaviest edge of $T[x,y]$ separates $x$ and $y$. Every walk between them
crosses this cut, where no edge can be strictly lighter than the
deleted edge by minimality of $T$. The tree path attains the bound.
\end{proof}

\subsection{Disjoint Sets for Ordered Merge Events}
Initially each of $k$ items is a root of rank zero. A find returns
the root of its item's parent chain and resets the parent of every
nonroot on this chain to that root. To unite two sets, find their
roots and do nothing if they agree. Otherwise attach the root of
smaller rank to that of larger rank. On a tie attach either root
and raise the survivor's rank by one. If merge events have a given
order they are processed in that order. Ordering the events, when
necessary, is charged by the calling construction.

\begin{lemma}\label{lem:unions}
On $k\ge1$ items, initialization and $Q$ union or find operations
take $O(Q+k\tau(k))$ time in total.
\end{lemma}
\begin{proof}
The parent pointers preserve the represented sets. Ranks increase
strictly along parent chains. A root first attaining rank $r$ has
at least $2^r$ original items, by induction on promotions.

The sets
at first promotions of distinct roots to $r$ are disjoint. Once
a member reaches a rank-$r$ root, it stays with that root or belongs
to a future root of rank at least $r$ that has already reached rank
$r$. Consequently at most $k/2^r$ vertices reach rank $r$ and all
ranks are at most $\lfloor\log_2 k\rfloor$.

Put $b_3=1+L_k$, $b_2=\lceil\log_2(b_3+1)\rceil$, and
$b_1=\lceil\log_2(b_2+1)\rceil$. Divide ranks into the three
half-open bands with boundaries $0,b_1,b_2,b_3$. During one find,
charge the last edge and every edge crossing a band boundary to
the operation. There are at most three such edges.

Each other
traversal from a vertex $u$ goes to a parent in its own band that
is not the root. Compression strictly increases the rank of $u$'s
parent. Once $u$ is a nonroot its own rank is fixed, so for a band
of width $w$ there are at most $w$ such charges to $u$.

The sum in the first band is at most $kb_1$. For either higher band
$[b_i,b_{i+1})$ it is at most $(k/2^{b_i})b_{i+1}\le k$. Initialization
and the at most two finds of each union give the result. This elementary
fixed-band analysis suffices here for the disjoint-set operations
studied by Tarjan~\cite{DBLP:journals/jacm/Tarjan75}.
\end{proof}

\subsection{Lowest Common Ancestors and Level Ancestors}
\label{sec:tree-queries}

The \emph{lowest common ancestor} of $u,v$ is their common ancestor of
greatest depth. A \emph{level ancestor} query gives a vertex and a
desired ancestor depth. Harel and Tarjan studied the first problem~\cite{DBLP:journals/siamcomp/HarelT84}.
Bender and Farach-Colton established linear preprocessing and constant
query time for the second~\cite[Theorem~13]{DBLP:journals/tcs/BenderF04}.
We give both constructions since their linear initialization is charged
each time a hierarchy is built.

Augment the input forest by a common root and record the original component
of every vertex. Increase every original target depth by one when querying
the augmented tree. For this subsection let $k$ count the augmented vertices,
take $B=\max(2,\lfloor\log_2(k+1)/8\rfloor)$, and fix an order of the
children. Constant-size inputs can be tabulated directly.

List the vertices of a depth-first tour, starting at the root and
recording the reached endpoint after each traversal of an edge.
Store the first occurrence of each vertex. Divide the list into blocks
of $B$ consecutive positions. The $B-1$ signs of the changes of depth
encode a block's relative depths. Complete a short final block with
arbitrary signs. For every sign string tabulate the first minimum-depth
position of each subinterval. Also tabulate the first minimum on each
sequence of $2^i$ whole blocks at each possible start by using the two
half-length answers at the preceding level. Store floor logarithms of
integers through $2k$ in an array as in \tref{Section}{sec:prelim}.

To answer a query in one tree, examine the tour interval between the
two first occurrences. Use the sign table for its one or two partial
blocks. For $h>0$ whole blocks in the middle, inspect the first and
last intervals of $2^{\lfloor\log_2 h\rfloor}$ whole blocks. Return the
vertex of minimum depth among these constantly many answers.

\begin{lemma}\label{lem:lca-primitive}
The tour tables return lowest common ancestors in constant time after
$O(k)$ preprocessing time and space.
\end{lemma}
\begin{proof}
The relevant tour interval stays in the subtree of the lowest common
ancestor and visits that ancestor on the way between the two vertices.
Its minimum depth is therefore attained precisely at that vertex.
Each tour step changes depth by one, so the sign tables are valid
independently of a block's initial depth. The two whole-block intervals
cover their union of $h$ blocks because $2\cdot2^{\lfloor\log_2 h\rfloor}\ge h$.
Tie-breaking by the first position is consistent under taking minima.

The tour has $2k-1$ entries. The sign tables take $O(2^B B^3)=O(k)$
operations by direct enumeration. The whole-block tables have
$O((1+k/B)\log(k+1))=O(k)$ entries. Constructing the codes and mapping
positions back to vertices takes one scan.
\end{proof}

Process vertices in postorder. Return to a vertex's parent the size of
its residual subtree, consisting of the vertex and its children's
residuals. If the sum reaches $B$, mark the vertex and return size zero.
Mark the augmented root as well. Deleting the marks partitions unmarked
vertices into rooted ordered trees. Encode the shape of each such tree
by writing one entering bit and one leaving bit for each vertex during
its depth-first traversal. Prefix a leading one to distinguish lengths.
For every valid shape on fewer than $B$ vertices, tabulate the ancestor
of every preorder position at every allowable relative depth. Save
the map between local preorder positions and original vertex indices.

At each mark $p$, store its ancestor $2^i$ edges above it whenever this
ancestor exists. During a depth-first traversal, a stack indexed by
depth gives all such ancestors in one lookup each. Independently,
compute the height of each vertex in postorder, measured in edges.
Select a child of maximum height at every nonleaf. The selected
edges partition the vertices into paths. Store each path in depth
order, extending its array upward by its own number of vertices or
as far as the root, whichever is shorter. Each vertex stores its
position in its own unextended path's extended array.

Given $(v,j)$, if $v$ is unmarked and $j$ is at least the depth of
its component root, use its shape table. Otherwise start at $p=v$
when $v$ is marked, or let $p$ be the marked parent of its component
root. Put $d=\operatorname{depth}(p)-j$. At $d=0$ return $p$.
Otherwise jump up $h=2^{\lfloor\log_2 d\rfloor}$ edges using $p$'s
table and call the result $x$. Return the array entry $d-h$ positions
above $x$ in its extended path.

\begin{lemma}\label{lem:marked-queries}
Deleting the marks leaves components of fewer than $B$ vertices. Their
shape tables and the tables and extended paths at marks take $O(k)$
time and space altogether. The indicated two lookups from a mark return
the requested ancestor.
\end{lemma}
\begin{proof}
Each postorder mark pays for at least $B$ residual vertices removed
from future sums. There are at most $k/B$ such marks. For each unmarked
component, its top vertex returned its size, which is strictly less
than $B$. Thus its shape needs fewer than $2B$ bits before the leading
length delimiter. Enumerating strings of length at most $2B$, checking
balanced traversals, and tracing parent chains for their valid entries
costs $O(2^{2B}B^3)=O(k)$. Depth comparison detects whether the desired
ancestor remains in the component. If it lies above the component,
its top vertex's parent is marked and lies on the query path.

The mark tables occupy $O((1+k/B)\log(k+1))=O(k)$ entries. The original
selected paths partition the vertices, so their upward extensions at
most double the total array length and can be constructed by walking
parent pointers. After the jump, $p$ is a descendant exactly $h$ edges
below $x$. Following maximum-height children from $x$ reaches a leaf
at the height of $x$, at least $h$. Its selected path therefore has at
least $h+1$ vertices.

The extension covers any existing ancestor of
$x$ within $d-h<h$ edges. Array indexing gives precisely the target.
\end{proof}

\begin{lemma}\label{lem:trees}
A rooted forest on $k$ vertices supports lowest common ancestor and
level ancestor queries in $O(1)$ time after $O(k)$ preprocessing time
and space. For the former query a component test detects inputs in
different original trees. For the latter, the requested depth must
lie between zero and the original query vertex's depth.
\end{lemma}
\begin{proof}
The augmentation adds one vertex. Apply \tref{Lemmas}{lem:lca-primitive}
and~\ref{lem:marked-queries} and translate back from the augmented depths.
\end{proof}

\subsection{Interval Minima Under Splits and Decreases}
\label{sec:interval-minima}

Initially a fixed array of $k\ge1$ records is one interval. A split
cuts an interval at a prescribed boundary into two nonempty intervals.
A decrease lowers a key. A query takes a record and returns the minimum
record of its present interval. Ties are resolved by original index.
Pettie and Ramachandran give inverse-Ackermann bounds on this
split--findmin interface~\cite[Theorem~4.2]{DBLP:journals/siamcomp/PettieR05}.
The following fixed number of blocking steps suffices for the desired
bound.

Pad the array to a power of two with unused positions and build a
balanced binary tree of intervals. Represent each live interval by
the ordered maximal tree nodes contained in it. Ignore unused positions.
Each such node stores its minimum and its enclosing interval descriptor.
Each element points to its node. The descriptor stores the minimum of
its nodes. On a decrease, compare with both stored minima. A lookup
from a record follows its node's descriptor pointer.

At a split, at most one represented node straddles the cut. If there
is one, follow the cut path to expose the maximal nodes on its left
and right and scan their member records to initialize their pointers
and minima. Split the old ordered list, splicing in the exposed lists.
Scan the two new lists to assign interval descriptors and recompute
their minima.

\begin{lemma}\label{lem:dyadic-minima}
The dyadic representation initializes the array and performs all its
splits in $O(k(1+\log k))$ aggregate time. A decrease, minimum query,
or descriptor lookup costs $O(1)$ time.
\end{lemma}
\begin{proof}
A maximal node can occur at each depth only at the ends of a run of
fully contained nodes because two siblings fully contained in the interval
are replaced by their parent. There are at most two maximal nodes per
depth, one beside each of the interval's two boundary paths. Thus
each list operation and scan takes $O(1+\log k)$ time apart from the
element rescans. Each rescan moves its element to a strict descendant
of its old node, so a fixed element pays for at most $O(1+\log k)$
rescans. At most $k-1$ splits make nonempty intervals. The maintained
minima and pointers give the individual operation bounds.
\end{proof}

Divide the array into consecutive blocks of size
$B=\max(2,\lceil\log_2(k+1)\rceil)$, allowing a final short block.
Use a lower structure in each block for the intersections of live
intervals with the block. An upper structure has one proxy per original
block. While the block is whole its proxy key is the pair consisting
of its least original key and the minimizing original index.

Each live interval consists of at most two proper end fragments and
one run of whole blocks. Keep that run as one upper interval. Once
a block is split internally its proxy is permanently isolated and
ignored. Each proper fragment's lower descriptor and each live
whole run's upper descriptor point to their owning interval. That
interval's descriptor gives the endpoints, the descriptors of its at
most three pieces, and the minimum of their minima.

\begin{enumerate}
\item On a lookup from a record in a split block, use its lower
descriptor to reach the owner. In a whole block use the upper
descriptor of its proxy. Return the owner's descriptor or minimum.
\item On a decrease, update the lower structure. In a whole block also
decrease the proxy if the minimizing pair has decreased. Refresh the
owner's minimum from its pieces.
\item At a cut between blocks, split the upper whole run if necessary.
At an internal cut, split the relevant lower interval. If the block
was whole, isolate its proxy by splitting off its nonempty left and
right neighbors in the upper run. Form the two new owner descriptors
from their endpoint indices. These identify any proper end fragments
and the first and last whole blocks. A lookup at an element or proxy
of each nonempty piece gives its descriptor. Reset its owner pointer
and recompute the two minima.
\end{enumerate}
An interval lying wholly inside one block uses one fragment if it is
proper and one proxy if it is the entire block. Two neighboring partial
blocks can have a common interval without any whole block between them.

\begin{lemma}\label{lem:sequences}
On $k$ records, initialization, all splits, $q$ decreases and $q'$
queries take $O(q+q'+k\tau(k))$ total time. Queries, decreases and
descriptor lookups each take worst-case constant time.
\end{lemma}
\begin{proof}
The pieces partition each logical interval. Its smallest key is their
least minimum. After a cut the only potentially broken whole run
is the run containing the cut. After a decrease the only changed
piece is the one containing the decreased record. The updates therefore
maintain the representation, with a bounded number of calls to the
upper and lower structures per operation.

Use \tref{Lemma}{lem:dyadic-minima} for the upper array of
$k'=\lceil k/B\rceil$ proxies. Its initialization and all splits take
$O(k'(1+\log k'))=O(k)$. If the lower structures are also dyadic,
their initialization and splits take $O(k(1+\log B))$ altogether.
This first blocking step gives $O(k(1+\log\log(k+4)))$ in aggregate
in addition to constant time per decrease or query.

Now keep the dyadic upper structure, but use the first blocking step
in each lower structure. Their preparation and splits sum to
$O(k(1+\log\log(B+4)))=O(k\tau(k))$. The construction makes only
two blocking steps. Thus the total number of primitive lookups and decreases
is $O(q+q'+k)$ and each top-level lookup or decrease has a constant
number of primitive operations. This proves the stated bounds.
\end{proof}

\section{Evaluation of an Ordered Hierarchy}\label{sec:eval}

The evaluator works in a directed graph of states. Its input certifies short connections in an undirected physical metric. This distinction permits a state to receive a finite upper bound even when no state path from the seed reaches that particular copy.

\begin{definition}\label{def:hier}
An \emph{ordered hierarchy} consists of a finite directed graph $D$ with nonnegative arc lengths, a map $\pi$ from states to points of an undirected physical graph, and a rooted forest whose leaves are the states. Each state arc $xy$ projects to a physical walk from $\pi(x)$ to $\pi(y)$ of length at most its arc length. Each internal node $C$ has at least two children and a positive height $h(C)$, nondecreasing towards the root. The arcs between distinct children of length strictly below $h(C)$ form an acyclic quotient. The quotient of the roots on all interroot arcs is acyclic. For each internal node, its children are connected by a tree whose links specify two endpoint states and certify physical walks between their images of length at most $c_0h(C)$, for an absolute $c_0$. These walks can be traversed in either direction. A physical link used solely for this certificate never becomes a relaxed arc.
\end{definition}
\begin{definition}[Seeded ordered hierarchy]\label{def:seeded}
A \emph{seeded ordered hierarchy} is an ordered hierarchy with weighted physical sources $(v_i,c_i)$ and initial state offers. Its target distance is
\begin{align*}
 \delta(v)=\min_i\{c_i+d(v_i,v)\}.
\end{align*}
An offer is a finite value. $+\infty$ marks the absence of an offer. A feasible offer at state $x$ has value at least $\delta(\pi(x))$. All initial offers must be feasible. A \emph{tight witness} is a directed state path whose initial offer plus its arc lengths equals $\delta$ at its last image. A state with such a witness is \emph{protected}. For every requested finite target, at least one state mapping to it must be protected. An ordinary source $s$ corresponds to the single pair $(s,0)$.
\end{definition}
Every prefix of a tight witness is tight. Otherwise, a shorter physical route to its prefix endpoint followed by the projected suffix would contradict the definition of $\delta$. We use this observation whenever we follow a witness into a child.

\subsection{Normalization}
For each internal node $C$, take the supplied tree of physical links on
its children. For each link, store an undirected edge between its two
named endpoint states with its certified length bound (at most
$c_0h(C)$). The union of these edges is a forest $F$ on the states. At each node,
the supplied links connect its disjoint child trees into one tree. A node with $d$ children contributes
$d-1$ edges, so this takes $O(N)$ time for $N$ states. In particular,
a zero-length physical link between two copies contributes to $F$.
The directed state arcs supply the child-order constraints below.

We call an internal hierarchy node a primary node. A leaf has mass zero. Give a node $C$ mass $S(C)$ equal to the total lengths of its selected
forest edges and the heights of internal nodes throughout its
subtree. The forest path between two states below $C$ has length at most $S(C)$, and the summand $h(C)$ ensures $S(C)\ge h(C)$.
Join each internal child $C$ to its parent $P$ when $4S(C)\ge h(P)$.
Call a parent--child link satisfying this test a \emph{normalization link}. Within a connected set of such links start a table of powers of two at
its least height and round each $h(C)$ down on this table,
to $\nu(C)\in(h(C)/2,h(C)]$. The table need only reach the largest
height in the set. At an unlinked boundary
$\nu(P)>2S(C)$. In particular the units are monotone, and either the
child unit divides the parent unit or twice the child's mass is smaller
than the parent unit.

\begin{lemma}\label{lem:mass}
Given an ordered hierarchy with $N$ states and $M$ arcs, normalization takes $O(M+N)$ time. It assigns a unit $\nu(C)$ and a physical diameter bound $S(C)$ to each surviving internal node. Parent units are at least twice child units, and $\sum_C S(C)/\nu(C)=O(N)$. At a child $C$ with parent $P$, their units are dyadically aligned or $2S(C)<\nu(P)$. Here dyadic alignment means that the parent unit is a power-of-two multiple of the child unit.
\end{lemma}
\begin{proof}
Write $M$ for the number of arcs and $N$ for the number of states. These tables and their indices take $O(N)$ operations. For a table point
$t$ crossed on a link from $C$ to $P$, charge every mass item $z$ below
$C$ the amount $4z/t$. The total charge is at least one.

Along the ancestors
of a fixed item in one table the crossed points at least double. After
a table break, the first higher point charged exceeds four times the
previous subtree mass. The first charged point is greater than the height
at the item's node, which is at least its value divided by
$\max(1,c_0)$.

Thus each item pays $O(1)$. Doubling from the minimum node
to the root of a linked set generates its whole table. Moving downward
from the root table index, stopping when the child's height is reached,
uses at most the counted crossings plus $O(1)$ operations per link.

Contract links between primary nodes of equal unit. An arc of length
less than the common unit still runs forward in a topological order of
the resulting children. Indeed a cycle would project to a directed
closed walk in the highest original sibling quotient it crosses. The
forest $F$ still connects the vertices of every surviving node. Use its
topmost original mass. Units now at least double from child to parent.
Charging each item to all surviving ancestors gives
\begin{align*}
\sum_C S(C)/\nu(C)=O(N).
\end{align*}
Discard state self-arcs in the initial scan. Removing a traversal from a state back to itself can only decrease its cost, and a zero-length self-arc can be omitted from any tight witness. Send an arc between different roots directly to the root quotient.
Locate each remaining original arc at its new lowest common ancestor. Topological
sorting there supplies the child ranks for arcs of length less than
$\nu(C)$, in total $O(M+N)$ time. The same sorting argument applies to
the roots. One-state roots just process their best finite offers.
\end{proof}
\subsection{Ordering the Children}
Fix a preorder of $F$. In a primary $X$ of degree $d$, provisionally
choose one state in each child. Run directed depth-first search on
the light child DAG, launching searches at unvisited children in the
$F$ preorder of the provisional representatives. A DFS root $i$ is
entered at its provisional representative $x_i$. A DFS tree arc
enters its head child $i$ at its actual head state $x_i$. At $i$,
inspect the outgoing light arcs in the cyclic $F$ preorder of their
tail states, starting at $x_i$. Ties at a state may be arbitrary.
The final representative of child $i$ is $x_i$.

\begin{lemma}[Tour of the child representatives]\label{lem:tour}
At an internal node $X$ of a normalized ordered hierarchy, let $d$ be the number of children, $\nu(X)$ its unit, and $S(i)$ the certified forest mass of child $i$. The ordered DFS traversal in the preceding algorithm admits a physical walk through the representatives in reverse finishing order of length at most $2\sum_i S(i)+2d\nu(X)+2S(X)$.
\end{lemma}
\begin{proof}
The reverse DFS finishing order is topological. It also admits short
certified physical walks between consecutive final representatives.
Charge each traversal of a forest edge its stored connector bound and each traversal of a selected state arc its arc length. In either direction the charged amount bounds the length of a physical walk between the endpoints. Traverse
only the selected DFS tree arcs, returning over each after its head's
search finishes. Within a child move between the selected exit tails
along $F$, starting and ending at its entry $x_i$. The cyclic preorder
restricted to the connected subtree $F[i]$ follows a closed tree tour,
so these moves cost at most $2S(i)$ together.

Mark $i$ on its final
return to $x_i$. At successive DFS roots, move between their provisional
representatives along $F$. These roots occur in the original preorder,
so those moves cost at most $2S(X)$. Every DFS tree arc costs less
than $\nu(X)$ per traversal and is used twice. The total charge is at most
\begin{align*}
2\sum_{i\in\mathrm{children}(X)}S(i)
          +2d\nu(X)+2S(X).
\end{align*}
Reverse the traversal and use its segments between successive marks.
For consecutive representatives $i,j$ in this order, let $g(i,j)=g(j,i)$ be the total charge between their marks. Concatenating the certified walks gives a physical walk of length at most $g(i,j)$ in either direction.
\end{proof}
\begin{lemma}\label{lem:gaps}
In a normalized ordered hierarchy with $N$ states and $M$ arcs, consider the child graph of each internal node $X$ restricted to arcs shorter than its unit $\nu(X)$. The traversal above finds a topological order, a representative per child, and a physical walk bound $g(i,j)$ between consecutive representatives. Over all these consecutive pairs, $\sum_X\sum_{i,j}g(i,j)/\nu(X)=O(N)$, and the orders and gaps can be computed in $O(M+N)$ time.
\end{lemma}
\begin{proof}
The segments between successive marks use disjoint portions of the traversal. Their gap bounds therefore sum to at most the total charge in \tref{Lemma}{lem:tour}. At each $X$, the child masses sum to at most $S(X)$. Divide by $\nu(X)$ and sum over $X$: \tref{Lemma}{lem:mass} bounds the mass terms by $O(N)$, and the child counts sum to the number of hierarchy edges.
The construction does not walk long physical paths explicitly. One
lowest-common-ancestor query and root sums give each $F$-distance.
Accumulate the costs in DFS event order and take the differences of
consecutive mark totals.

To order the DFS starts, counting-sort one provisional representative per child by parent identifier and then forest preorder as a lexicographic pair. Scanning each parent's resulting list and skipping already visited children gives its DFS roots in the required order in $O(N)$ total time.

To locate the cyclic adjacency list start at
an arbitrary entry $x_i$ in constant time, sort outgoing tails,
possible incoming entry positions, and provisional representatives
by the integer triple consisting of the parent identifier, child
identifier, and $F$ preorder. For each entry position in a child with outgoing light arcs, store the
successor tail index, wrapping at the end of the list. For a child with
no outgoing light arcs, store an end marker at every entry position.
On reading that marker, DFS finishes the child immediately. Each arc is
assigned only to its lowest-common-ancestor quotient. Counting sorts,
DFS, and all the walk certificates therefore use $O(M+N)$ work.
\end{proof}
\subsection{Visiting Slots and Expiry}

Process roots in condensation order. Each node has an unopened, live, or retired status. An unopened interval in the leaf order stores its minimum tentative value. A leaf accepts decreases until it emits once or retires. Minima are ordered lexicographically by value and leaf identifier. Nominal windows are half open. Permanent deadlines are inclusive. A timestamp is due precisely when it belongs to the current window and satisfies all permanent deadlines.

A primary node first opens when visited with a finite tentative minimum
$D_0$ in the parent's actual window. More precisely, in a nominal
slot $[a,b)$ let $c$ be the minimum inherited physical deadline,
using $+\infty$ when none is present. A candidate $z$ is due precisely
when $a\le z<b$ and $z\le c$. \tref{Lemma}{lem:evalcorrect} proves the lower bound for pending candidates. The node's absolute expiry time is
$D_0+S(C)$. All physically tight arrivals into $C$ occur no later than
this time, by the physical triangle inequality. Keep its scheduler active
until all leaves are settled or retired, or it has covered its expiry,
inclusive. An empty queue may receive later offers. The local
slot origin can be chosen by the following form of the alignment rule of
Pettie and Ramachandran~\cite{DBLP:journals/siamcomp/PettieR05}. If $D_0+S(C)<b$, start at $D_0$. Otherwise
start at the largest $b-j\nu(C)$ no larger than $D_0$. The necessary
subtractions can be performed as signed pairs of sums. Finding the
integer $j$ by successive steps costs $O(1+S(C)/\nu(C))$ in this case.
For a root use the first alternative. At an unlinked boundary the mass
is less than half the parent unit and the child unit is at most half
the parent unit, so this origin is inside the parent's slot. At a linked
boundary divisibility gives the same conclusion. No future boundary cuts
a local slot while physically relevant events remain beyond that boundary.
Divisibility gives alignment, or the node expires before the next parent
boundary. If the last nominal slot overshoots a bound on a subtree that
expires in this call, propagate the bound as an additional timestamp cap
and suppress emissions beyond it. Always align to the nominal slot end,
keeping this cap separate from the grid calculation. This keeps all actual emissions inside
the incoming window. An expiry value itself is included. A parent window's
right endpoint is excluded. The number of local slots and unsuccessful
last checks over all primary nodes is $O(N)$ by the mass bound.

Use a hierarchy-leaf DFS layout respecting these topological child
orders. At an open primary $X$ keep a linked list containing its
registered live children and the maximal nonempty contiguous runs
of its still unregistered children, in the original order. An expired
or completed child is unlinked. A run is kept at its own position
even if the markers that created it later disappear. Adjacent runs
separated by a removed child need not be merged. Child intervals,
including the union of the children in one run, are intervals in
the leaf layout. Maintain the current minimum tentative leaf in each
such unopened interval by split--findmin. On first opening $X$, its
whole child list is one run with the known first finite minimum.

Scan the live list in increasing topological order during a nominal
slot $[a,b)$ of $X$. Visit a registered child with the usual incoming
window. When the cursor reaches a run with no due minimum, pass it
for this slot. Otherwise take a minimizing state with feasible label
$z<b$ and let $i$ be its immediate child of $X$. Replace the run,
in its current list position, by its nonempty left subrun, child $i$,
and its nonempty right subrun in that order. At most two interval
splits isolate the states below $i$. An integer level-ancestor query
identifies $i$ from the minimizing leaf. Keep the cursor on the first
new piece so all earlier ranks have their opportunity before $i$ runs.
The child may receive further decreases before that visit.

For each newly created subrun, let $j$ be its child adjacent to $i$.
Post a supplemental feasible offer at the state $x_j$ of value
\begin{align*}
u=\max\{b,\ z+S(i)+g(i,j)\}.
\end{align*}
The offer is sound. By feasibility, some source $(v_k,c_k)$ has a path to the image of the minimizing leaf whose length plus $c_k$ is at most $z$. Travel to $x_i$ inside $F[i]$ at
cost at most $S(i)$ and use the gap walk to $x_j$. An upper bound may
be padded up to $b$. In particular the supplemental offer cannot
create a due event at an earlier rank in this same slot. No state
has been settled simply by posting it. An original relaxation remains
eligible immediately, including a decrease at a child that has been
registered and is still awaiting its rank.
\begin{algorithm}[ht]
\caption{Scanning a live node in topological order}\label{alg:visit}
\Fn{\textsc{Scan}($X,I,c$)} {
\KwIn{A live node $X$ of a normalized ordered hierarchy (\tref{Lemma}{lem:mass}), a portion $I$ of one nominal slot $[a,b)$, inclusive deadline $c$.}
\KwOut{Emissions and pending offers after this portion of the slot.}
\ForEach{list piece in cursor order (\tref{Lemma}{lem:tour}), including pieces inserted at the cursor}{
 \eIf{the piece is an unregistered run}{
  \If{its interval minimum $z$ (\tref{Lemma}{lem:sequences}) lies in $I$ and $z\le c$}{
    Split at the child containing the minimizing leaf and register that child\;
    Post the two boundary offers from the gap certificates (\tref{Lemma}{lem:gaps}) when the corresponding subruns exist\;
    Resume at the first piece just created\;
  }
 }{
  Open the child if necessary, with its least due offer\;
  Visit its live slots intersecting $I$ through its inclusive deadline\;
  At a leaf emit its due value once and relax all outgoing arcs\;
 }
}
Retire completed pieces and pieces whose permanent deadlines have been covered\;
}
\end{algorithm}

For an ordinary invocation $I$ is the nominal slot intersected with the window supplied by its parent. When a child has already opened, its clock persists across visits. The first check at a registered leaf compares its current tentative value with the permissible window. At a root the window is initially unbounded and the origin is the minimum seed or incoming offer. Skip a root without any finite offer. Place settled leaves in singleton intervals. At expiry, walk only the pieces still linked to the expiring node. Recurse on linked registered children, and traverse the untouched child subtrees of each remaining unregistered run. Mark each newly retired leaf and unlink each disposed piece. Later offers check the leaf's settled or retired flag before updating its key. Every visited live node advances through all its local slots intersecting the current visit, even if no new offers arrive. At temporary boundaries the resumption rule of \tref{Lemma}{lem:clip} applies.

\begin{algorithm}[ht]
\caption{Evaluating all roots of a seeded ordered hierarchy}\label{alg:evaluate}
\Fn{\textsc{Evaluate}($D,\mathcal H,\mathcal Q$)}{
\KwIn{State graph $D$, ordered hierarchy $\mathcal H$, and finite initial offers $\mathcal Q$.}
\KwOut{The state emissions.}
Normalize the hierarchy, construct the topological child orders and gaps, and initialize the leaf layout with infinite keys (\tref{Lemmas}{lem:mass} and~\ref{lem:gaps})\;
Split the layout at every root boundary and post the least initial offer at each state that receives one\;
\ForEach{root $X$ in topological order of the root quotient}{
 Skip $X$ if its least pending offer is infinite\;
 \eIf{$X$ is a leaf}{
  Emit its minimum and relax its outgoing arcs\;
 }{
  Open $X$ at its least pending offer $D_0$, with expiry $c=D_0+S(X)$\;
  \ForEach{successive width-$\nu(X)$ slot $I$ beginning at $D_0$ or later, through the slot containing $c$}{
   Run \algcall{alg:visit}{Scan} on $(X,I,c)$, stopping if all its children finish\;
  }
 }
}
\Return{the emitted state--value pairs}\;
}
\end{algorithm}

\begin{lemma}\label{lem:evalcorrect}
In a seeded ordered hierarchy, \tref{Algorithm}{alg:evaluate} emits $\delta(\pi(x))$ at every protected state $x$. Every other emission at $x$ has value at least $\delta(\pi(x))$.
\end{lemma}
\begin{proof}
Upper bounds are preserved by addition along arcs and by the gap certificates. For correctness, use induction over roots, then chronological visit portions, and within each portion over child rank and recursively over remaining height. The inductive assertion applies to every protected prefix in the current portion. The last entry of its tight path into the current node has been posted, or the continuation after that entry has already been posted by an emission. The induction has processed tight predecessors in earlier roots and earlier portions of the same node. Within the current portion it has processed tight predecessors in children below the current rank. At a root, the seed and all entries from earlier roots are available.
\begin{enumerate}
\item In a portion $[u,v)$ at $C$ (possibly also subject to a permanent cap), follow a tight path to its last entrance into the target child. For an interchild entrance with predecessor time below $u$, use the induction over earlier portions. If the predecessor time is at least $u$, the crossing costs less than $v-u\le\nu(C)$. Its child has smaller rank and its tight predecessor has emitted. A seed or an entry from outside $C$ was available by the call invariant. Every prefix timestamp is at most the target timestamp, hence respects its cap.
\item Continue on this path past any already emitted tight states. The first outstanding state on the remaining internal suffix has a due offer or was due earlier. An unopened child cannot have such an unresolved earlier tight arrival, since the same last-entry argument applies at the earlier portion. Its run thus has a due minimum by the time its rank is needed.

Nonnegative costs ensure that all live offers respect elapsed windows in the ordinary traversal. A backward child arc of a slot forces any newly generated value to be at least the nominal right endpoint. Boundary offers on a split are padded to that endpoint. Splitting at a later child first retains and scans the left subrun.

Inside a child apply the same argument one level lower. At a leaf a tight offer and the upper-bound invariant force equality.
\item At first opening, a strictly earlier tight arrival than the current minimum would already have a tight offer at the first outstanding state of its internal suffix. Thus skipping to the prescribed origin loses no required state. The physical diameter bound puts every required timestamp in $C$ at or below $D_0+S(C)$. Retire only after covering that value, including equality, or reaching an inherited permanent cap known by the same argument. An excluded upper endpoint is processed in the next portion. This closes the induction and proves that each target has a tight output.
\end{enumerate}
\end{proof}
\begin{lemma}\label{lem:evaltime}
Given a seeded ordered hierarchy on $N$ states and $M$ arcs, ordinary evaluation, including normalization and all supplemental offers, takes $O(M+N\tau(N)+Q)$ time for a list of $Q$ initial offers. An input scan keeps the minimum initial offer at each state.
\end{lemma}
\begin{proof}
A new subrun has a finite actual tentative value at its boundary
representative of at most $u$. It either splits by the slot containing
$u$ or disappears under an ancestor cap first. Since $z<b$, its
unsuccessful slot visits number at most
\begin{align*}
O\bigl(1+(S(i)+g(i,j))/\nu(X)\bigr).
\end{align*}
The original run splits on first opening. A child creates at most
two subruns. A gap in the topological list is charged at most twice.
The sum of these visit bounds is $O(N)$, as is the total number of
splits. Visits to a registered child $i$ cost
$O(1+S(i)/\nu(X))$ in its parent's list and hence $O(N)$ in all.

Disposal visits each still-linked piece once. Removed children are absent from subsequent ancestor traversals, and an untouched subtree is traversed only when its enclosing run is discarded. There are $O(N)$ pieces and hierarchy nodes in total, so retirement costs $O(N)$ and rejecting a later offer costs $O(1)$. There are $O(N)$
supplemental decreases and $O(M)$ original relaxations.

Standard
unopened-interval supervision stores the interval identifier at its
chosen minimum, breaking ties by state name. Querying before and
after a decrease updates that identifier, and splitting creates only
constantly many new interval records. Registered but unopened nodes
keep their own interval until visited. No ordered integer dictionary,
real queue of children, or multiplicative factor for scanning light
arcs is needed in the evaluation bound.

\tref{Lemma}{lem:sequences} gives the split--findmin bound for all operations on these intervals. The leaf layout makes every interval split a split of one sequence. Physical leaves need no searches for their containing run on ordinary offers. It suffices to update the tentative leaf and its split--findmin key. At the cursor the run owns a pointer to a sample leaf in its interval for the next minimum query. All full-list scans, including the traversal over a freshly split run, are included in the indicated successful and unsuccessful visits.
\end{proof}

\section{A Local Hierarchy for Arbitrary Positive Lengths}\label{sec:base}

This section constructs a representation that will be used only inside small subgraphs. Write $N$ and $m$ for their local sizes. The cost $O(m+N\log\log(N+4))$ is affordable there. We give the unrestricted construction because a bound dependent on the ratio of the extreme edge weights would be insufficient even on a four-vertex input. The owner decomposition also applies at the outer scale, with a different choice of $b$.
\subsection{Owner Decomposition}
Write $L_N=\lceil\log_2(N+2)\rceil$. Let $G=(V,E)$ be the connected positive graph, $N=|V|$, and $T$ its MST.
Zero-length components may be contracted before this construction.
Write \(D_T\) for tree path length and \(b_T\) for the largest edge
weight on a tree path. The MST cut property gives
\(b_T(u,v)\le w(uv)\) for every graph edge. Delete each non-tree edge
of weight strictly greater than \(D_T(u,v)\), since its tree path is a replacement
of no greater length. Tree distances use a lowest-common-ancestor
structure and root distances. All remaining physical edges obey
\(w(uv)\le D_T(u,v)\).
Take \(b=L_N^4\). Choose a nonempty set \(A_0\) meeting every
connected \(b\)-vertex subtree of \(T\). One linear-time construction
processes a rooted tree in postorder, sums the residual sizes of the
children plus one, and marks the vertex whenever this sum reaches \(b\).
A mark returns size zero to its parent. The marked sets paid for by
successive marks are disjoint, so there are at most \(\lfloor N/b\rfloor\)
marks. If no mark was made, use one pivot. Thus
\(|A_0|\le\max\{1,\lfloor N/b\rfloor\}\). Each remaining component has fewer than
\(b\) vertices. Add the branch vertices of the minimal subtree
spanning \(A_0\), obtaining \(A\) with \(|A|=O(1+N/b)\).
Every component of \(T-A\) has at most two distinct pivot neighbors.
Indeed, three such neighbors would have a Steiner branch vertex
inside the component, and every branch vertex of the subtree spanning
\(A\) already belongs to \(A\).

\begin{definition}[Owner decomposition]\label{def:owner}
An \emph{owner decomposition} of a connected positive-weight graph with a minimum spanning tree $T$ and separator size $b$ consists of the pivot set $A$ just constructed, its pivot-deleted components, and the owner cells obtained by the following corridor cuts. Each pivot-deleted component has fewer than $b$ vertices and at most two pivot neighbors. We use $p(v)$ for the unique pivot of the owner cell of $v$, $R_v=D_T(v,p(v))$ for its tree distance, and $r_v=b_T(v,p(v))$ for its tree bottleneck. Write $h=b+1$ and $\sigma(p,q)=b_T(p,q)$ for pivots $p,q$. The reduced graph omits each nontree edge heavier than its tree path.
\end{definition}
For the outer invocation replace the displayed $b$ by its specified separator size. For every two-pivot component, delete a maximum-weight tree edge
on the corridor between its two pivots. For two adjacent pivots,
delete their common edge. The components after these deletions are
connected owner cells, each with one pivot. Write \(p(x)\) for the
pivot of \(x\)'s cell and set \(r_x=b_T(x,p(x))\), with \(r=0\)
at pivots. The tree route from \(x\) to \(p(x)\) has at most
\(h=b+1\) edges, so \(D_T(x,p(x))\le h r_x\).
The combinatorial construction is linear. Prune unmarked leaves from the tree to find the minimal subtree on the marks. Add its vertices of degree at least three, of which there are at most as many as its leaves. Each corridor is found by following degree-two paths in that subtree and is scanned once. Removing the chosen edges breaks the minimal subtree into one connected part per pivot, with all its pendant subtrees still attached. Each route from a nonpivot to its owner is contained in its pivot-deleted component up to the final edge. The deleted corridor edges induce a tree \(T_A\) on \(A\).
For pivots \(p,q\), the largest weight on \(T_A[p,q]\), denoted
\(\sigma(p,q)\), equals \(b_T(p,q)\). Set \(\sigma(p,p)=0\).
The tree and its sorted merge hierarchy cost
\(O(N+|A|\log(|A|+1))\) time.

\begin{lemma}\label{lem:owners}
In the reduced graph of an owner decomposition (\tref{Definition}{def:owner}), for an intercell edge $xy$ of weight $w$, its two owner radii and the bottleneck between its pivots are at most $w$. For an intracell edge with different endpoint radii, both radii are at most $w$. Every edge satisfies $w\le h(r_x+r_y)+N\sigma(p(x),p(y))$.
\end{lemma}
\begin{proof}
For an intercell graph edge \(xy\), put \(p=p(x)\), \(q=p(y)\).
Then
\begin{align*}
w(xy)\ \ge\ \max\{r_x,r_y,\sigma(p,q)\}.
\end{align*}
The path \(T[x,y]\) crosses all the deleted edges on \(T_A[p,q]\).
If this path passes through \(p\), it contains \(T[x,p]\).
Otherwise it exits \(x\)'s own two-pivot component across its
selected edge \(f\). The portion connecting the same-side endpoint
of \(f\) back to \(p\) is part of the corridor and has bottleneck at
most \(w(f)\). These facts bound \(r_x\) by the bottleneck of
\(T[x,y]\). The argument at \(y\) is identical. Within one owner
cell, a change of \(r\) along an edge also implies
\(w(xy)\ge\max\{r_x,r_y\}\), by the tree bottleneck property.
For every edge,
\begin{align*}
w(xy)\le h(r_x+r_y)+N\sigma(p(x),p(y)).
\end{align*}
This follows by routing through the two owner pivots in \(T\).
\end{proof}
\begin{lemma}[Tree access to any pivot]\label{lem:anypivot}
In an owner decomposition with minimum spanning tree $T$, let $v$ be a vertex of owner $p(v)$ and owner bottleneck $r_v$. For every pivot $q$, the tree path $T[v,q]$ has bottleneck at least $r_v$.
\end{lemma}
\begin{proof}
If $T[v,q]$ goes through $p(v)$ it contains the path defining $r_v$. Otherwise its first corridor exit is the cut edge of a two-pivot component. From the endpoint of that edge on $v$'s side, the corridor back to $p(v)$ uses no larger edge, by the choice of the cut. Concatenating the initial path up to this endpoint and that corridor gives a walk to $p(v)$ whose bottleneck is at most that of $T[v,q]$. Removing repetitions leaves $T[v,p(v)]$ and proves the bound.
\end{proof}
\subsection{The Local State Graph}
For this local construction only, a macro edge is an intercell edge with \(w(xy)\le N^3\sigma(p,q)\).
All other edges are radial. Every deleted corridor tree edge is macro.
For a macro edge, \(\sigma\le w\le N^3\sigma\), with a named
corridor maximum from the pivot tree. For a radial edge, \tref{Lemma}{lem:owners} and
\(\sigma<w/N^3\) in the intercell case give
\begin{align*}
w(xy)\le 3h\max\{r_x,r_y\}.
\end{align*}
An intracell edge obeys the same inequality directly. In particular
there is no positive radial edge between two pivots. The tests for all
edges take linear time after computing \(N^3\sigma\) once per pivot-tree
maximum, with \(O(|A|\log N)\) additions. To form $N^3\sigma$ exactly, multiply by the integer $N$ three times, each time using its binary expansion and $O(\log N)$ additions and doublings.
Set $K$ to the least power of two at least $8h$. Call two positive radial endpoints near when their radii differ by at most a factor $K^2$, and far otherwise. Retain radial edges for which both endpoint radii are positive and
their ratio is at most \(K^2\) to define near connected components.
An isolated positive vertex forms a component. In a near component
with minimum radius \(a>0\), give each \(v\) the native shell
\(i(v)\) determined by
\begin{align*}
aK^{i(v)}\le r_v<aK^{i(v)+1}.
\end{align*}
Precompute \(K^2r_v\) by \(O(\log K)\) doublings per vertex.
Adjacent native indices along a near spanning tree differ by at
most two. Starting at the minimum, walk the tree and cache successive
powers of \(K\) when needed. There are \(O(N)\) cached powers in
total and all indices require \(O(m+N\log K)\) time.

A vertex's native inward and outward copies use its own shell index. The shell bottom at index $j$ is $aK^j$, where $a$ belongs to that near component. The inward state at index \(j\ge0\) contains vertices in native
shells \(j,j+1\). The outward state at \(j\ge0\) contains those in
\(j-1,j\). In each case use separate copies and add both orientations
of all near radial edges whose endpoints are present. A \emph{block}
is a connected component of this undirected intrastate graph.
Add the following further arcs, always with the original physical
length when an edge is used.
\begin{itemize}
\item In an inward copy at \(j\), a near edge to a native shell
\(i<j\) transfers to the inward copy at \(i\). An outward copy at
\(j\) similarly transfers to the outward copy at native index \(i>j\).
\item A far positive radial edge transfers from every inward copy
at its higher-radius endpoint to the native inward copy of its
lower-radius endpoint. The reverse transfers from every outward
copy at the lower endpoint to the native outward copy at the higher
endpoint.
\item Every vertex may switch for zero cost from either inward
copy to its native outward copy. A pivot has one inward and one
outward copy with this switch. Its incident radial arcs point into
its inward copy from positive inward states, and away from its
outward copy to positive native outward states.
\item In each direction of a macro edge, allow every copy of the
tail to transfer to the \emph{native inward} copy of the head.
Initialize the native inward copy of each prescribed source at
its supplied cost.
\end{itemize}
\begin{definition}[Radial state representation]\label{def:radial}
A \emph{radial state representation} is the directed graph formed by the inward and outward two-shell copies, bidirectional blocks, phase switches and macro resets above, together with the projection to physical vertices and the distinguished native inward entry at each vertex. Its parameters are the nonnegative vertex coordinate, adjacent-shell ratio $K$, and classification into radial and macro edges. In this section the coordinate is $r_v$; \tref{Section}{sec:full} uses $J_v$.
\end{definition}

\begin{lemma}[A forbidden excursion]\label{lem:nopeak}
Consider the local radial state representation of an owner decomposition on $N>2L_N^4$ vertices, with local macro test $w\le N^3\sigma$ on intercell edges and $K\ge8h$. On a contiguous radial segment of a shortest path, let $M>0$ be the greatest radius or edge weight. Its two endpoint radii cannot both be at most $M/K$.
\end{lemma}
\begin{proof}
Consider any
contiguous radial segment of a globally shortest path. Let \(M\)
be the maximum of its vertex radii and edge weights. If its endpoint
radii are both at most \(M/K\) and \(M>0\), its cost is at least \(M\).
This holds immediately when an edge attains \(M\). Otherwise, reaching
a vertex of radius \(M\) from a smaller radius requires an edge of
weight at least \(M\), by \tref{Lemma}{lem:owners} or the intracell change-of-radius
inequality.

On the other hand, the pivots visited by the segment
all belong to one \(T_A\) component using edges below \(M/N^3\).
The two endpoint pivots consequently have tree distance at most
\(M/N^2\). The two endpoint spokes together cost at most \(2hM/K\).
Their concatenation has cost below \(M\), contradicting optimality.
\end{proof}
\begin{lemma}\label{lem:coarsepeak}
In the local radial state representation of \tref{Definition}{def:radial}, take any maximal radial run of a globally shortest path and split it at a vertex of minimum radius. Then a prefix radius never exceeds $K$ times its running minimum. On the suffix a radius never falls below the running maximum divided by $K$.
\end{lemma}
\begin{proof}
Apply \tref{Lemma}{lem:nopeak} to segments of the given path.
Split any radial run of the shortest path at a vertex of minimum
radius. On the prefix ending at this vertex, every current radius
is at most \(K\) times the smallest radius already seen. Otherwise
the subpath from the earlier small radius to the splitting vertex
violates the preceding paragraph. On the suffix, every current radius
is at least \(1/K\) times the largest radius already seen, by applying
the prefix statement in reverse. The assertion includes runs starting
or ending at a pivot.
\end{proof}
\begin{lemma}\label{lem:lift}
For a connected positive-weight graph with $N>2L_N^4$ vertices and $m$ edges, the local radial state representation of \tref{Definition}{def:radial} has $O(N)$ states and $O(m+N)$ arcs. It realizes every physical geodesic from its initial native inward state. Removing macro arcs and contracting the bidirectional blocks gives a directed acyclic graph.
\end{lemma}
\begin{proof}
Each vertex and edge have a constant number of copies. With the macro
arcs deleted, contracting the blocks leaves a DAG. Inward near
transfers strictly lower \(j\). Outward near transfers strictly raise it.
Across independently anchored grids, an inward far transfer strictly
lowers the numerical shell bottom because its source bottom exceeds the
source radius divided by \(K^2\), hence exceeds the target radius.
For an outward far transfer the target native bottom exceeds the
target radius divided by \(K\), hence exceeds the source state bottom,
which is at most \(K\) times the source radius. Pivot arcs and
phase switches respect the same two-phase order.

Each physical geodesic has a state realization of the same cost.
On its inward prefix track the lowest native shell seen in the
current grid and on its outward suffix track the highest one.
The radius bound keeps the current vertex in the indicated two shells.

A far inward transfer is descending and puts the new vertex below
every prior radius on that prefix. A far outward transfer is
ascending and sets a new radius record on its suffix. This resets
the local grid index. The switch at the minimum-radius vertex and
the resets on macro edges splice the runs.

Conversely, each state path
projects to a physical walk, after omitting switches. Therefore the
minimum over the copies of a vertex is its physical shortest-path
distance, also for arbitrary prescribed source offers. A particular
phase copy can have a larger constrained distance.
\end{proof}
\subsection{Threshold Components and Joins}
For a state vertex \(z\), let \(I_0(z)\) be the least weight of a
macro edge whose head reset can reach \(z\) using radial arcs and
switches. Let \(O_0(z)\) be the least weight of a macro edge that
can be traversed after such a path from \(z\) to an eligible tail
copy. Empty minima are \(+\infty\). Compute these numbers by min
propagation in the block DAG and its reverse, ignoring radial weights.
For \(I_0\), seed only the native inward head copy. For \(O_0\), seed
every eligible tail copy. Keep an attaining macro edge with each
finite value. Write \(B_0(z)=\max\{I_0(z),O_0(z)\}\).

Define \(I_*(z)\) and \(O_*(z)\) with the same admissible paths,
now minimizing the largest physical arc weight, counting the
preceding or succeeding macro edge. Put
\(\theta(z)=\max\{I_*(z),O_*(z)\}\). Compute \(I_*\) in block DAG
order. Within a block, solve the undirected multisource bottleneck
problem from the currently posted costs. For a block MST \(S\)
and posted costs \(c_x\), the answer at \(v\) is
\(\min_x\max\{c_x,b_S(x,v)\}\). A postorder sweep takes the minimum of the local seed and the child offers, each raised to its incident edge weight. A preorder sweep sends the resulting parent offer back to each child with the same maximum rule. Retracing a tree edge in the return sweep cannot improve a minimum beyond the value of a simple path, since deleting a repeated segment only decreases its bottleneck. The sweeps compute all
these values. Relax each outgoing transition by taking a maximum
with its length. Reverse all directions for \(O_*\). The block
MSTs and the two sweeps use total \(O(m+N\log\log N)\)
deterministic time by \tref{Lemma}{lem:mst}.
Introduce one auxiliary vertex for each pivot. For every block MST
edge join its endpoints at its physical weight. For every finite
\(\theta(z)\), join \(z\) to the auxiliary vertex of its owner at
weight \(\theta(z)\). Join the auxiliary pivot vertices across their
corridors at the corridor weights. 
These weighted undirected joins will specify the hierarchy. The following two lemmas justify using them even though the state transitions are directed. A \emph{core} at threshold $W$ is the strongly connected component containing the pivot states of a nonsingleton component of the pivot tree restricted to corridor weights at most $W$.

\begin{lemma}\label{lem:supportbase}
In a local radial state representation with owner-path hop bound $h$ and shell ratio $K$, let $B_0$ be the maximum of the earliest attainable incoming and outgoing macro weights at a state, ignoring radial weights. Let $\theta$ be the corresponding maximum when the entire access paths are measured by their bottlenecks. For every state with finite $B_0$, one has $r\le\theta$ and $B_0\le\theta\le 3hK^2B_0$. Finiteness is equivalent for $B_0$ and $\theta$.
\end{lemma}
\begin{proof}
If \(B_0(z)<\infty\), concatenate its two attaining radial portions
at \(z\). This is a single admissible run between two macros of
weight at most \(B_0(z)\). In its inward portion the shell bottom
is nonincreasing and each radius is less than \(K^2\) times its
bottom. The maximum radius there is at most \(K^2\) times the
radius of the macro head. In the outward portion the same estimate
holds from the terminal macro tail in reverse, since the bottom is
nondecreasing, radii are below \(K\) times the bottom, and the final
radius is at least its bottom divided by \(K\). \tref{Lemma}{lem:owners}
and the radial edge bound \(w(xy)\le 3h\max\{r_x,r_y\}\)
therefore give
\begin{align*}
B_0(z)\le\theta(z)\le 3hK^2B_0(z),\qquad r_z\le\theta(z).
\end{align*}
For the last inequality, a macro endpoint has
radius at most its edge weight. Along a physical path the radius can
increase only across an edge of at least the new radius. Both
quantities are infinite together. Each finite \(B_0(z)\) has an
attaining macro edge with named corridor scale \(\sigma\), so
\(\sigma\le\theta(z)\le 3hK^2N^3\sigma\).
\end{proof}
\begin{lemma}\label{lem:join}
In the local radial state representation, fix $W>0$. Retain block MST edges, state-to-owner joins at $\theta$, and pivot corridor edges of weight at most $W$. The resulting undirected components, restricted to states, are exactly the strongly connected components of the directed graph on arcs of length at most $W$.
\end{lemma}
\begin{proof}
Retain state arcs of cost at most \(W\), and let \(\mathcal P_W\)
be the components after retaining in \(T_A\) the corridors of
weight at most \(W\). A component with two or more pivots gives a
strongly connected set of state vertices, called its core.
To see this, a pivot can travel outward along its owner tree to the
native outward copy of any endpoint of a corridor edge \(f\) in
the component. That tree path has nondecreasing radii and bottleneck
at most \(w(f)\). Traverse \(f\), resetting to the native inward
copy, and descend the other owner tree to its pivot. The reverse
traversal works as well. The zero switch permits a concatenation of
these routes and a round trip joins the two copies of each pivot.

The same argument from an arbitrary macro tail's owner reaches its
native outward copy before traversing the macro edge.
Thus each eligible macro head reset is reachable from the core and
can reach the core. A nonnative tail copy can always reach the
core by traversing the eligible edge.

For a positive-radius copy \(z\), membership in a core is equivalent
to \(\theta(z)\le W\). The component is then the one containing
\(p(z)\). Indeed, a witnessing preceding macro supplies a path from
a core, and a succeeding macro supplies a return path to a core.
All physical edges of the combined path are at most \(W\).
The owner pivots visited are in the same member of \(\mathcal P_W\)
by \tref{Lemma}{lem:owners}, and the macro endpoints' owner spokes also have bottleneck
at most \(W\).

Conversely, in any SCC containing a macro arc, a
directed walk through \(z\) between a last preceding and a first
succeeding macro witnesses both inequalities. A threshold SCC
containing no macro arc stays inside one block, because the radial
block quotient is a DAG. It is exactly an undirected threshold
component of that block.

The statement holds at a pivot as well,
with a singleton local component for either phase copy. Strict
thresholds are obtained by changing all comparisons consistently to
strict ones.
 The preceding two cases exhaust the strongly connected components. The auxiliary vertices only identify the appropriate
cores. They may be suppressed when components with no state vertices
are removed and unary hierarchy nodes are contracted. The description
has \(O(N)\) union events. It never requires a threshold rank for
each of the \(O(m)\) physical or state arcs.
\end{proof}
\subsection{Assembling the Local Hierarchy}\label{sec:local-assembly}
For a singleton block keep its individual finite attachment join and omit all internal scale tests and merges. In a nonsingleton positive-radius block \(U\), let \(s_U\) be the lower boundary
of the lowest native shell present in its two-shell window. Thus
\(s_U\le r_x<K^2 s_U\) throughout \(U\). For the outward window
at index zero use the actual minimum positive boundary. Every
intrastate edge has weight at most \(3hK^2s_U\). Contract, for
hierarchy purposes only, the components of the intrastate edges
of weight less than a boundary \(t_U\in[s_U/2,s_U]\), chosen
below. Each such component \(Y\) has one
owner and one identical radius \(r_Y\ge s_U\), since changing either
along an edge would require weight at least \(s_U\).
The MST path between the endpoints of each of its edges has
bottleneck less than \(s_U\). It avoids the pivots because a path from
an endpoint to any pivot has bottleneck at least its owner radius,
by \tref{Lemma}{lem:anypivot}. Thus \(Y\) lies within one component of \(T-A\)
and has fewer than \(b\) vertices.
Temporarily treating each $Y$ as indivisible preserves the outer SCC calculation. Below $r_Y$ any new internal zero connection preserves the common radius and owner. A macro of weight below $r_Y$ still cannot reach radius $r_Y$, because increasing the radius requires an edge at least as large as the new radius. At thresholds at least $r_Y$, $Y$ is genuinely connected by its defining edges. A macro-free cycle through different blocks remains impossible. For every component $Y$ so obtained, use its exact local Kruskal merge tree. Sort the fewer than
\(b\) edges of its own MST, then process equal-weight edges
simultaneously. The aggregate extra cost is \(O(N\log(b+2))\).
Every such local height is below \(t_U\).
Only the \(O(N)\) union events need comparable classes. The
\(O(m)\) macro arcs are used by raw minimum comparisons and
retain the pointer to their corridor maximum. For such a nonsingleton block \(U\),
an infinite minimum of \(\theta\) means that its entire hierarchy
is local. Otherwise choose \(z_U\) minimizing \(\theta\) in \(U\)
and use the named corridor key \(\sigma_U\) for \(B_0(z_U)\).
\tref{Lemma}{lem:supportbase} gives
\begin{align*}
\sigma_U\le\theta(z_U)\le N^{O(1)}\sigma_U,\qquad
 s_U\le\theta(z_U).
\end{align*}
Fix $c=10$. In the nontrivial range $N>2L_N^4$ we have $N\ge32$, $h\le N$ and $K<16h$. Hence $3hK^2\le768N^3\le N^5$ and $N\cdot3hK^2\le N^{c-2}$.
Call \(U\) aligned if \(s_U\ge\sigma_U/N^c\). All significant local
weights of an aligned block, meaning those at least \(s_U/2\),
are then within a fixed polynomial factor of \(\sigma_U\).
In an unaligned block, even its total MST length is below
\(\sigma_U/2\). It forms one local tree below its first macro
attachment, and all its members have the same value of
\(\theta\) because the minimizer \(z_U\) can reach and be reached from
every member using edges below its attachment threshold.
Its local scales can therefore use a separate grid starting at
\(s_U\). A block without any finite attachment uses the same rule.

The alignment test costs no large multiplication per block. Group
the values \(s_U\) by named key \(\sigma_U\). In one group choose
successive medians to partition the list at the cutoff
\(\sigma_U/N^c\), determining the side of each median by
\(O(\log N)\) additions. Each search uses \(O(\log N)\) medians
and the total partitioning work is linear in the group size.
All groups together cost \(O(N+|A|\log^2 N)\).
For each named corridor key \(\sigma\), choose the minimum of
\(\sigma\) and all aligned \(s_U\) assigned to it as the start
of an interval. Choose its upper endpoint as $N^{16}\sigma$. Indeed $\theta(z_U)\le3hK^2N^3\sigma\le N^8\sigma$ and $s_U\le\theta(z_U)$; even the greatest relevant block weight is at most $3hK^2s_U\le N^{13}\sigma$. For any fixed integer exponent these products take $O(\log N)$ additions per name by the preceding binary expansion. Include every finite
\(\theta(z)\) in the interval of its own named key.
There are \(O(|A|)\) intervals, each of polynomial aspect.
Sort the endpoints, merge overlaps, and start a doubling grid
at the minimum of each resulting component. Stop after covering
its maximum. Over all components, this takes
\(O(|A|\log N)\) additions because each input interval contributes only
\(O(\log N)\) octaves. Intervals close enough for one terminal
doubling step to overlap can instead be merged first.
Merge the already sorted corridor keys into the endpoint scan and
record two indices for each support $[l,u]$, namely the largest grid point at most $l$ and the first grid point strictly above $u$. Use one extra doubling at the top of a merged grid if necessary. The upper index serves only as a search sentinel. For every aligned block assigned to the name, $l\le s_U$, so the lower index is at or below $t_U$, even when merging supports puts $t_U$ below $l$.
A labelled value in its interval is classified by binary search
among \(O(\log N)\) consecutive grid positions, in \(O(\log\log N)\)
comparisons. For aligned \(U\) choose \(t_U\) as the grid lower
boundary of the class of \(s_U\). The remaining block MST weights
at least \(t_U\) use this same grid, starting with its named class.
In a separate-grid block set \(t_U=s_U\). Its significant weights
occupy \(O(\log(hK+2))\) local doubling classes.

Construct merge nodes by processing the globally aligned union
events in class order, using a simultaneous merge for each class.

\tref{Algorithm}{alg:construct-local} collects the construction for use on an atom. The unrounded join records recover the exact threshold components after restriction to states by \tref{Lemma}{lem:join}. A simultaneous merge for a class $[h_0,2h_0)$ instead records at height $h_0$ the components generated by all weights in and below that class. Its strict light-arc order uses the lower boundary, and its connecting arcs use the upper bound $2h_0$. Computing an ordinary shortest-path tree directly in a small input would omit the maps needed by a later multi-entry search.
\begin{algorithm}[ht]
\caption{Constructing a local hierarchy for later external offers}\label{alg:construct-local}
\Fn{\textsc{LocalHierarchy}($H$)}{
\KwIn{A connected undirected graph $H$ on $k$ vertices with positive weights.}
\KwOut{State arcs, physical projections, native entry states, and an ordered hierarchy.}
\If{$k\le2L_k^4$}{
 Keep one native entry state per vertex, with its physical image, and both orientations of every original nonloop edge\;
 Find an MST after retaining a lightest edge per vertex pair, and merge its equal-weight edges simultaneously to form the hierarchy, using the MST edges as physical connectors (\tref{Section}{sec:prelim})\;
 \Return{these states, arcs, maps, and hierarchy}\;
}
Find an MST with the three growth passes of \tref{Lemma}{lem:mst}, remove strictly tree-dominated edges, and form its owner cells at separator size $L_k^4$\;
Classify radial and macro edges by their pivot-tree bottlenecks (using tree queries, \tref{Lemma}{lem:trees})\;
Make the inward and outward windows and their macro resets\;
Propagate the two adjacency minima and two bottleneck minima through the blocks\;
Generate the linear list of join records and classify the significant weights on the grids\;
Combine equal-class joins (\tref{Lemma}{lem:unions}), splice in exact local trees, and order the light child arcs\;
\Return{the state representation and its hierarchy with physical connectors}\;
}
\end{algorithm}

\begin{lemma}\label{lem:base}
Given a connected undirected graph with $N\ge1$ vertices and $m$ edges of positive real length, \tref{Algorithm}{alg:construct-local} returns an ordered hierarchy with a native entry at every vertex. Every geodesic has an equal-length state realization from its native entry, and every state arc projects to a feasible physical walk. Construction takes $O(m+N\log\log(N+4))$ time, and the connection constant is at most two.
\end{lemma}
\begin{proof}
The base case is given in \tref{Section}{sec:prelim}. For the radial construction, \tref{Lemma}{lem:lift} gives path realization. A merge in the class \([h_0,2h_0)\) has height \(h_0\). All its
connecting arcs can be realized in the actual state graph below
\(2h_0\), and every arc of weight below \(h_0\) between distinct
children belongs to an acyclic quotient. These statements are
exactly the threshold-SCC characterization at the two boundaries
of the class.

For connectivity after contracting the children, use
the actual state arcs of weight below the upper boundary. Paths
joining states in the same threshold SCC stay in that SCC, since
every intermediate state has both an entering and a returning path.
Their projection gives a connected undirected child quotient even
when an auxiliary union event itself has no corresponding direct
state arc.

Separate local trees are spliced at their attachment
events. The intervening strict inequalities ensure their heights
precede the enclosing merge.

In particular at a first global attachment
of height \(h_0\), an aligned atom has
\(t_U\le h_0\) because \(s_U\le\theta(z)\) and the two values
use the same underlying global doubling grid wherever their intervals
overlap. An unaligned block has largest local MST edge below
\(\sigma_U/2\le h_0\). Consequently the local children already lie
in genuine strict-threshold components at an enclosing global merge.

For a global grid component boundary
that is not an exact power continuation, stop a class at the next
component start if necessary. For a tiny \(Y\), place the separately
sorted exact tree below \(t_U\). If the final state graph has
several SCCs, its maximal components are processed in their
condensation order.

Discard empty auxiliary nodes and compress unary nodes. Send an arc
between distinct roots to their condensation quotient. For the
remaining arcs, a lowest-
common-ancestor query and a level-ancestor query assign each original
state arc to the pair of children it joins at their common parent.
At each parent of height $h_0$, place the assigned arcs of weight strictly below $2h_0$ in an undirected child adjacency list. Store the original tail and head states and the length with each incidence. The connectivity argument above guarantees a spanning tree in this list. A depth-first search supplies one physical connector per discovery edge, with its recorded endpoints and length. This also works at an exact merge, where the selected MST edges have weight $h_0$. Each state arc occurs in at most one list and the sum of the child counts is linear. At each parent, topologically sort the child arcs of weight strictly
below its height. Over the whole hierarchy this takes \(O(m+N)\)
time. \tref{Lemma}{lem:unions} fits the \(O(m+N\log\log(N+4))\)
total bound. Including local MSTs, local sorts and the state
construction, the hierarchy uses \(O(m+N\log\log(N+4))\) time in
the comparison-addition model. It has \(O(N)\) nodes and the
state graph has \(O(m+N)\) arcs.
\end{proof}

\subsection{Replacing Connected Subgraphs by Local Hierarchies}\label{sec:substitute}
\begin{definition}[Replaceable atom]\label{def:atom}
A \emph{replaceable atom} at cutoff $c>0$ is a set of states with a connected undirected internal graph $H$. Each internal edge has positive length, is usable in both directions, and projects to a physical walk of the stated length. A given spanning tree of $H$ has every edge shorter than $2c$. An outside arc, a seed, or a target may be incident to any state. When the atom is a leaf of a larger hierarchy, its parent's height is at least $c$.
\end{definition}

\begin{definition}[Partition hierarchy]\label{def:partition}
A \emph{partition hierarchy} on a directed state graph with nonnegative arc lengths, a physical image for each state, and a physical walk of no greater length for each arc is a rooted forest whose leaves partition its states. Each part is a singleton or a replaceable atom. Internal heights, light-arc orders, root orders and physical connecting trees satisfy \tref{Definition}{def:hier} on these parts: compare two children using all arcs between their underlying states. Every state keeps its own physical image, and every certificate names its actual endpoint states. A replaceable atom with a parent has cutoff at most its parent's height. All path lengths, offers and tight witnesses refer to the underlying state graph.
\end{definition}

\paragraph{Replacement procedure.}
Run \tref{Algorithm}{alg:construct-local} on $H$. Retain the maximal local subtrees of height strictly below $c$, including singleton leaves not covered by any such subtree. If several remain, give them a common root of height $c$. Redirect an old entering arc or seed to its endpoint's native inward copy. At every copy of an old tail install its old outside exits. Each copy of an old target retains that target's physical image.

\begin{lemma}[Capping a local hierarchy]\label{lem:cap}
For a replaceable atom at cutoff $c$, the capped local hierarchy is a single rooted ordered hierarchy. Its heights are at most $c$. Every physical link added at its new root has length less than $2c$.
\end{lemma}
\begin{proof}
There are two checks if a new root is needed.
\begin{enumerate}
\item \emph{Order.} A directed cycle of arcs shorter than $c$ crossing retained children would cross original roots or have a highest original node whose distinct children it visits. The original root quotient is acyclic. In the second case the highest node was removed and had height at least $c$. Projecting the cycle to its children gives a directed closed walk with an interchild arc in its acyclic light quotient. Both cases are impossible.
\item \emph{Connection.} Join copies of each old state by zero-length physical certificates. For each tree edge of $H$, add a certificate between copies of its two endpoints. After the retained children are contracted, these links still connect the quotient. Choose a spanning tree of the quotient. Each selected link has length less than $2c$.
\end{enumerate}
A unique retained subtree already supplies the required root. At any enclosing old node, replacement duplicates endpoints only inside a single old child, preserving every child-order constraint and physical connection.
\end{proof}
\begin{lemma}[Native entries and arbitrary exits]\label{lem:splice}
Replacing a replaceable atom preserves a tight witness through that atom, with any number of visits. An external offer enters the native inward local copy and every local copy can use each old exit of its projected tail.
\end{lemma}
\begin{proof}
Cut the witness at every entrance to and exit from the atom. A nonempty internal piece must be shortest between its endpoints in $H$. Otherwise replacing it by a shorter internal path and following the projected suffix contradicts tightness. \tref{Lemma}{lem:lift}, or the direct bounded-size construction, realizes a shortest piece from the native inward entry to some copy of its exit tail.

The next outside arc is installed at that copy. An empty piece can leave directly at its native entry. This covers the first and last pieces as well as two visits to the same atom. Replacing the pieces in path order preserves total length. Conversely all new arcs project to feasible walks, so replacement introduces no underestimate.
\end{proof}
\begin{lemma}[Substituting disjoint atoms]\label{lem:substitution}
Let a partition hierarchy have $N$ underlying states and $m$ underlying arcs. If each replaceable atom has fewer than $b$ states, substituting its capped local hierarchy at every nonsingleton part produces an ordered hierarchy and preserves every required tight witness. The substitutions cost $O(m+N\log\log(b+4))$ beyond the construction of the partition hierarchy. The same single-atom construction supports outside offers before an enclosing hierarchy is specified.
\end{lemma}
\begin{proof}
\tref{Lemmas}{lem:cap} and~\ref{lem:splice} give order and path preservation independently. Each old state has a constant number of new copies. Each old exit is replicated only at the copies of its tail, and every copy of that arc has one fixed head. Internal edges belong to disjoint inputs. Summing \tref{Lemma}{lem:base} over these inputs proves the work bound. The construction invokes the local algorithm once per atom.
\end{proof}

\section{Short Pivot Routes from Random Samples}\label{sec:routes}

\begin{definition}[Pivot route certificate]\label{def:route-certificate}
A pivot-routing instance is a loopless undirected multigraph on $t$ pivots.
Each of its $M$ edges $e=xy$ carries the positive length $a_e$ of a
physical walk from $x$ to $y$. For a common factor $P\ge1$, a route
certificate for $e$ is an input weight $\lambda_e\le a_e$ for which a
physical $x$--$y$ walk of length at most $P\lambda_e$ exists. The
\emph{dictionary} lists the distinct weights used as certificates.
\end{definition}

The next construction returns a sorted dictionary and one index per
input edge. The subsequent algorithm uses only the upper bound
$P\lambda_e$. It can use the same index for many pivot pairs even if the
certifying walks differ. In the application $t=O(1+N/L_N^{64})$, where
$N$ is the physical vertex count. Loops may be discarded before forming
the instance, since their endpoints coincide.

\subsection{Sampling and Certificate Assignment}
Replace each group of parallel edges by a representative of minimum
weight. Keep a pointer to it from every original edge. A certificate
for the least weight works for all edges of the group. Hence the
remaining graph is simple and has at most $t^2$ edges. This reduction
only limits the number of sampling rounds. We charge the initial and
final scans to the original $M$.

\tref{Algorithm}{alg:name} samples the current residual edge set $U$.
Its estimator, \tref{Algorithm}{alg:estimate}, reports disconnection or
returns an index into a table of upper bounds on distances in the
sample. A group with a common table index has a single bound $Z$.
Finding the smallest input weight $a_*$ in that group satisfying
$Pa_*\ge Z$ lets every edge with $a_e\ge a_*$ use the same certificate.
To find $a_*$, test a median by repeated doubling. On success retain
that median as a candidate and continue with the strictly smaller keys.
On failure continue with the strictly larger keys. The least successful
candidate is the answer, provided one exists. Partitioning around the
median discards its ties in both cases.

\begin{algorithm}[ht]
\caption{Assigning certificates to pivot edges}\label{alg:name}
\Fn{\textsc{NameRoutes}($G_P$)}{
\KwIn{A pivot-routing instance $G_P$ as in \tref{Definition}{def:route-certificate}.}
\KwOut{A sorted dictionary, a dictionary index per input edge, and the factor $P$.}
$u\gets\lceil\log_2(t+2)\rceil$, $g\gets4u+3$\;
$P\gets$ the least power of two at least $8\cdot256\,g u^2$\;
$U\gets$ the least-weight representative for each distinct endpoint pair\;
\While{$|U|>32t$}{
 Choose a uniformly random $16t$-element subset $S$ of $U$\;
 Run \algcall{alg:estimate}{Estimate} on $S$ for the endpoint pairs of all edges in $U$\;
 Group all connected pairs by returned index, and start with no qualified edges\;
 \ForEach{nonempty group with bound $Z$}{
  Find its smallest $a_*$ satisfying $Pa_*\ge Z$ by median pruning, using linear-time median selection (\tref{Lemma}{lem:basic-primitives}) and repeated doubling for each threshold test, if one exists\;
  \If{$a_*$ exists}{qualify each edge $e$ in the group with $a_e\ge a_*$ and propose $\lambda_e=a_*$\;}
 }
 \If{at least half the edges of $U$ qualify}{
  Append one dictionary record of weight $a_*$ per qualifying group, point each qualified edge to its group's record, and remove those edges from $U$\;
 }
}
Append a record for each remaining representative with its own weight and store its pointer\;
Sort the dictionary records by weight, coalesce ties, and use the record pointers to map original edges to the resulting indices\;
\Return{the dictionary, the indices, and $P$}\;
}
\end{algorithm}

Each accepted round preserves the invariant that all removed edges
already have a certificate. A rejected round discards its proposals
and draws a fresh sample with fresh estimator randomness. When $t<2$
the loopless instance has no edges and its dictionary is empty.

\begin{lemma}[Critical edges]\label{lem:critical}
Fix $g=4\lceil\log_2(t+2)\rceil+3$ in a loopless undirected weighted
graph on $t$ vertices. Call an edge \emph{critical} if, after its deletion,
its endpoints have no path of at most $g$ edges each no heavier than the
deleted edge. There are at most $2t$ critical edges.
\end{lemma}
\begin{proof}
Two parallel edges cannot both be critical. On any cycle of at most
$g+1$ critical edges, the other edges witness the noncriticality of a
heaviest edge. The critical subgraph is therefore simple with no cycle
of length at most $g+1$. Repeatedly delete a vertex of degree at most
two, charging its remaining edges to that vertex. If a subgraph remains,
its minimum degree is at least three. In a breadth-first search of depth
$j=\lfloor g/2\rfloor$, a collision of branches would give a cycle
of length at most $2j+1$. Thus at least $3\cdot2^{j-1}>t$ distinct
vertices occur at the last level. This contradiction shows that every
edge was charged, at most twice per vertex.
\end{proof}

\begin{lemma}[Distances in a uniform sample]\label{lem:bad}
Let $U$ be a loopless residual graph with $h>32t$ edges and let
$S$ be a uniform $16t$-edge subset. In expectation at most $h/8$
edges $e=xy$ satisfy $d_S(x,y)>g a_e$, including disconnected pairs.
\end{lemma}
\begin{proof}
Draw a uniformly random edge $e$ from $U\setminus S$. Conditional on
$S\cup\{e\}$, every edge of this union is equally likely to be $e$.
The edge has no witness of the kind in \tref{Lemma}{lem:critical} in $S$
precisely when it is critical in the union. Thus the expected number
of outside edges without a witness is at most
$(h-16t)2t/(16t+1)<h/8$. A witnessed edge has distance at most $ga_e$
and a sampled edge supplies its own one-edge path.
\end{proof}

\subsection{Estimating Distances in a Sample}
The estimator receives $S$ with $O(t)$ edges and a list of $h$ pairs
of distinct pivots. It works separately in each connected component.
It uses the randomized partitions of tree-metric constructions
such as Fakcharoenphol, Rao and Talwar~\cite{DBLP:journals/jcss/FakcharoenpholRT04}.
The finite set of radii and the band construction below allow direct
implementation with comparisons and additions on input lengths.

\begin{algorithm}[ht]
\caption{Estimates of sampled distances}\label{alg:estimate}
\Fn{\textsc{Estimate}($S,\mathcal Q$)}{
\KwIn{A positive-weight sample graph $S$ with $O(t)$ edges, and $h$ distinct-endpoint demands $\mathcal Q$.}
\KwOut{A table of bounds and, for each demand, a table index or a disconnection flag.}
Build the sample merge forest by the sorted Kruskal sweep below and flag demands whose endpoints lie in different trees\;
Group connected demands by the weight band of their merge (using tree queries, \tref{Lemma}{lem:trees})\;
\ForEach{band anchor $a$ with nonempty demand group, in increasing order}{
 Build the contracted quotient $Q_a$ and map the demand endpoints to it\;
 \ForEach{scale $\Delta$ of this band}{assign every quotient vertex its first covering center\;}
 Build the partition trees from the largest scale to the smallest\;
 Store one index of value $4\Delta+a$ per part at scale $\Delta$\;
 Record for each demand in the band the index at the lowest common ancestor of its leaves (\tref{Lemma}{lem:trees})\;
}
\Return{the table, indices, and disconnection flags}\;
}
\end{algorithm}

\paragraph{Bands and quotients.}
Let $T$ be the least power of two at least $4t$. Sort the sample edges
by weight. The least unassigned weight $a$ starts a band containing
the remaining weights below $Ta$. Process edges in sorted order,
joining distinct current components as in Kruskal's algorithm. Each
successful join creates a binary parent labeled with its weight and
band identifier. Keep the selected forest edges. The merge of a
connected demand $xy$ is the lowest common ancestor of $x$ and $y$
in this merge forest.

For band $a$, contract all edges of weight strictly below $a/T^2$.
Retain the edges in $[a/T^2,T^2a]$ with distinct contracted endpoints.
The resulting multigraph $Q_a$ stores only the vertices incident to a
retained edge. To construct successive quotients, sweep the bands in
increasing order. Insert a selected forest edge as soon as its weight
becomes less than the contraction cutoff. Maintain a component label
at every original vertex, relabeling the smaller list at each union.
Two cursors delimit the band window in the sorted sample. Compare
$T^2w$ with $a$ at its lower end and $w$ with $T^2a$ at its upper
end, using shared doublings. Assign temporary local identifiers to the
components encountered on retained edges. Map the endpoints of all
demands of the band by their current component labels.

\paragraph{Centers, partitions, and table indices.}
Use scales $a,2a,\ldots,2^{\ell_a}a$, taking the least $\ell_a$
for which the last scale is at least $tT^2a$. Let $H$ be the least
power of two at least $(16t)^5$. At each scale $\Delta$, choose
independently a uniform permutation of quotient vertices and a
uniform integer $j\in\{0,\ldots,H-1\}$. A vertex receives the
first center in that order within distance $R=\Delta(1+j/H)$.
Implement the comparison using $H$-scaled edge weights and the
threshold $(H+j)\Delta$. These scaled quantities use binary doubling
and addition on the input lengths.

Run one radius-limited Dijkstra search per center in permutation
order. At each vertex store the best distance offered by earlier
centers within $R$, initially infinity. A search maintains its own
tentative distances using timestamps. When it settles a vertex,
discard that visit unless its distance is strictly smaller than the
stored value. On an improvement, update the value and scan its
incident edges. The first improvement sets the permanent center
label at this scale.

In each quotient component form a rooted tree from the coarsest
scale to the finest. A scale refines each existing part by its
center labels. Within each quotient component, number its centers consecutively from zero and give the parts at each scale their own consecutive numbering from zero. Stable counting sorts on the parent part and center identifiers group equal pairs using count arrays bounded by the component size. Assign each new part a separate index in the global table for demand lookups. Store $4\Delta+a$ at each new part
of scale $\Delta$, and append one distinct leaf per vertex below the
finest parts. Preprocess the tree for lowest common ancestor queries.
\tref{Algorithm}{alg:estimate} uses the invariant that all vertices of
a part at scale $\Delta$ have the same center at that scale.

\begin{lemma}[Lifting a quotient path]\label{lem:quotlift}
Let $x,y$ be distinct connected sample vertices, let $\beta$ be the
weight of their merge, and let $a$ be its band anchor. The images of
$x,y$ in $Q_a$ are distinct, present, and connected. For their
quotient distance $d$,
\begin{align*}
 a\le\beta\le d\le d_S(x,y)\le d+a.
\end{align*}
\end{lemma}
\begin{proof}
When the merge at $\beta$ is processed, $x,y$ are connected using
edges of weight at most $\beta$ and were disconnected below $\beta$.
This also holds for ties, since all strictly lighter edges precede
them. Thus $\beta$ is the minimum path bottleneck. We have
$a\le\beta<Ta$ and a simple path using at most $t-1$ such edges.
Hence $d_S(x,y)\le(t-1)\beta<T^2a$. A shortest sample path cannot
use an edge above $T^2a$ and projects to a walk in $Q_a$. The images
are distinct since no path with bottleneck below $a/T^2$ connects
them. They are therefore incident to retained edges on the projected
walk, which proves $d\le d_S(x,y)$.

Every contracted component has a spanning tree with edges lighter
than $a/T^2$. Lift a simple shortest quotient path by using this tree
between each entry and exit, including at the endpoints. There are
at most $t$ components, each requiring fewer than $t$ edges, so the
added length is at most $t^2a/T^2\le a$. Moreover a quotient walk with
bottleneck below $\beta$ would lift to a sample walk with bottleneck
below $\beta$. The quotient distance is thus at least $\beta$.
\end{proof}

\begin{lemma}[Separation at one scale]\label{lem:separate}
In a metric on at most $t$ points, two vertices at distance $d$
receive different centers at scale $\Delta$ with probability at most
$H_t(d/\Delta+1/H)$, where $H_t=\sum_{i=1}^t 1/i$.
\end{lemma}
\begin{proof}
Order possible centers by the lesser of their distances to the two
vertices, breaking ties by index. Center $i$ is within exactly one
of the two closed balls only if $R$ lies in a half-open interval of
length at most $d$. Its probability is at most $d/\Delta+1/H$,
because the possible radii are evenly spaced by $\Delta/H$. Given
such a radius, the first $i$ centers in the distance order all belong
to the union of the two balls. The independent permutation makes
center $i$ the first in this union with probability at most $1/i$.
Different labels require that the permutation's first center in the
union cover exactly one endpoint. Sum over $i$.
\end{proof}

\begin{lemma}[Work to assign one scale's labels]\label{lem:record-min}
In a quotient with $v$ vertices and $e$ edges, the pruned searches
return each vertex's first covering center in expected
$O((v+e)\log^2(t+2))$ time, even conditional on the chosen radius.
\end{lemma}
\begin{proof}
After any prefix of the center order the stored value at each vertex
is the least distance within $R$ from that prefix, or infinity if
no such center exists. To prove the invariant for the next center
$c$, consider a vertex $z$ whose true distance from $c$ strictly
improves this stored minimum and is at most $R$. Every vertex $x$
on a shortest $c$--$z$ path must also strictly improve. Otherwise an
earlier center could reach $x$ as cheaply and follow the remaining
path to $z$ within $R$, contrary to strict improvement there. The
path to $z$ survives pruning, so Dijkstra's extraction rule settles
$z$ at its true distance. Conversely any new strict improvement by
a discovered walk implies a strict improvement by the true distance.
This proves the invariant and the first-center assignment by induction.

At any fixed vertex, order the centers within $R$ by their occurrence
in the random permutation. The $i$th improves the record only if its
distance is a strict minimum among the first $i$ distances. By symmetry
this has probability at most $1/i$, including ties. The expected number
of adjacency scans at that vertex is at most $H_t$. Every heap insertion,
relaxation, and extraction is charged to an initial center or a scanned
incidence. \tref{Lemma}{lem:searches} contributes a logarithm per heap
operation. Linearity of expectation proves the bound.
\end{proof}

\begin{lemma}[Sample estimator]\label{lem:estimate}
For a positive-weight undirected sample on $t\ge2$ pivots and $O(t)$
edges, \tref{Algorithm}{alg:estimate} uses $O(t\log(t+2))$ table entries
and expected time $O(h+t\log^3(t+2))$ on $h$ demands of distinct
endpoints. It identifies disconnected pairs. For every connected
demand the value $Z$ at its returned index satisfies
\begin{align*}
  d_S(x,y)\le Z,\qquad
  \mathbf E[Z\mid S]\le256u^2 d_S(x,y),\qquad
  u=\lceil\log_2(t+2)\rceil .
\end{align*}
\end{lemma}
\begin{proof}
\begin{enumerate}
\item At the largest scale $\Delta\ge tT^2a$, even a simple path
of $t-1$ quotient edges has length at most $\Delta$. The first
center therefore covers its whole component. At a part of scale
$\Delta$ the two demand vertices have a common center at distance
less than $2\Delta$ from each. \tref{Lemma}{lem:quotlift} gives the
unconditional bound $d_S\le4\Delta+a$. This argument permits center
paths to leave the part.

\item Let $d$ be the quotient distance and let $I_\Delta$ indicate
that the endpoints have different labels at scale $\Delta$. If all
labels agree, their lowest common part is at $a$. Otherwise it is
at twice the largest separating scale, because refinement preserves
every separation at coarser scales. Consequently
\begin{align*}
 Z &\le 5a+8\sum_\Delta\Delta I_\Delta,\\
 \mathbf E[Z\mid S]
 &\le 5a+8H_t\left((\ell_a+1)d+\sum_\Delta \Delta/H\right).
\end{align*}
Here $H_t\le2u$, $T<8t$, and $\ell_a+1\le7u$. The sum of scales
is less than $4tT^2a$, so $H\ge(16t)^5$ gives
$\sum_\Delta\Delta/H\le a\le d$. Substitution in the last display
and $d\le d_S$ give the advertised constant.

\item Band anchors increase by at least $T$. For an edge of weight $w$,
the anchors whose windows contain $w$ lie in $[w/T^2,T^2w]$, a range
of ratio $T^4$. Thus every edge is retained in at most five windows.
Since only incident quotient vertices are stored,
$\sum_a(|V(Q_a)|+|E(Q_a)|)=O(t)$. Computing $T^2$ multiples costs
$O(\log t)$ additions per sample weight or anchor. Sorting the sample by repeatedly merging adjacent sorted runs
and building the merge forest cost $O(t\log(t+2))$, using smaller-list
relabeling on each union. Each relabeled vertex enters a set at least
twice as large as before, so the contraction sweep also costs
$O(t\log(t+2))$. Each demand is mapped once by its original merge and
its two current component labels. For each of the $O(\log(t+2))$
scales per quotient, \tref{Lemma}{lem:record-min} bounds the labeling
work, and integer grouping is linear in the quotient size. Scaling
by $H$ and forming the radii requires $O(\log(t+2))$ additions per
scaled number. Finally there are at most as many new parts at a scale
as quotient vertices. \tref{Lemma}{lem:trees} gives linear preprocessing
and one constant-time partition-tree query per demand. Summing yields
both complexity bounds.
\end{enumerate}
\end{proof}

\subsection{Analysis of Certificate Assignment}
\begin{lemma}[Pivot route certificates]\label{lem:names}
On a loopless pivot-routing instance with $t=O(1+N/L_N^{64})$,
\tref{Algorithm}{alg:name} gives every edge a certificate as in
\tref{Definition}{def:route-certificate}. The dictionary has
$O(t\log^2(N+2))$ entries and $P$ is a power of two of order
$\log^3(t+2)$. The algorithm terminates almost surely in expected
time $O(M+N)$, including the sorted dictionary.
\end{lemma}
\begin{proof}
\begin{enumerate}
\item For an accepted edge $e=xy$, the proposed weight is an input
weight with $a_*\le a_e$ and $d_S(x,y)\le Z\le Pa_*$.
Concatenating the physical walks represented by a shortest sample
path certifies $a_*$. A terminal edge certifies its own weight.
The map to minimum-weight parallel representatives preserves these
two inequalities for every original demand.

\item Fix the residual $U$ of a trial and put $h=|U|$. Conditional
on $S$, a demand with $d_S\le ga_e$ has
$\Pr[Z>Pa_e\mid S]\le1/8$ by Markov's inequality and
\tref{Lemma}{lem:estimate}. If $Z\le Pa_e$, its own input weight is a
feasible candidate in its group, so the minimum $a_*$ exists and the
edge qualifies. By \tref{Lemma}{lem:bad}, the expected total number
unqualified is at most $h/8+h/8=h/4$, including disconnected
demands. Rejecting requires more than $h/2$ unqualified edges, and
therefore has probability at most $1/2$.

\item Creating the representative graph costs $O(M+t)$ by stable
sorting of integer endpoint pairs. A partial uniform shuffle produces
the sample in $O(h)$ expected time. In a group of $j$ edges the median
tests take $O(j+\log(j+1)\log P)$ time by \tref{Lemma}{lem:basic-primitives},
since each test halves the number of remaining weights and uses
$\log_2 P$ doublings. Grouping by table indices and applying
\tref{Lemma}{lem:estimate} costs $O(h+t\log^3(t+2))$ in expectation
per trial. Indeed there are $O(t\log(t+2))$ groups, $j\le t^2$, and
$\log P=O(\log(t+2))$. At fixed residual, the expected cost through
the first accepted trial is at most twice this expectation. Reaching
trial $i$ has probability at most $2^{1-i}$ and does not affect its
fresh random choices. Summing the expected costs of the reached trials
gives the bound even when the cost of a trial depends on its success.

\item Accepted rounds halve the residual, whose initial size after
simplification is at most $t^2$. Their sizes sum to at most $2M$.
There are $O(1+\log(t+2))$ accepted rounds, recording at most one
name per qualifying group, and at most $32t$ terminal edges.
Sorting the resulting $O(t\log^2(t+2))$ names by comparison costs
$O(t\log^3(t+2))$. Coalescing equal names provides an old-to-new
index map for the final scan of the assignments. All terms other than
the $O(M)$ scans are $O(t\log^4(t+2))=O(N)$ under the pivot-count
hypothesis. The per-residual geometric tail and the finite number of
rounds prove almost-sure termination.
\end{enumerate}
\end{proof}

\section{Auxiliary Distances to the Owner Pivots}\label{sec:aux}
\subsection{Edge Classification and the Auxiliary Metric}

Return to the connected positive input with $N$ vertices and use the owner construction of \tref{Section}{sec:base} with $b=L_N^{64}$. Write $r_v=b_T(v,p(v))$, $R_v=D_T(v,p(v))$, and $h=b+1$. The macro rule of this section replaces the one in the local hierarchy.

Apply \tref{Lemma}{lem:names} to the projected interowner edges, whose lengths
are $A_e=R_x+w+R_y\le(2h+1)w$. For each such edge obtain its name
$\lambda_e\le A_e$ and a certified upper bound $d_e$ for a physical
walk between its two pivots, with $d_e\le P\lambda_e$ for one fixed
polylogarithm $P$ in $N$. Using its certified total length, take $d_e=P\lambda_e$ with a
power of two $P$. Compute this once for each name by doubling. An
individual projected edge is itself a physical walk, so concatenation
of projected routes gives the claimed physical certificate.
Call an interowner edge radial if $2d_e\le w$, and macro otherwise.
Make intraowner edges radial and put $d_e=0$ on them.

\begin{lemma}\label{lem:macro}
In an owner decomposition with a certified pivot-pair assignment $\lambda_e\le R_x+w(e)+R_y$ and $d_e=P\lambda_e$ bounding the pivot distance, classify an intercell edge as macro when $2d_e>w(e)$. Every cut corridor edge is macro. Each macro of weight $w$ has a name $\lambda$ satisfying $\lambda/(2h+1)\le w<2P\lambda$.
\end{lemma}
\begin{proof}
Each projected weight is $R_x+w+R_y\le(2h+1)w$ by \tref{Lemma}{lem:owners}. For a macro the radial test fails, so $w<2d_e=2P\lambda_e$. A deleted corridor edge $f$ attains the bottleneck on the tree path between its owner pivots. Their distance in the physical graph is at least that bottleneck by the MST cut property. Thus $d_f\ge w(f)>0$, ruling out $2d_f\le w(f)$.
\end{proof}
\subsection{The Owner Coordinate}
Identify all pivots into a common vertex $a$. Keep the radial edges
with auxiliary length $w'_e=w_e+d_e$, and retain all owner MST spokes.
Those spokes form an MST $T'$ of this graph. Within an owner the original
MST path satisfies the bottleneck test, and between different owners the
new tree path through $a$ has bottleneck $\max(r_x,r_y)\le w_e$.
Parallel edges are harmless. Positive loops at $a$ can be discarded.
Define $J_v=d_{w'}(v,a)$, using $J=0$ on every original pivot.

\begin{lemma}[Physical interpretation of the owner coordinate]\label{lem:coordinate}
Given an owner decomposition, identify its pivots and retain the radial edges with weights $w'_e=w(e)+d_e$, where $d_e$ certifies a pivot walk and $2d_e\le w(e)$. Let $J_v$ be distance to the identified pivot. Then $J_v$ bounds the length of a physical route to its own pivot. It satisfies $r_v\le J_v\le R_v\le h r_v$ and $|J_x-J_y|\le 3w_e/2$ on every radial edge. The retained owner forest becomes an MST $T'$ after pivot identification.
\end{lemma}
\begin{proof}
An auxiliary optimum starting at $v$ stops on its first visit to $a$.
Before identification its radial edges form a physical walk to some
pivot $q$. In reverse order concatenate the certified owner-to-owner
routes on those edges, from $q$ to the original owner $p(v)$. The full
walk has length at most $J_v$. The original MST bottleneck property
and the available owner tree path now give
\begin{align*}
r_v\le J_v\le R_v\le h r_v .                 
\end{align*}
The auxiliary triangle inequality gives for every radial edge
\begin{align*}
|J_u-J_v|\le w'_e\le 3w_e/2 .                 
\end{align*}
Thus the same numerical value provides a feasible owner spoke and a
Lipschitz coordinate across changes of owner.

For the tree assertion, the tree path for an intraowner edge has bottleneck at most its edge weight by the original MST. For an interowner edge the new tree path passes through the common root, and its bottleneck is at most $\max(r_x,r_y)\le w_e$. Adding $d_e\ge0$ preserves the cycle test.
\end{proof}
\subsection{Width Classes, Calendars, and Atoms}
In $T'$
the root bottleneck and root distance at $v$ are exactly $r_v,R_v$.
Form tree pieces $P$ of $T'-a$ by retaining parent--child edges whose
endpoint intervals $[r_v/4,4R_v]$ overlap. Form graph pieces $C$ by
the same overlap test on auxiliary edges of the graph minus $a$.
Each $P$ is a subset of a single $C$. Root $P$ at the vertex closest to the common pivot.

Generate a dyadic grid in each graph piece. Its origin is the smallest positive $r$ in the piece. Extend by at least seven octaves below the origin and two above the maximum $R$. Put $q_v$ equal to the largest grid value at most $r_v/32$, so $32q_v\le r_v<64q_v$, and put $U_v^+$ equal to the first grid value strictly above $R_v$. For each occurring pair $(C,q)$ generate consecutive cells $[iq,(i+1)q)$ from zero up to the largest assigned $U_v^+$. Merge the marker sets within $C$. A cell is visited at its starting marker. For an edge crossing graph pieces orient the dependency from the endpoint $u$ of smaller range to the endpoint $v$ of larger range, and add the marker arc $U_u^+\to16q_v$. Include every consecutive-marker arc inside each merged marker set. Process the resulting DAG in any topological order. Other widths' markers never subdivide a width-$q$ cell.
Form auxiliary atoms from the edges on nonpivots of auxiliary weight
$w'<2\max(q_u,q_v)$. Such an edge has equal endpoint $r$ because
changing that bottleneck in $T'$ requires weight at least the larger
bottleneck. The edge lies in one $C,q$, and its $T'$ path has all weights
strictly below $r_u$. It cannot go through $a$, so each atom $Y$ lies
in one component of $T'-a$ and $|Y|<b$. Give $Y$ a physical spanning
tree using its defining edges and write $L_Y$ for its total weight.
$L_Y\le 2q |Y|$. All interatom edges have length at least twice
either endpoint's cell width. At initialization post the feasible
upper bounds $R_v$ and the source's incident edge offers.

\begin{lemma}[Octave spans of overlapping ranges]\label{lem:span}
Let $T'$ be the owner-spoke tree of \tref{Lemma}{lem:coordinate} with root ranges $r_v\le R_v\le h r_v$, $h=b+1$. Use the tree and graph pieces formed by overlap of $[r_v/4,4R_v]$ off the root. Across all graph pieces, the number of dyadic grid points between the least $r_v$ and greatest $R_v$, including a constant number of points at either end, is $O(N)$. Each vertex obtains its width and upper-marker indices in $O(1+\log\log(b+4))$ time.
\end{lemma}
\begin{proof}
At a child $x$ of the common pivot, $R_x=r_x$. At any other tree-piece root $x$, $R_x\le17r_x/16$, since a cut from a nonroot parent $u$ requires
$r_x>16R_u$ and the incident tree edge then has weight $r_x$.

On a retained edge from parent $u$ to child $v$, overlap gives $r_v\le16R_u$, so $R_v=R_u+w(uv)\le17R_u$. The sum
of octave spans of all $P$, measured from their smallest $r$ to their
largest $R$, is $O(N)$.

Merging their ranges along graph overlaps
adds at most a constant number of octaves per joining edge. In each
$C$ create a common dyadic grid, based at its minimum $r$ and extending
a constant number of octaves beyond its members' ranges in both
directions. The lowest few boundaries can equivalently be represented
after scaling all quantities by a fixed power of two.

For each $v$
choose the greatest grid width $q_v\le r_v/32$ and the least grid
marker $U_v^+>R_v$.

Root a spanning tree of the retained overlap edges of each graph piece
at a vertex of minimum radius, whose radius is the grid origin.
On each tree edge $uv$, overlap gives $r_v\le16R_u\le16h r_u$
and the same inequality with $u,v$ exchanged. Visit parents before
children and store the grid index immediately below or equal to each radius.
From the parent's stored index, binary search among $O(1+\log(h+2))$
nearby grid indices for the child's radius. The grid value five indices below its
radius index is $q_v$, and $r_v\le R_v\le h r_v$ puts $U_v^+$
within $O(1+\log(h+2))$ indices above it. The two searches take
$O(1+\log\log(b+4))$ time per vertex.
\end{proof}
\begin{lemma}[Counting width cells]\label{lem:cells}
For the graph pieces and dyadic widths $32q_v\le r_v<64q_v$ of \tref{Lemma}{lem:span}, generate one width-$q$ calendar in each occupied graph-piece/width pair from zero through the largest assigned upper marker. The total number of cells and the time to merge their markers are $O(N)$.
\end{lemma}
\begin{proof}
In detail, for a tree piece $P$ rooted at $x$ and a nonempty width $q$,
every $q$-representative's path inside $P$ uses tree edges of length
less than $64q$. Let $W_P(64q)$ be the total weight of these eligible
tree edges in $P$. The normalized horizon is
$O(1+(R_x+W_P(64q))/q)$. Over nonempty $P,q$ the unit terms count at
most $N$.

At a fixed $P$, the geometric series for $R_x/q$ is $O(1)$
because $q>r_x/64$. A tree edge of weight $w$ contributes
$\sum_{q>w/64} w/q=O(1)$. The $C,q$ maximum is at most the sum of
its nonempty tree-piece bounds.

Within $C$, merge the marker grids into
an increasing chain. Reading them from coarse to fine and scanning the
old chain only to the new grid's endpoint costs linear time in the
number of cells because every previously present point in that range is also
a point of the finer grid.
\end{proof}
\begin{lemma}[Atom size at an auxiliary width]\label{lem:aux-atom}
In the owner-spoke tree $T'$ and its auxiliary graph, connect nonroot endpoints $u,v$ when $w'(uv)<2\max(q_u,q_v)$. Each resulting atom has a common graph piece, a common width and fewer than $b$ vertices. Its defining edges supply a spanning tree of weight less than $2q$ per edge.
\end{lemma}
\begin{proof}
A defining edge has weight below $2\max(q_u,q_v)$, which is below the larger endpoint bottleneck. A change of bottleneck across it would violate the MST cycle property in $T'$. Thus $r_u=r_v$, their ranges overlap, and $q_u=q_v$. The tree path between its ends avoids the common pivot, as a route from an endpoint to that pivot must cross an edge of weight $r_u$. Each atom therefore lies in one pivot-deleted tree component.
\tref{Lemma}{lem:owners} and the construction of the pivot set give fewer than $b$ vertices in that component.
\end{proof}
\begin{lemma}\label{lem:calendar}
Let $T'$ be the owner-spoke MST of \tref{Lemma}{lem:coordinate} on $N$ vertices, with hop bound $h=b+1$, and let $m$ be the auxiliary edge count. The grids and width calendars constructed from the overlapping intervals $[r_v/4,4R_v]$ contain $O(N)$ entries altogether and can be constructed in $O(m+N\log\log(b+4))$ operations. The components used as atoms contain fewer than $b$ vertices. All marker arcs increase numerical time.
\end{lemma}
\begin{proof}
\tref{Lemmas}{lem:span}, \ref{lem:cells} and~\ref{lem:aux-atom} give the size and construction bounds. It remains to order the markers. If an edge crosses graph pieces, label its
ends $u,v$ with $r_v>16R_u$. Only $u$ can precede $v$ on a shortest
auxiliary path since $J_u\le R_u<r_v\le J_v$. Connect the marker
$U_u^+$ in the first chain to $16q_v$ in the other. The first is at
most $2R_u$, while $16q_v>r_v/4$.

The source cell of $u$ ends by the
first marker and the cell containing $J_v$ starts at or beyond the
second. All marker arcs advance physical time, giving a DAG in total
$O(m+N)$ size. Cells of coincident starts may be processed in any
order consistent with those chains and arcs.

Each width calendar is stored on its own grid, with pointers into the common chain, so visiting a marker activates only lists starting cells there. The sum of all such registrations equals the number of cells.
\end{proof}
\subsection{Local Searches, Clipping, and External Arrivals}
\paragraph{Family gaps and activation lists.}\label{par:family-gaps}
To schedule the first appearance of each atom, assign it to the $P,q$
of one representative. Within each such family order the representatives
by the preorder of $T'$, obtainable by integer sorting. The walks between
consecutive representatives along $T'$ have total length at most
$2W_P(64q)$, since their root paths inside $P$ use only these eligible
edges. Obtain each consecutive gap $g(i,j)$ from tree distances. Schedule
a scan of the family in every cell of its $C,q$ calendar from zero
through the inclusive horizon
$\max_{Y\text{ in family}}(R_{\rm rep(Y)}+L_Y)$, valid for all its
vertices. Indeed, for each defining edge of $Y$, its path in the MST $T'$ uses only edges lighter than $2q$. Along that path the root bottleneck stays at the common endpoint value $r\ge32q$, since a change of bottleneck requires an edge at least as heavy as the larger value. Thus every path vertex belongs to the same graph piece and width, and each path edge belongs to $Y$. It follows that $T'[Y]$ is connected. It is an MST of the internal graph of $Y$, since a lighter spanning tree there could replace it in $T'$. Its total weight is therefore at most $L_Y$. For $v\in Y$, the tree triangle inequality gives $R_v\le R_{\rm rep(Y)}+D_{T'}(\mathrm{rep}(Y),v)\le R_{\rm rep(Y)}+L_Y$.

The sum of these scan counts is $O(N)$ by the preceding
tree-piece calculation and $\sum_Y L_Y/q=O(N)$. For a horizon beyond
the existing $C,q$ marker chain one may stop at its original end,
which already strictly exceeds every actual distance in that class. For each family, walk its width calendar from the first cell, stopping after the last cell whose start is at most the family horizon or when the calendar ends. Append one family record to the activation list of each visited cell's starting marker using the calendar pointer. The bound on the sum of the family scan counts pays for all records and the work to create them. At a marker traverse just its activation list.

Within a family maintain a linked list of registered atoms and its
contiguous still-unregistered runs. A run splits at the atom $i$
containing its first due minimum $z$ in a cell $[a_0,b_0)$. For each
of its at most two new nonempty subruns, let $v$ represent its boundary atom $j$ adjacent to $i$. Propose $u=\max\{b_0,z+L_i+g(i,j)\}$ at the native inward copy of $v$. Discard the proposal if $u>R_v$ or it fails to decrease the pending key. This copy already has an initial offer $R_v$ and its atom is still unregistered, so its key after the test is at most $u$ in every case. The auxiliary physical tree walk certifies feasibility. The padding precludes a new due event
behind the current cursor. The initial run has feasible minimum at
most the largest representative $R$ in its family. Thus all its
unsuccessful initial checks are covered by the family scan count.
While the family remains active, a surviving run with boundary key at most $u$ must split by the first scan whose cell ends strictly after $u$. If the scans end sooner, the run disappears at that point. Its unsuccessful check count is at most
$O(1+(L_i+g(i,j))/q)$. Each atom gives rise to at most two new
runs, each gap is charged at most twice, and the normalized tour
lengths have sum $O(N)$. Therefore the extra processing is linear
apart from split--findmin. As in the contiguous-run evaluator, a
decrease updates the sequence containing its state, a run split
produces at most two other runs, and each original relaxation costs
only constantly many sequence operations. The online split--findmin cost is
$O(m+N\tau(N))$ globally by \tref{Lemma}{lem:sequences}.

Each atom $Y$ uses the full induced graph $G'[Y]$ with auxiliary lengths. The light edges determine membership and supply a discovery tree. Once atom identifiers are known, a single edge scan collects every internal edge, including heavier chords, and lists each interatom exit at its tail. For evaluation, the physical graph is the full $G'$ with source $(a,0)$ and target distances $J_v$. Every internal walk projects into $G'$, and an offer may be feasible by a walk that enters from outside $Y$. Post input offers to native inward copies. Allow every local copy to relax the outside physical exits of the vertex it represents. Suppress such an exit from $u$ if its emitted timestamp exceeds $R_u$, and discard an external offer to $v$ if it exceeds $R_v$. Both comparisons are against precomputed real upper bounds on the protected times. For an edge between graph pieces keep only the direction from the smaller range to the larger one. In each family the registration time $z$ together with $L_Y$ sets a lasting inclusive deadline $z+L_Y$. Start the local representation once there is a due offer.
After registration keep each atom live through at most
$O(1+L_Y/q)$ cells, using the smaller of its own and inherited physical
deadlines. The internal auxiliary edges of $Y$ form a connected undirected graph, and its given tree has edge weights below $2q$. Thus $Y$ is a replaceable atom at cutoff $q$ (\tref{Definition}{def:atom}). \tref{Lemmas}{lem:cap} and~\ref{lem:splice} give a single capped root, native entries and arbitrary-copy exits. Normalize this root by \tref{Lemma}{lem:mass}. Every internal unit is at most $q$.

\paragraph{Visiting an active atom.}\label{par:atom-cell}
At a singleton root, emit the due minimum once, relax the allowed exits through the endpoint caps, and retire the leaf. At an internal root, open on the first due key $D_0$ and anchor its clock at $D_0$. Keep the smaller of the atom deadline and the root expiry $D_0+S(\mathrm{root})$ as the inclusive cap $c$. In each cell visit the successive nominal root slots whose intersections with the cell contain times at or below $c$. For each such intersection $I$, call \algcall{alg:visit}{Scan} on the root with $(I,c)$ and relax the allowed exits on emission. On resuming any clipped slot, start at its first live child rank with its registrations, offers, and deadlines intact. Stop once all children finish or the cap has been covered, including equality.
\begin{algorithm}[ht]
\caption{Visiting the auxiliary width calendars}\label{alg:aux}
\Fn{\textsc{OwnerDistances}($T',G'$)} {
\KwIn{Positive auxiliary graph $G'$ and its owner-spoke MST $T'$ rooted at the common pivot $a$.}
\KwOut{Exact distances $J_v$ from the common root.}
Compute $r_v,R_v$ down $T'$ and build the tree and graph pieces, calendars and interpiece marker arcs (\tref{Lemma}{lem:calendar})\;
Find the components $Y$ of the nonroot edges with $w'<2\max(q_u,q_v)$, retaining a discovery tree of total length $L_Y$ and a representative per component\;
Group the atoms by the tree piece and width of their representative\;
Construct the \hyperref[par:family-gaps]{family gaps and activation lists} in representative preorder using the tree queries of \tref{Lemma}{lem:trees}\;
Collect the induced graphs $G'[Y]$ and interatom exits in one edge scan. Run \algcall{alg:construct-local}{LocalHierarchy} on each $G'[Y]$ and cap it at its width\;
Normalize each hierarchy and construct its child orders, representatives, and gaps (\tref{Lemmas}{lem:mass} and~\ref{lem:gaps})\;
Initialize one leaf layout in family order, then the preorder of atom representatives in $T'$ within each family, then local child order, with infinite keys and a split at every family boundary\;
Post $R_v$ at each native inward copy and all direct offers from the root\;
\ForEach{marker in topological order}{
 \ForEach{family in the activation list of this marker}{
  Register due atoms by splitting the family's live runs (\tref{Lemma}{lem:sequences})\;
  Visit each active atom by the \hyperref[par:atom-cell]{cell procedure above}, emitting a due singleton or scanning the intersecting root slots\;
 }
}
\Return{the minimum emitted value at each vertex, with zero at the root}\;
}
\end{algorithm}

\begin{lemma}\label{lem:clip}
Let a seeded ordered hierarchy have one root and units at most $q>0$. Visit it in consecutive cells $[jq,(j+1)q)$, splitting local slot visits at cell boundaries. Open the root on its first due key; an empty cell leaves an unopened root unopened. Assume each external offer is feasible, arrives before the visit to its timestamp's cell, and is never inserted below the current cell's left endpoint. The first cell starts at or before every finite initial timestamp. Assume the visited cells cover every protected timestamp, including any equal to an inclusive deadline. Then every protected prefix emits in its cell. On $N$ states and $M$ arcs, with $Q$ initial offers, $E$ external offers, and $C$ cell visits, the total work is $O(M+N\tau(N)+Q+E+C)$. A tight witness here may start at an external offer.
\end{lemma}
\begin{proof}
\tref{Figure}{fig:clipped} illustrates a cut. Intersect each live nominal slot $[a_1,b_1)$ with the current external cell and the inherited visit. Process that portion with the usual permanent inclusive cap. At the end of an external cell, retain registrations, cursors' nominal slot identities and unapplied offers. In the next cell restart each intersected slot at its first live child rank. An unopened nonroot node $C$ uses the two-case origin rule of \tref{Section}{sec:eval}, with $b$ equal to its parent's nominal right endpoint. If its first due key $D_0$ satisfies $D_0+S(C)<b$, start at $D_0$. Otherwise start at the largest $b-j\nu(C)$ no larger than $D_0$. In this second case $b-D_0\le S(C)$, so successive steps take $O(1+S(C)/\nu(C))$ time. The root starts at its first due key as specified above. Intersect the resulting nominal visit with the external cell. New subruns use $b_1$ for padding, even when the portion ends sooner. No temporary cut retires a node. Implement this lazily on the recursive calls actually visited in this atom.
\begin{enumerate}
\item The last entrance to a child of a protected path has a predecessor strictly before the current portion, a predecessor in a lower-ranked child within the portion, or an available outside entry. These are precisely the three cases in \tref{Lemma}{lem:evalcorrect}. A late nontight key smaller than the current left endpoint could hide a due tight key in the same unopened interval. The arrival hypothesis excludes this case for every inserted key.

An external entry at the shared boundary belongs to the new portion. Rescanning from the first rank makes it available even at an earlier rank. The length of the portion is at most the nominal unit, so the strict light-arc test applies.
\item A nominal slot has at most one interior boundary of this atom's external grid. Internal parent and child slots align while active across a parent boundary, by \tref{Lemma}{lem:mass} and the origin rule. Thus each local nominal scan, including scans of persistent children and unsplit runs, occurs at most twice. A cursor resumes from the first live piece, so the charged unit is the same list-piece visit as in \tref{Lemma}{lem:evaltime}. Markers belonging to other widths cause no calls into this atom. Each external offer requires one key update and each cell call has constant overhead beyond its scans, giving the stated bound by \tref{Lemma}{lem:evaltime}.
\end{enumerate}
\end{proof}

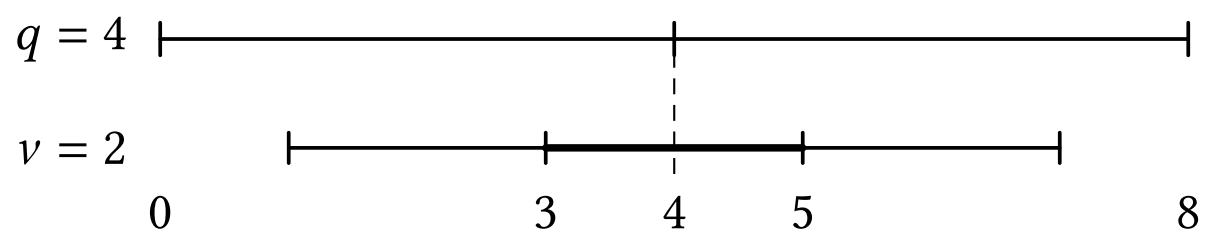
\begin{figure}[ht]
\centering
\begin{tikzpicture}[x=.85cm,y=.72cm,font=\small]
\draw[paper graph edge] (0,1)--(8,1);
\foreach \x in {0,4,8}{\draw[paper graph edge] (\x,.85)--(\x,1.15);}
\node[left] at (-.1,1) {$q=4$};
\draw[paper graph edge] (1,0)--(7,0);
\foreach \x in {1,3,5,7}{\draw[paper graph edge] (\x,-.14)--(\x,.14);}
\node[left] at (-.1,0) {$\nu=2$};
\draw[paper highlighted edge] (3,0)--(5,0);
\draw[densely dashed] (4,-.24)--(4,1.2);
\foreach \x in {0,3,4,5,8}{\node[below] at (\x,-.25) {$\x$};}
\end{tikzpicture}
\caption{An external width-four boundary cuts the local nominal slot $[3,5)$ at time $4$. The visits process $[3,4)$ and $[4,5)$, each in child-rank order. A relaxation at time $4$ is available in the second visit. A supplemental offer generated in either visit is padded to at least $5$. Each local slot encounters at most one such cut.}\label{fig:clipped}
\ifdefined\Description
\Description{Two horizontal grids have widths four and two. The upper grid has a boundary at time four. The local slot between three and five crosses that boundary and is visited in two pieces.}
\fi
\end{figure}
\begin{lemma}[Time order of external offers]\label{lem:chronology}
In \tref{Algorithm}{alg:aux}, retain an interatom offer of value $z+w'$ only in an allowed direction and only when $z\le R_u$ and $z+w'\le R_v$. Each effective decrease at a state still accepting offers is available before the starting marker of its destination cell. Every tight interatom step passes the direction and endpoint tests. At a pending head an equal tight key already suffices, and a settled or retired head ignores later proposals.
\end{lemma}
\begin{proof}
Fix a tail $u$ and head $v$ on a crossing. A tight step has timestamps $J_u,J_v$ within the two endpoint caps by \tref{Lemma}{lem:coordinate}.
\begin{enumerate}
\item \emph{Different graph pieces.} In the permitted direction $r_v>16R_u$. The tail cell ends at or before its marker $U_u^+$ since $z\le R_u<U_u^+$. The crossing's weight is at least $r_v$, so the head cell starts at or after $16q_v$. The arc $U_u^+\to16q_v$ puts the tail visit first. No tight step goes in the discarded direction, since $J_u\le R_u<r_v\le J_v$.
\item \emph{One graph piece.} The two atoms are distinct, so $w'\ge2\max(q_u,q_v)\ge q_u+q_v$. The tail cell ends by $z+q_u$ and the head cell starts strictly after $z+w'-q_v$. Their order follows from the common chain of markers.
\end{enumerate}
The head cap puts the arrival at or below $R_v<U_v^+$, so its cell lies in the width calendar. By the \hyperref[par:family-gaps]{family horizon bound}, $R_v\le R_{\rm rep(Y)}+L_Y$ for the destination atom $Y$. Its cell's starting marker therefore also activates the destination family, including equality at the horizon. All initial offers are available before the first marker. For a family proposal that decreases the pending key at $v$, padding and the representative cap give $b_0\le u\le R_v$. Its cell is therefore later on the same width calendar and within the family horizon. A discarded proposal causes no scheduled arrival. Thus the arrival hypothesis of \tref{Lemma}{lem:clip} holds also at the first visit and after a family splits.
\end{proof}
\begin{lemma}\label{lem:aux}
\tref{Algorithm}{alg:aux} computes all $J_v$ in deterministic $O(m+N\tau(N))$ time once the names are given, where $\tau(N)=1+\log\log\log(N+16)$.
\end{lemma}
\begin{proof}
By \tref{Lemma}{lem:chronology}, all retained external arrivals meet the timing conditions of \tref{Lemma}{lem:clip}. The calendars start at zero. Before registration the family checks its unopened runs for a due minimum. After registration the inclusive deadline covers every $J_v$ in the atom by the triangle inequality, and the family horizon covers their cells.

Inside a cell, take the last edge entering its target atom on an optimal path. It was posted from the common root or an earlier cell. The remaining segment is an internal geodesic. \tref{Lemma}{lem:splice} realizes it from its native entry and \tref{Lemma}{lem:clip} processes it, proving exactness by induction in the marker order.

For work, \tref{Lemma}{lem:calendar} bounds calendar construction. The family horizons, consecutive gaps and atom diameters were charged above, giving $O(N)$ family and atom visits beyond the local evaluations. Their local graphs have $O(N)$ states and $O(m+N)$ arcs in total, because their physical domains are disjoint and the number of copies per vertex is bounded.

Their construction costs $O(m+N\log\log(b+4))$. Collecting induced edges and exits, normalizing the hierarchies, and building their child orders and gaps cost $O(m+N)$ in total by \tref{Lemmas}{lem:mass} and~\ref{lem:gaps}. One split--findmin layout can respect family order, atom order within each family and the local hierarchy order within each atom, since these groupings are nested. Operations on runs can therefore be paid once globally. \tref{Lemmas}{lem:mass}, \ref{lem:evaltime} and~\ref{lem:clip} bound the remaining work. Since $b\le L_N^{64}$, the asserted bound follows.
\end{proof}

\section{The Outer Hierarchy and the Single-Source Algorithm}\label{sec:full}
\subsection{The Outer State Graph and Threshold Components}
After \tref{Algorithm}{alg:aux} has returned $J$, delete any radial edge with
$w_e>J_x+J_y+d_e$. There is a strictly shorter physical route via the
two owner spokes and their pivot detour, so no physical geodesic uses
such an edge. An owner tree edge remains because on the edge from a parent
to a child $v$ we have $J_v\ge r_v\ge w_e$. Every surviving radial
edge satisfies
\begin{align*}
w_e\le 2(J_x+J_y).                       
\end{align*}
The $J$ and detour certificates remain valid in the original metric.
Use the two-shell construction of \tref{Section}{sec:base} on the coordinate
$J$ with $K=16$. Near positive radial endpoints are within a factor
$K^2$, and each near component has a base-$K$ grid from its minimum
positive $J$. A vertex occurs in at most two inward windows (each
covering shells $j,j+1$) and two outward windows (covering $j-1,j$).
Intrawindow near edges are bidirectional. Lowering the inward shell
record or raising the outward record goes to the new native copy.
Far inward transfers descend to native inward and far outward transfers
ascend to native outward. The zero switch goes from an inward copy to
native outward. At zero use the two pivot copies. A macro may leave
any ordinary copy and resets to native inward. The running-extremum
argument gives a full state realization of every physical geodesic.
All grid construction and radial copying take $O(m+N)$ since $K$ is
absolute. With macros absent the ordinary two-phase block quotient is
acyclic, by monotonicity of shell bottoms and the one-way switch.

The coordinate $J$ need not be monotone on the original owner tree.
Add to each nonpivot $v$ two special states $v_{\rm in},v_{\rm out}$.
An ordinary inward copy of $v$ may enter $v_{\rm in}$ at zero.
The inward tree states move to their parents along physical owner
edges, ending at the ordinary inward pivot copy. From an ordinary
outward pivot copy one may move down the tree of outward tree states.
At $v$ that state may enter the ordinary native outward copy at zero.
The only ways to enter these outward tree states are from their parent
tree state or its owner pivot. Macro arcs have only ordinary tail and
head copies. These additional arcs project to physical walks and
therefore preserve all distance lower bounds. They preserve the
macro-free block DAG because the extra inward-to-outward routes pass from
ordinary inward through the inward tree, pivot switch and outward
tree, terminating in ordinary outward. Ordinary inward-to-pivot and
pivot-to-ordinary-outward radial arcs respect that order as well.

Make each special state a singleton block. Compute the two unweighted and two bottleneck propagation values in macro-free block order as in \tref{Section}{sec:base}. Build the auxiliary pivot vertices, weighted corridor joins, local block MST joins and state-to-owner joins at $\theta$. In a positive ordinary block $U$ put $s_U=\min_{v\in U}r_v$. Choose the local cutoff $t_U$ as specified below. The connected subgraphs of edges lighter than $t_U$ will be replaced by local representations.

\begin{lemma}\label{lem:peak}
In the outer radial state representation with coordinate $J$ from \tref{Lemma}{lem:coordinate}, shell ratio $K=16$, and radial deletion test $w>J_x+J_y+d_e$, the surviving radial edges satisfy $w\le2(J_x+J_y)$. Every globally shortest radial run, split at a minimum coordinate, has inward and outward oscillation at most a factor of $16$. The ordinary graph realizes every geodesic, and the augmented macro-free quotient is acyclic.
\end{lemma}
\begin{proof}
Deletion requires $w>J_x+J_y+d_e$. The spokes and certified detour then give a strictly shorter replacement. For a surviving edge $d_e\le w/2$, so $w\le2(J_x+J_y)$.

Consider a geodesic radial segment with endpoint coordinates $J_x,J_y$
and peak coordinate $M>0$. If $J_x,J_y\le M/16$, then \tref{Lemma}{lem:coordinate} gives
segment length $L\ge (2/3)(M-M/16)=5M/8$. The two owner spokes and
the chain of pivot detours along the segment together cost at most
$J_x+J_y+L/2\le M/8+L/2<L$. Thus a geodesic cannot have these two
small endpoints.

Splitting a radial run at its minimum $J$ gives an
inward prefix on which $J$ never exceeds 16 times the running minimum,
and an outward suffix on which $J$ never drops below $1/16$ of the
running maximum. 
In detail, if an inward prefix ever exceeds $16$ times its running minimum, the subpath from that earlier minimum to the splitting vertex has both endpoint coordinates below one sixteenth of its peak. Applying the same argument in reverse proves the outward bound.

The running-shell proof of \tref{Lemma}{lem:lift} now applies with the constant $K=16$. Each added tree arc points towards the pivot in the inward phase or away from it in the outward phase. Entering the inward tree is irreversible before the pivot. Leaving the outward tree is irreversible after the pivot. Thus no new macro-free cycle between blocks is introduced.
\end{proof}
\begin{lemma}[Bottleneck of access to a macro edge]\label{lem:outer-bottleneck}
Consider an outer radial state representation with shell ratio $K=16$, hop bound $h$, and the additional directed owner-tree states. At any state let $B_0$ denote the larger of the minimum attainable incoming and outgoing macro weights, ignoring radial weights. Let $\theta$ use instead the maximum weight along each entire access walk, minimized separately on each side. The two values are finite together and, when finite, satisfy $r\le\theta$ and $B_0\le\theta\le4K^2hB_0$.
\end{lemma}
\begin{proof}
Put $B_0=\max(I_0,O_0)$ and
$\theta=\max(I_*,O_*)$. If $B_0$ is finite, concatenate its two
attaining walks through the state under consideration. This gives
one macro-free walk between a preceding macro head and a following
macro tail, both ordinary states with $J\le hB_0$ by \tref{Lemma}{lem:coordinate}.

The inward ordinary part has $J\le K^2 h B_0$ by nonincreasing shell
bottoms. The outward part has the same bound read from the terminal
macro tail in reverse, by nondecreasing bottoms. These assertions also
cover walks staying in a single phase, or passing through a pivot
without using its corresponding optional tree states. 

Walks using the special states have at most one inward tree segment, going
toward an owner, and at most one outward tree segment, going away
from an owner. In the former all tree-edge weights are at most $r$
at the entering ordinary vertex. In the latter they are at most $r$
at the terminal ordinary vertex. Both incident ordinary vertices
occur on the just-bounded parts, with $r\le J$. If either tree
segment is empty its bound is vacuous. \tref{Lemma}{lem:peak} then gives
\begin{align*}
B_0\le\theta\le 4K^2h B_0,\qquad r_z\le\theta(z). 
\end{align*}
For the last bound append physical edges from the preceding macro,
whose ordinary head has $r\le w$. An increase in the original tree
bottleneck requires an edge of at least its new value. All added
zero transitions preserve the physical vertex. The inequalities
remain valid for a special state because the
complete concatenation still begins and ends in ordinary copies.
Both quantities are infinite together.
\end{proof}
\begin{lemma}\label{lem:support}
Augment the outer radial state representation by the directed owner-tree states described in \tref{Section}{sec:full}. For every $W>0$, the block MST edges, state-to-owner joins at $\theta$ and corridor edges up to weight $W$ describe exactly the state components strongly connected by arcs of length at most $W$. Using only joins and arcs strictly below $W$ gives the corresponding strict-threshold components.
\end{lemma}
\begin{proof}
At a physical threshold $W$, components of the pivot corridor tree
using edges of weight at most $W$ still describe the macro-containing
SCCs. Each such corridor is macro. Its owner endpoint is reachable
from the pivot outward copy via the special tree path and the zero
transition to the native outward state, with every weight at most
the corridor weight. From the native inward macro head the special
inward tree reaches its owner with the same bottleneck bound.

Apply
this also to any macro of weight at most $W$. A state with
$\theta(z)\le W$ thus has a route from a core and back to a core.
They are the same core because every visited interowner physical edge has
corridor bottleneck at most its own weight.

Conversely, a closed walk
in a macro-containing SCC exhibits its previous and next macro at
threshold $W$. A macro-free SCC belongs to one ordinary undirected
block or one singleton special state. The ordinary local minimum
spanning trees, the joins at $\theta(z)$ to the owner auxiliary, and
the pivot corridor joins therefore give the same compact union
description of \tref{Lemma}{lem:join}. For the strict version, choose a positive threshold below $W$ and above every arc and join weight strictly below $W$. Such a threshold exists because the lists are finite. The inclusive characterization there is the strict characterization at $W$.
\end{proof}
\subsection{Numerical Scales and Atom Substitution}

For each macro edge replace its assigned name by the least physical weight of any macro with that name. \tref{Lemma}{lem:macro} shows that the weights of macros sharing an old name differ by a factor of at most $2P(2h+1)$. A finite $B_0$ stores the pointer of an attaining macro, so the new name for a join is at most its $\theta$ and within a fixed polylogarithmic factor of it. Take the least power of two $F$ strictly larger than the product $2P(2h+1)(8K^2h)^2$. Choose it using integer parameters once for the graph. Initialize the support $[a,F^3a]$ for every distinct finite join name $a$ and every pivot corridor weight $a$, including names used only by singleton blocks. Every singleton block, including each pivot or special state, contributes its own finite $\theta$ join when one exists. Such a block has no internal merge. It skips the block-scale test, block grid, MST classification and atom substitution. For each nonsingleton positive ordinary block with a finite attachment, pick a minimizing state and its name $a_U$. Call the block aligned when $F^2s_U\ge a_U$.
If $F^2s_U<a_U$, every edge of its MST has weight below $a_U/2$, hence below
half its earliest join. For each aligned block lower its name's interval start to the smaller of its current start and $s_U$. These intervals cover the significant aligned weights and joins. Sort the interval endpoints and merge overlaps and gaps
smaller than a factor two, as in the baseline proof. Each input
interval adds $O(\log\log(N+4))$ octaves. Pointers from each name
into the resulting table give binary search on that many positions.
For an aligned block take the floor $t_U$ of $s_U$ on its name's global grid.
Every other nonsingleton positive ordinary block takes this floor on the separate grid described below.
Both choices give $s_U/2<t_U\le s_U$.
Consequently unaligned blocks and aligned atoms precede their
first outer merge. Processing unions by their integer grid classes,
suppressing auxiliary nodes and topologically ordering arcs lighter
than each node gives a partition hierarchy in the sense of \tref{Definition}{def:partition}. Use the child-quotient searches of \tref{Section}{sec:local-assembly} to extract the physical connecting trees, assigning each arc by the parts containing its endpoints and retaining its actual endpoint states. For a merge class $[a,2a)$, the strict version of \tref{Lemma}{lem:support} supplies connectivity using arcs of weight strictly below $2a$. Replacing its atom leaves gives the ordered hierarchy required by \tref{Section}{sec:eval}.

For a block using a separate local grid, pick $v$ attaining $s_U$. Every $J$ in the block is at most $K^2J_v$, because the two native shells defining its window have this aspect. Consequently the largest local edge is at most $4K^2R_v$ and the smallest significant edge is greater than $r_v/2$. Use the auxiliary grid pointer of $v$ for $r_v$ and $R_v$, extending each piece's table at both ends by a fixed number of octaves just once. This supports the block-specific binary search in $O(1+\log\log(b+4))$ comparisons.

\begin{lemma}\label{lem:outercost}
For the augmented outer radial state representation on an $N$-vertex graph with $m$ edges, take $b=L_N^{64}$ and use pivot names satisfying \tref{Lemma}{lem:names}. Its significant local events (block MST weights at or above their block cutoff) and finite state-to-pivot joins can be classified in $O(m+N\tau(N))$ work, including all global grids. Each subthreshold atom lies in one pivot-deleted component and has fewer than $b$ vertices.
\end{lemma}
\begin{proof}
\begin{enumerate}
\item Every edge inside a two-shell window has weight at most $4K^2J_v\le4K^2h s_U$ using a minimum-radius $v$. In a block below the alignment cutoff this bound is less than $a_U/2$, which in turn is at most half its earliest attachment. Thus the whole block is joined before any of its states meets a core. Its members have equal attachment threshold, since the connecting edges are all lighter than the minimum. The block can use its own grid below that attachment.

For an aligned block, \tref{Lemma}{lem:outer-bottleneck} gives $s_U\le\min_{z\in U}\theta(z)\le Fa_U$. Hence $a_U/F^2\le s_U\le F a_U$ and significant edge weights are covered by the expanded support for its name, except possibly the lower rounding boundary. For a support $[l,u]$ assigned to this block, $l\le s_U$, so the grid floor of $l$ is at or below $t_U$. Starting the search at that floor includes every significant weight below $l$. Weights below the designated floor are local.
\item Sort interval endpoints. Merge intervals that overlap or have a gap of ratio less than two. The total number of octaves is $O(k\log F)$ for $k$ names, by summing interval widths on the logarithmic line and charging at most one octave per bridged gap. The dictionary size in \tref{Lemma}{lem:names} gives $k=O(t\log^2(N+2))$ including the $O(t)$ corridors. Sorting and generating the grids costs $O(N)$.

Merge the sorted endpoints and names with the generated grid coordinates once. For each support $[l,u]$, store the indices from the grid floor of $l$ through the first grid point strictly above $u$, adding one final doubling per merged grid if needed. The upper sentinel is used only for lookup. These indices span $O(\log F)$ positions per name. Threshold tests for the alignment grouping can use median pruning with $O(\log F)$ doublings per tested median as in \tref{Lemma}{lem:names}. There are $O(N)$ events, each requiring $O(1+\log\log F)=O(\tau(N))$ comparisons.
\item If $w<t_U$, neither the owner nor the original radius can change by \tref{Lemma}{lem:owners}. Its MST path has bottleneck below $r_v$ and avoids all pivots. \tref{Lemma}{lem:anypivot} gives this conclusion for every pivot. Transitivity puts all vertices of one atom inside one component of $T-A$. That component has fewer than $b$ vertices.

The low local unions may be replaced by a single atom before making the outer hierarchy. Below $t_U$ they are within one bidirectional block.

A cycle through different blocks must use a macro. Joining a macro to such an atom at a threshold below $t_U\le r_v$ is impossible by \tref{Lemma}{lem:outer-bottleneck}, also after zeroing its internal links, since these links preserve the common radius and owner. This justifies the temporary contraction at every outer event.
\end{enumerate}
\end{proof}
\subsection{The Single-Source Algorithm and Main Proof}
\begin{algorithm}[ht]
\caption{Single-source distances and a predecessor tree}\label{alg:sssp}
\Fn{\textsc{ShortestPaths}($G,s$)} {
\KwIn{An adjacency-list undirected multigraph with indexed vertices, finite nonnegative weights, and a source $s$.}
\KwOut{A distance for each vertex and a predecessor tree on the reachable vertices.}
Contract zero-edge components, discard all resulting loops, retain the source component, and keep original edge identifiers\;
Let $N$ be the number of retained vertices\;
\If{$N\le L_N^{64}$}{
 Retain a lightest edge per vertex pair and compute exact source distances by the scan-based Dijkstra search (\tref{Lemma}{lem:searches})\;
 Restore the original vertices, assign unreachable ones infinity, and traverse the finite tight original arcs from $s$ (\tref{Lemma}{lem:restore})\;
 \Return{restored distances and the tree of first-discovery arcs}\;
}
Build its MST with the three growth passes of \tref{Lemma}{lem:mst}, form its owner cells, and delete strictly tree-dominated nontree edges\;
Run \algcall{alg:name}{NameRoutes} on the projected interowner edges\;
Classify macro and radial edges. Identify all pivots at a common root to form $G'$ from the radial edges with lengths $w_e+d_e$, discarding loops. Let $T'$ be the identified owner forest\;
Compute the owner coordinates by calling \algcall{alg:aux}{OwnerDistances} on $(T',G')$\;
Delete each radial edge $e=xy$ with $w_e>J_x+J_y+d_e$ from the graph used to construct the outer states\;
Construct the outer states, threshold joins, grids, and capped local hierarchies via \algcall{alg:construct-local}{LocalHierarchy}\;
Seed the native inward outer copy of $s$ with zero, using its native inward local subcopy if substituted\;
Evaluate the ordered hierarchy with \algcall{alg:evaluate}{Evaluate} and take the minimum emitted value for each vertex\;
Restore the original vertices, assign unreachable ones infinity, and traverse the finite tight arcs from $s$\;
\Return{restored distances and the tree of first-discovery arcs}\;
}
\end{algorithm}

\begin{lemma}[Restoring a predecessor tree]\label{lem:restore}
Let $G$ be a finite nonnegative undirected graph, and contract every zero-edge component. Given exact distances in the contracted source component, all original distances and a shortest-path tree of the reachable vertices can be recovered in $O(m+n)$ time, using only original edge weights in the output.
\end{lemma}
\begin{proof}
Zero edges identify vertices at equal distance, and every path in the contracted graph lifts through a zero path inside each contracted component. On the restored graph orient $uv$ from $u$ to $v$ whenever $d(u)+w(uv)=d(v)<\infty$.

Every reachable vertex has a shortest path of tight arcs from $s$, since each prefix of any shortest path is shortest. A traversal thus reaches every such vertex, and the first-discovery arcs give a tree even with zero cycles.

Compute the distance stored in the output by summing original edge weights down the tree. These sums equal the restored labels by induction from the root. This last operation also returns ordinary real values if intermediate differences were represented formally.
\end{proof}
\begin{proof}[Proof of \tref{Theorem}{thm:main}]
Contraction turns every zero edge into a loop. Deleting the loops leaves only positive edges. The base-case convention of \tref{Section}{sec:prelim} applies here. For the hierarchy construction, \tref{Lemma}{lem:mst} supplies the required spanning tree.

 \tref{Lemma}{lem:names} supplies valid names for every random outcome accepted by \tref{Algorithm}{alg:name}. \tref{Lemma}{lem:aux} supplies the exact owner coordinates, and \tref{Lemmas}{lem:peak} and~\ref{lem:support} then give the tight-path representation and its threshold components. \tref{Lemmas}{lem:outercost} and~\ref{lem:substitution} construct an ordered hierarchy for these paths. \tref{Lemma}{lem:evalcorrect} supplies at least one exact copy per vertex and no underestimate. The minimum over the copies is therefore the desired distance.

\tref{Lemma}{lem:restore} now gives the claimed output on the original graph.

For the work bound, $N\le n$ is the size of the contracted reachable graph.
\begin{enumerate}
\item \emph{Preprocessing and sampling.} Graph traversal takes $O(m+n)$ time. \tref{Lemma}{lem:mst} constructs its MST in $O(m+N\tau(N))$ time. Routing is the only randomized phase and costs $O(m+N)$ in expectation by \tref{Lemma}{lem:names}.
\item \emph{Representations.} There are $O(N)$ states, $O(m+N)$ state arcs and $O(N)$ merge events. \tref{Lemmas}{lem:aux}, \ref{lem:outercost} and~\ref{lem:substitution} charge $O(m+N\tau(N))$ for constructing the two representations. Here $b\le L_N^{64}$ gives $1+\log\log(b+4)=O(\tau(N))$.
\item \emph{Evaluation and output.} \tref{Lemma}{lem:evaltime} charges $O(m+N\tau(N))$ for the final evaluation. \tref{Lemma}{lem:restore} adds one linear scan. Summation gives the asserted expectation; \tref{Lemma}{lem:names} also gives almost-sure termination.
\end{enumerate}

\end{proof}

\section{Minimum-Cut Approach and Techniques}\label{sec:techniques}

Write $\lambda$ for the minimum cut value. We first remove zero
capacities and test connectivity; a disconnected positive-capacity
support gives an immediate zero cut. Thus the discussion below
assumes a positive minimum cut. A cut \emph{two-respects} a tree
when at most two tree edges cross it. The tree need not be a
subgraph of the input for the local cut algorithm, although the
recursive construction returns an original-edge spanning tree.

\subsection{From a tree to its best two-respecting cut}

Fix a reference tree $T$. Deleting one tree edge partitions the
vertices into two components and thus defines a one-respecting cut.
Its capacity is the total input capacity whose tree path uses that
edge. All one-respecting cut capacities can be computed together in
linear time.
Deleting two tree edges determines another cut: an input edge
crosses it precisely when its tree path uses exactly one of
the two selected tree edges.

The local algorithm uses a structural witness for the optimum.
If a one-respecting cut is not already optimal, choose an optimum
pair with the fewest tree edges strictly between it. Every edge
strictly between the pair has larger one-respecting cut capacity
than either endpoint. Otherwise, replacing an endpoint
by such an interior edge would improve the cut or give an equally
good pair with a shorter connecting path.

Sample input edges independently at a rate inversely proportional
to the least one-respecting cut capacity. For each tree edge, consider
the sampled input paths containing it. With constant probability
the optimum pair has the same nonempty sampled-path set. A sampled
path common to the pair orders their entire equivalence class.
Interior edges of that class may remain, but their one-respecting
cut capacities are larger than both endpoint capacities. This is
exactly the condition needed by the nearest-no-larger selection rule.
The algorithm recovers all such classes from the two-edge-cut
cactus of a graph obtained by contracting sampled input edges.
It selects only linearly many candidate pairs using nearest
neighbors whose one-respecting cut capacities are no larger.

The candidate cuts are pairwise noncrossing. Choose a root vertex
and take the shore of each cut that avoids it. These shores form a
laminar family: any two are nested or disjoint. A tree represents
this containment relation. One scan of the original edges, with least-common-ancestor
queries in that tree, evaluates every candidate exactly.
Thus sampling reduces the candidate family without approximating
the final cut capacities. Section~\ref{sec:block3-local} proves
the linear work bound and a success probability greater than one tenth.

\subsection{Why three recursive guarantees are necessary}

The tree distribution is built by processing random edge arrivals
through several forests. An arrival goes to the first forest in
which its endpoints are disconnected. The resulting component
partitions are nested: a later forest's partition refines every
earlier one. We stop when even the last forest has sufficiently
few components, choose one forest uniformly, contract its
components, and complete it recursively using original capacities.

Fix a minimum cut $C$, meaning its original crossing-edge set.
Let $F$ be the selected forest. If $F$ contains no edge of $C$,
each contracted component lies entirely on one side. The cut
therefore remains a minimum cut of the quotient: it has its
original capacity, and every quotient cut pulls back to an
original cut. If $F$ already contains one or two edges of $C$,
some contracted components may contain vertices from both sides.
The surviving edges of $C$ then need not define any quotient cut.
A guarantee solely about minimum cuts cannot be applied to them.

The recursive distribution consequently preserves three statements.
It avoids every fixed original-edge set of capacity at most
$49\lambda/100$ with positive probability, uses at most one edge
of every fixed set of capacity at most $37\lambda/50$ with positive
probability, and crosses every fixed minimum cut at most twice
with positive probability. These events are asserted separately
for each fixed set; no simultaneous guarantee over all sets is needed.
The same distribution satisfies all three, without knowing the set.

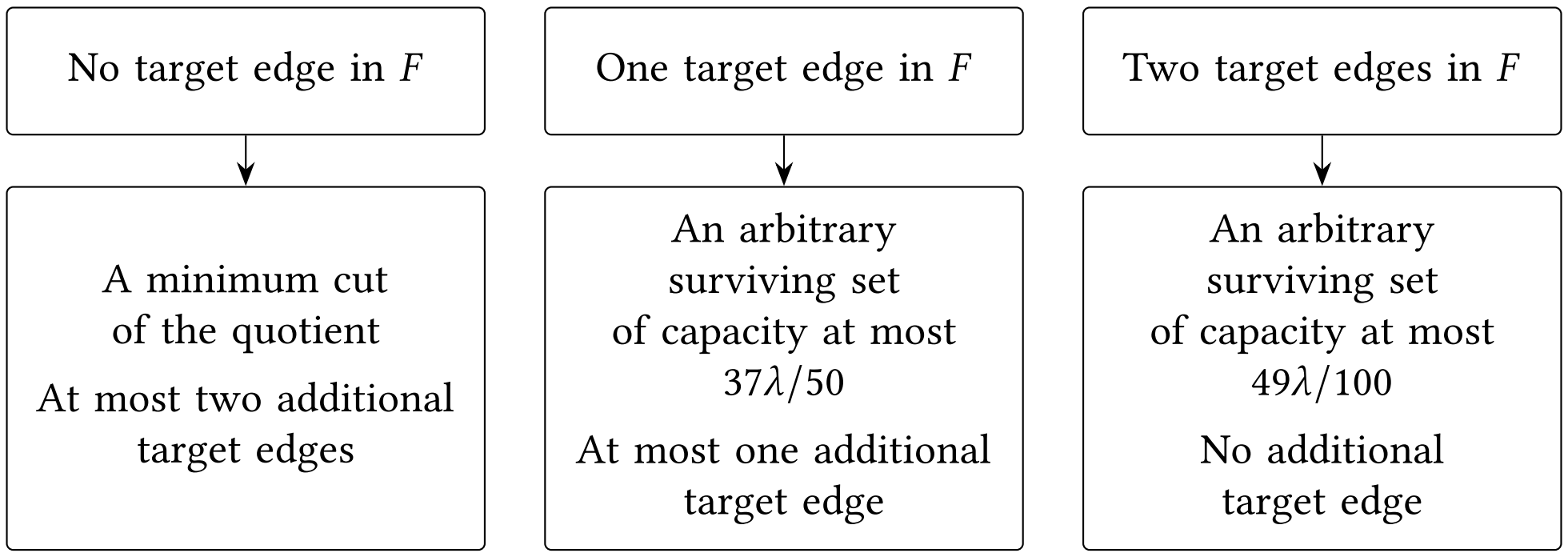
\begin{figure}[ht]
\centering
\begin{tikzpicture}[x=1cm,y=1cm,
 every node/.style={font=\small,align=center},
 box/.style={draw,rounded corners=1.5pt,text width=3.6cm,
             minimum height=1.05cm,inner sep=5pt}]
\node[box] (a) at (0,1.45) {No target edge in $F$};
\node[box] (b) at (4.45,1.45) {One target edge in $F$};
\node[box] (c) at (8.9,1.45) {Two target edges in $F$};
\node[box,anchor=north,minimum height=3cm] (d) at (0,.5)
 {A minimum cut\\of the quotient\\[4pt]At most two additional\\target edges};
\node[box,anchor=north,minimum height=3cm] (e) at (4.45,.5)
 {An arbitrary surviving set\\of capacity at most\\$37\lambda/50$\\[4pt]At most one additional\\target edge};
\node[box,anchor=north,minimum height=3cm] (f) at (8.9,.5)
 {An arbitrary surviving set\\of capacity at most\\$49\lambda/100$\\[4pt]No additional\\target edge};
\draw[-{Stealth[length=2mm]}] (a.south)--(d.north);
\draw[-{Stealth[length=2mm]}] (b.south)--(e.north);
\draw[-{Stealth[length=2mm]}] (c.south)--(f.north);
\end{tikzpicture}
\caption{The three completion cases for a fixed minimum cut and
the selected partial forest $F$. The capacity conditions in the
last two columns are established for a constant fraction of forest
indices. Every quotient minimum cut is at least the original
minimum cut value $\lambda$, so these original-capacity thresholds
are sufficient for the recursive guarantees.}
\label{fig:completion-cases}
\end{figure}

The three cases for $F$ now have appropriate completion guarantees.
With no target edge already used, invoke the genuine-cut guarantee.
With one target edge, sufficiently small surviving capacity permits
one more. With two, the smaller surviving-capacity threshold permits
none. The analysis shows that a constant fraction of forest indices
fit one of these cases. Figure~\ref{fig:completion-cases} summarizes
the argument; it does not require a target cut to survive contraction
in the last two cases.

\subsection{The extra structure of a genuine cut}

In the unrounded conceptual process, for a marked edge set the total
accepted count in an initial segment of forests has an instantaneous
rate equal to the marked
capacity still crossing the last partition in that segment.
The rounded implementation establishes an integrated version of this
comparison, accounting for its timing error. Small accepted counts
therefore imply small surviving capacity,
except with an exponentially small probability. This implication
handles the two arbitrary-set guarantees.

A genuine cut has a stronger property. Before a forest receives
its first target edge, every component lies on one side of the
cut. Every target arrival is then eligible for that forest.
The indices with positive accepted target counts form an initial
segment, and the first several target arrivals are forced into
the corresponding initial segment of forests. One can subtract
this forced initial contribution before estimating the remaining
capacity. The resulting bound is strong enough to leave the
necessary crossing budget for recursive completion.

A three-piece linear potential counts the useful forest indices.
Its breakpoints are the two surviving-capacity thresholds above.
As the index advances, an excessive accepted count supplies
enough increase to pay for an otherwise unusable index; a useful
index is charged only a constant. The ratio controlling the active
piece is monotone, so only two exceptional transitions occur.
This yields a constant useful fraction even when useful and
unusable indices are interspersed. The proof handles the boundary at which the algorithm actually stops,
including its dependence on the arrival history.

\subsection{Generating and routing arrivals in almost linear work}

The forest process uses independent Poisson arrivals with rates
equal to original capacities. Generating every arrival until the desired stopping boundary can be
expensive. The algorithm instead
introduces connections along one random maximum spanning tree.
Each aggregate pair receives its capacity multiplied by a random
power of two, generated with fair bits and doubling. For an
input pair, the minimum priority on its tree path determines
when its endpoints become connected by the initial tree edges.
All arrivals after that time can be omitted: every forest
already connects their endpoints.

The key estimate says that the expected sum, over all pairs, of
capacity divided by this minimum priority is linear in the
number of vertices. It follows by coupling the random priorities
to exponential edge arrivals and charging one expected unit to
each component merge. The estimate holds for the whole tree,
so it controls all scales together.

We do not sort the ordinary arrivals. Divide each pair's lifetime
into geometrically arranged intervals and round arrivals upward
to their interval's right endpoint. The ratio of consecutive
endpoints is one plus the reciprocal of a sufficiently large
multiple of the number of forests. Group arrivals at each rounded
time. Inside a group, process
original edges in a fixed deterministic order. Exact Poisson
counts for each edge can be generated without accessing capacity
bits or forming real quotients. A tree-path minimum query assigns
each pair to an attaining tree edge, so the groups are constructed
in one scan. Repeated exact minimum spanning tree calls recover
the forests for the resulting ordered list; comparison keys can
be evaluated using only integer scaling and comparison of
opaque priorities.

Rounding changes the arrival order, so its effect must be included
in the probability proof. We analyze a continuous expansion of
each processed group. After fixing the priorities, the cumulative
expected number of arrivals on an original edge differs from its
unrounded value by at most a fixed relative mesh error times that
value. Two integration
estimates show that this error is absorbed by fixed slack in the
stopping and surviving-capacity bounds. This is the point at which
the fixed ordering within a group becomes useful: it removes
random tie-breaking within a time group while preserving the
needed conditional Poisson structure.

\subsection{Fixed depth and a uniform algorithm}

Let $K$ denote the number of forests at one recursive step, and
let $R$ denote the desired reduction in the number of vertices.
The rounded construction has expected work $O(m+nK^2)$ outside
one recursive call, whose graph has $O(1+n/R)$ vertices. Taking
$K$ proportional to the logarithm of $R$, with a constant depending
on the inherited success probability, preserves all three
guarantees. Choosing successive reductions from iterated logarithms
gives, after a fixed number $s$ of steps,
$O_s(m+nL_{s+1}(n)^2)$ time. The square is absorbed by the preceding
iterate, which yields Theorem~\ref{thm:main-fixed}.

For growing depth, we explicitly track the probability loss at
every step. A fixed absolute factor of loss per level yields an
exponentially small, but known, trial success probability.
The number of levels is $\log^*n$ up to an additive constant.
Choosing the number of forests at each level to account for that
probability, and repeating the whole trial sufficiently often,
gives Theorem~\ref{thm:main-uniform}. The work cap is imposed on
the aggregate repeated computation, so its failure is subtracted
once from the unconditional success probability. This accounting
explains both the factor $2^{O(\log^*n)}$ and why substituting a
growing depth into the fixed-depth theorem would be invalid.

\section{Minimum-Cut Preliminaries and Exact Sampling}\label{sec:block3-primitives}

We work with an undirected multigraph whose edges have nonnegative real
capacities. The number of vertices is denoted by $n$, and the number of
input edges, including parallel edges, by $m$. Self-loops can be deleted.
For a vertex set $S$, let $\delta_G(S)$ denote the edges with exactly one
endpoint in $S$. For an edge set $A$, let $c(A)$ denote the sum of its
capacities. A cut is a nonempty proper vertex set, identified with its
complement when convenient. Its capacity is $c(\delta_G(S))$.
The minimum cut value is denoted by $\lambda$.

The connected components of the positive-capacity support can be found
in linear time. If there is more than one component, any component gives
a cut of capacity zero, which is optimal by nonnegativity. We therefore
assume connected positive support whenever we discuss a randomized
construction. A tree-distribution subroutine on a single vertex returns
the empty tree. For real $x\ge 2$, define the truncated iterated logarithms by
\[
 L_1(x)=\max\{2,\lceil\log_2 x\rceil\},\qquad
 L_{j+1}(x)=\max\{2,\lceil\log_2 L_j(x)\rceil\}
 \quad(j\ge1),
\]
so that $\log^{(h)}n$ is asymptotically equivalent to $L_h(n)$.

\paragraph{Computational model.}
Input capacities are opaque real registers. Copying, addition,
subtraction, and comparison have unit cost. The algorithm does not
inspect their representations or apply division, logarithms, or floor
to them. Ordinary word operations are permitted on vertex and edge
identifiers, integer counters, and independent random bits. The word
length is a sufficiently large constant multiple of
$\log(n+m+2)$, using the original input sizes throughout recursive
calls. Each independent fair bit costs one operation.
Multiplying an opaque register by a positive integer $r$ takes
$O(1+\log r)$ additions and doublings; we include this work explicitly.
Subtraction can also be simulated by signed pairs of nonnegative
registers, with constant overhead.

We use four standard graph operations. Depth-first search takes linear
time. Stable bucket sorting takes time proportional to the list length
plus the integer key range; all bucket arrays below have one entry per
vertex or edge identifier. Static least-common-ancestor queries
on a rooted tree take constant time after linear preprocessing
\cite{DBLP:journals/siamcomp/HarelT84}. Given all endpoint pairs in advance, minima on
their tree paths can be found in linear time by the non-tree-edge
sensitivity algorithms of Dixon, Rauch, and Tarjan and of
Buchsbaum et al.\ \cite{DRT1992,BuchsbaumEtAl2008}; this is not the
full tree-edge sensitivity problem. The latter work's path-maximum
algorithm explicitly identifies an attaining tree edge.
Finally, the exact minimum spanning tree algorithm of Karger, Klein,
and Tarjan takes expected linear time and uses only comparisons of
weights and fair coin flips \cite{KKT1995}.
We break ties lexicographically by distinct integer identifiers.
Thus a call returns the same tree for every outcome of its internal
randomness; only its work is random. Reversing comparisons gives a
maximum spanning tree. Every use below charges the conditional
expected work for the actual graph supplied to the call.

\paragraph{Elementary distributions.}
For a parameter $p$ in the unit interval, a Bernoulli variable is one
with probability $p$ and zero otherwise. For a nonnegative integer $r$,
we write $\operatorname{Bin}(r,p)$ for the sum of $r$ independent such
variables. For a positive real parameter
$u$, the distribution $\operatorname{Pois}(u)$ assigns probability
$e^{-u}u^j/j!$ to each nonnegative integer $j$; at parameter zero
it is concentrated at zero. These are the binomial and Poisson
distributions, respectively.

\subsection{Ratio coins and weighted choices}

\begin{lemma}[Exact ratio coin]\label{lem:block3-ratio-coin}
Let $D$ be a positive opaque register and let $x$ be a register between
zero and $D$. An exact Bernoulli random variable of parameter $x/D$
can be generated using at most two iterations in expectation. Each
iteration uses a constant number of permitted operations and one
fair bit. The bound holds uniformly over the register values.
\end{lemma}
\begin{proof}
The endpoint cases return their deterministic answers. Otherwise, double
the current numerator. Record a bit indicating whether the result is
at least $D$, and subtract $D$ if it is. These are the successive binary
digits of the original ratio. Compare each digit with a fresh fair bit.
At the first difference, return whether the random digit is smaller.
The random digits represent a uniform point of the unit interval, so
the output has the required law. At each step the probability of
agreement is one half, independently of the earlier random digits.
The number of iterations is therefore geometric with mean two.
The comparison never evaluates a quotient of opaque registers.
\end{proof}

\begin{lemma}[Static weighted sampling]\label{lem:block3-categorical}
For a list of $P$ nonnegative opaque weights with positive total,
linear preprocessing and space support exact independent weighted
choices in expected $O(1+\log P)$ time and fair bits per choice.
The bound is uniform over the weights and over any earlier history
that fixes the current list.
\end{lemma}
\begin{proof}
Place the positive weights at the leaves of a balanced binary tree,
and store the sum below each node. Starting at the root, choose a
child with probability equal to its sum divided by the parent's sum,
using Lemma~\ref{lem:block3-ratio-coin}. The product of these conditional
probabilities along a root-to-leaf path telescopes to the leaf weight
divided by the total. The height is logarithmic in the list length,
and each decision has uniformly constant expected cost. Fresh bits
give independent choices. The argument remains valid after conditioning
on the complete preceding history.
\end{proof}

An exact uniform integer in any finite initial interval is obtained by
drawing the necessary number of fair bits and rejecting values outside
the interval. Acceptance probability is at least one half, so the
expected number of bits is logarithmic in the interval length.

\subsection{Poisson counts without division}

\begin{lemma}[Ratio-Poisson sampling]\label{lem:block3-ratio-poisson}
For a nonnegative opaque register $y$ and a positive opaque register
$d$, an exact Poisson variable of mean $y/d$ can be generated in
expected $O(1+y/d)$ time and fair bits.
\end{lemma}
\begin{proof}
We first generate a Poisson variable of mean one. Propose a
nonnegative integer $k$ by counting heads before the first tail.
Its proposal probability is $2^{-k-1}$. Accept the proposal with
probability $2^{k-1}/k!$, using a rational ratio coin. These acceptance
probabilities are at most one: the first three are one half, one,
and one, and each subsequent term decreases. The unnormalized
probability of accepting $k$ is
\[
  2^{-k-1}\frac{2^{k-1}}{k!}=\frac1{4k!}.
\]
The equality cancels the two powers of two. Summing over all
nonnegative integers shows that the acceptance probability is $e/4$;
normalizing therefore gives the Poisson law of mean one. Exact
integer arithmetic computes the rational coin in work polynomial in
$k$. A geometric proposal has finite moments of every fixed order,
so the expected work and number of bits are absolute constants.
This argument does not assume that an arbitrarily large factorial
fits in one word.

Starting with a copy of $y$, repeatedly subtract $d$ while the copy
is at least $d$, adding an independent mean-one Poisson count each
time. For the final remainder $x$, draw one more mean-one Poisson
count and retain each of its units with probability $x/d$.
Summing the Poisson probabilities over the discarded units shows that
the retained count has mean $x/d$ and a Poisson law. Adding independent
Poisson counts adds their means: expand the power of the sum of their
means in the convolution of their probabilities. This gives the
required total law. The number of whole
unit steps is at most $y/d$, and the remainder has constant expected
cost. An unbounded output counter can be incremented once per unit
using a binary counter with linear total bit-flip cost.
\end{proof}

\paragraph{Original edges and aggregation.}
Parallel endpoint pairs can be grouped by stable bucket sorting of integer keys and
their capacities added in $O(m+n)$ time. Retain the list of original
edges for each pair. A weighted choice from that list returns an
original edge with probability proportional to its capacity. Building
all list samplers takes linear total work. Making at most one choice
per list also takes expected linear work, since the logarithm of a
positive list length is at most that length. Ordinary arrivals in
the final construction are generated directly for original edges;
the lists are needed for the shared initial tree edges and for the
base distribution.

\paragraph{Work caps.}
All probability arguments refer to an uncapped exact process. Before
randomized work, store a valid spanning tree of each connected input,
and store a nontrivial cut and its exact capacity at the top level.
Couple the implementation to the uncapped process until a prescribed
work budget is exceeded, then return the stored valid output.
Let $A$ be a positive cap multiplier. If $W$ is the nonnegative
work and $B$ bounds its expectation, a cap at $AB$ fails with probability at most $1/A$, by Markov's inequality.
We subtract this failure probability unconditionally. We never
assert that a sample conditioned on avoiding a cap retains its
original law. The cap includes integer arithmetic, random-bit
generation, and all graph subroutines; its space is bounded by its
accounted work.

\section{A linear-time algorithm for a fixed tree}
\label{sec:block3-local}

The global algorithm produces a spanning tree that crosses a fixed minimum
cut at most twice with a controlled probability.  This section gives the
corresponding cut-finding algorithm.  Its linear bound is important:
repetition then costs only as much as producing the trees themselves.
Throughout this section, $G$ is the input graph and $T$ is a tree on the
same vertex set.  A cut \emph{two-respects} $T$ if at most two edges of
$T$ cross it, following the standard tree-packing terminology
\cite{Karger2000}.  We also call a cut crossing one tree edge
\emph{one-respecting}.  The tree is not required to be a subgraph of $G$.

\begin{theorem}[Minimum cut two-respecting a fixed tree]
\label{thm:block3-local}
There is a randomized algorithm that returns a nontrivial cut
two-respecting $T$, together with its exact original capacity, in
worst-case $O(m+n)$ time and space.  With probability greater than one
tenth, the returned cut is minimum among all cuts two-respecting $T$.
If a minimum such cut is one-respecting, the algorithm succeeds for every
outcome of its random choices.  These bounds charge each fair random bit
separately.
\end{theorem}

The proof has three steps.  Sampling makes a carefully chosen optimum
appear as a two-edge cut in an unweighted graph of linear size.  The
cactus representation of these two-edge cuts yields only a linear
number of candidates.  Their cuts form a laminar family, so all their
original capacities can be evaluated in one scan of $G$.

\subsection{Choosing an optimal pair with a shortest connecting path}

For an original edge $a$, let $c(a)$ denote its capacity and let $P(a)$
be its endpoint path in $T$. For a
tree edge $e$, let $F_e$ be the set of original edges whose paths contain
$e$, and let $A(e)$ denote the capacity of $F_e$.  The set $F_e$ is precisely the original edge boundary of the unique
one-respecting cut crossing $e$.  For distinct tree
edges $e,f$, let $I(e,f)$ denote the capacity of their common set
$F_e\cap F_f$.  The unique cut with tree boundary consisting of $e$ and
$f$ has capacity $D(e,f)$ defined by
\begin{equation}\label{eq:block3-local-pairvalue}
 D(e,f)=A(e)+A(f)-2I(e,f).
\end{equation}
This equality follows because its original edge boundary is the symmetric
difference of $F_e$ and $F_f$: edges in the intersection have been counted
twice and must be removed twice.  Define the minimum one-respecting and
two-respecting capacities, respectively, by
\[
 \mu=\min_{e\in E(T)}A(e),\qquad
 \lamtwo=\min\bigl\{\mu,\min_{e\ne f}D(e,f)\bigr\}.
\]
The first equality defines $\mu$ by the one-respecting cuts, and the second
defines $\lamtwo$ by all allowable tree boundaries.  The unequal-index
condition requires two distinct edges.  If $T$ has only one edge, the
inner minimum is omitted.

All values $A(e)$ can be computed in linear time.  Root $T$.  The least common ancestor of two vertices is their deepest
common ancestor; preprocess these queries in linear time
\cite[Section~5]{Karger2000}.  For each original edge, add its capacity at
both endpoints and subtract twice its capacity at their least common
ancestor.  A postorder sum gives the capacity of each one-respecting cut
at the child endpoint of its tree edge.  Save a minimum one-respecting cut
as a fallback.  If its capacity is zero, it is already optimal and the
algorithm terminates.

For a pair of distinct tree edges, its connecting path is the smallest
tree path containing both edges.  Its length counts its tree edges; its
interior consists of the edges strictly between the pair.

\begin{lemma}\label{lem:block3-local-span}
Suppose the best two-respecting cut is strictly better than the
one-respecting fallback.  Among optimal pairs, fix one with a shortest connecting path,
independently of the random choices.  Let $B$ be the larger of its two one-respecting
capacities.  For this pair $e,f$,
\[
 I(e,f)>B/2\ge\mu/2.
\]
The first inequality follows from strict improvement over the smaller
one-respecting cut, and the second follows from the definition of $\mu$.
Every interior tree edge has one-respecting capacity strictly greater
than $B$.
\end{lemma}

\begin{proof}
Name the endpoints so that $f$ has the larger one-respecting capacity.
The strict improvement over the one-respecting cut of $e$ and
\eqref{eq:block3-local-pairvalue} give the first claimed inequality.
Now take an interior tree edge $g$.  Every original path containing both
$e$ and $f$ also contains $e$ and $g$.  Consequently,
\[
 I(e,g)\ge I(e,f),
\]
because the latter intersection is contained in the former and capacities
are nonnegative.  If the one-respecting capacity of $g$ were at most
that of $f$, then
\[
 D(e,g)\le D(e,f),
\]
by the pair-capacity formula, the assumed capacity comparison, and the
preceding intersection inequality.  Strict improvement contradicts
optimality; equality contradicts the choice of a shortest connecting path.  Thus every
interior edge has the asserted strictly larger capacity, including when
some other edge capacities are tied.
\end{proof}

\subsection{Sampling and two-edge cuts}

Independently include each original edge $a$ with probability
\begin{equation}\label{eq:block3-local-sampling}
 p_a=\min\{1,c(a)/(2\mu)\}.
\end{equation}
This equality defines the sampling law.  It is implemented by the exact
ratio coin of Lemma~\ref{lem:block3-ratio-coin}, after directly including
edges whose capacities reach the denominator.  Let $S$ denote the
sampled set.

\begin{lemma}\label{lem:block3-local-capture}
For the pair fixed in Lemma~\ref{lem:block3-local-span}, with probability
greater than $\tfrac12(1-e^{-1/4})$, no sampled edge crosses its cut and
some sampled edge has a tree path containing both edges of the pair.
\end{lemma}

\begin{proof}
Let $C$ be the original edge boundary of the chosen cut, and let $J$ be
the set of original edges whose paths contain both $e$ and $f$.  These
sets are disjoint.  Since the capacity of $C$ is strictly less than
$\mu$, every edge in $C$ uses the unsaturated part of the sampling law.
Hence
\[
 \sum_{a\in C}p_a=\frac{\lamtwo}{2\mu}<\frac12.
\]
The equality follows by summing the sampling probabilities over the cut;
the inequality is its strict improvement over $\mu$.  The union bound
therefore gives probability greater than one half that the sample misses
$C$.

If an edge in $J$ is sampled with probability one, the sample hits $J$
surely.  Otherwise,
\[
 \sum_{a\in J}p_a=\frac{I(e,f)}{2\mu}>\frac14.
\]
The equality is again the unsaturated sampling law, and the inequality
uses Lemma~\ref{lem:block3-local-span}.  Independence then gives
\[
 \Prb[S\cap J=\varnothing]
 =\prod_{a\in J}(1-p_a)
 \le \exp\!\left(-\sum_{a\in J}p_a\right)
 <e^{-1/4}.
\]
The event's equality to the empty set defines missing $J$; the next
equality multiplies independent failure probabilities, the first
inequality uses the elementary exponential bound for each factor, and
the last uses the preceding lower bound on the sum.  Missing $C$ and
hitting $J$ depend on disjoint sets of independent coins, so their
probabilities multiply, proving the claim.
\end{proof}

For a tree edge $e$, let $\Sigma(e)$ denote the set of sampled edges
whose paths contain $e$; thus
\[
 \Sigma(e)=S\cap F_e.
\]
This equality expresses the set definition using the sampled set $S$
and the previously defined set $F_e$.  These sets are
used only in the proof and are never stored.  The event in
Lemma~\ref{lem:block3-local-capture} says that the fixed optimal pair
has the same nonempty set.

Compute the connected components of the sampled graph.  Contract these
components in a separate copy of $T$, retaining the identities of tree
edges and keeping parallel edges.  Discard loops and call the resulting
connected unweighted multigraph $Q$.  It has at most $n-1$ edges.  A
sampled edge that belongs to $T$ is distinct from the copy of that edge
used to form $Q$.

\begin{lemma}\label{lem:block3-local-series}
A tree edge $e$ has an empty set $\Sigma(e)$ precisely when its image
in $Q$ is a bridge.  Two distinct tree edges have the same nonempty
set precisely when their nonloop images form a two-edge cut of $Q$
containing no bridge.
\end{lemma}

\begin{proof}
A sampled edge crossing a one-respecting cut supplies an alternate path
around its tree edge.  Contracting the sample therefore makes that edge
either a nonbridge or a loop.  If no sampled edge crosses the one-respecting
cut, the edge remains a bridge.  This proves the first assertion, with
discarded loops understood as nonbridges.

Equality of the two sets means that no sampled edge crosses the unique cut
whose tree boundary is the given pair.  That cut survives contraction
and has exactly those two edges as its boundary.  Nonemptiness of the common set
makes each edge a nonbridge; surviving boundary edges cannot be loops.
Conversely, a two-edge cut of $Q$ containing no bridge lifts to a cut
crossed by exactly the given two tree edges and by no sampled edge.
The two sets consequently coincide, and nonbridging ensures
that they are nonempty.
\end{proof}

Remove the bridges of $Q$ and construct the cactus representation of
the two-edge cuts in each remaining component.  Reattach the bridges
to obtain a cactus with bridges, denoted by $K$; cycles of length two
are allowed.  This construction is deterministic and linear-time
\cite[Lemmas~11 and~15]{MNS2017}.  Each cycle edge retains an original
tree-edge identity, and two nonbridge edges form a two-edge cut
precisely when they lie on the same cactus cycle.  Thus each cycle is
one equivalence class of at least two tree edges whose sets $\Sigma(e)$
coincide and are nonempty.

The representation contracts the three-edge-connected vertex classes;
the paths defining these classes may leave the class.  Its structural
property follows because no cut of size at most two separates a class,
whereas every two distinct quotient vertices can be separated by such
a cut.  A connected graph with neither loops nor the cactus structure
contains two vertices joined by three paths that have only their
endpoints in common, contradicting that separation property.
The cited construction supplies the vertex map and the edge identities
in linear time.

\subsection{A laminar candidate family}

On every cactus cycle, retain the cyclic order of its tree-edge identities
and their one-respecting capacities.  For each edge, choose in each
direction the nearest \emph{other} edge whose capacity is at most its
own, if there is one.  Scan the doubled cyclic sequence in each direction with a stack.
Before processing an edge, repeatedly remove the top edge while its
capacity is larger than the current capacity. The last remaining edge
is the nearest one of no larger capacity.
For each direction, record answers only for the second copy of each
edge. Discard an answer whose cyclic distance reaches the cycle length,
thereby excluding the first copy of that same edge. After checking the
answer, push the current edge. Each copy enters and leaves the stack
at most once, so the total work is linear.  At most $2(n-1)$ pairs are generated.  Stable bucket
sorting on their integer edge identifiers removes duplicate unordered
pairs in linear time.

\begin{lemma}\label{lem:block3-local-selection}
On the event in Lemma~\ref{lem:block3-local-capture}, the candidate pairs
include the fixed optimum with a shortest connecting path.
\end{lemma}

\begin{proof}
Choose a sampled edge whose path contains both edges of that optimum.
Every edge in their equivalence class lies on this path.  Sampled
component contraction identifies the path's endpoints.  Its image in
the cactus is therefore a closed walk without repeated retained edges.
Such a walk traverses each visited cactus cycle in its cyclic order,
possibly with excursions through other cycles at articulation vertices.
The order of the class along the original path consequently agrees
with the cactus order, up to rotation and reversal.

One arc between the optimal pair follows their original connecting
path.  All its interior edge capacities exceed both endpoint capacities,
by Lemma~\ref{lem:block3-local-span}.  Starting at the larger-capacity
endpoint, the other endpoint is therefore the nearest edge with capacity
at most its own in that direction.  The algorithm selects this pair.
\end{proof}

Two cuts are \emph{noncrossing} if one of the four intersections of
their two shores is empty.  Equivalently, orienting both cuts away
from one fixed vertex makes their selected shores nested or disjoint.

\begin{lemma}\label{lem:block3-local-compatible}
The candidate cuts are pairwise noncrossing.  Their shores avoiding one
fixed vertex therefore form a laminar family.
\end{lemma}

\begin{proof}
Each selected pair has an open witnessing arc whose interior capacities
strictly exceed both endpoint capacities.  Two selected pairs cannot
alternate around a cycle.  Indeed, in cyclic order $a,b,c,d$, suppose
the pairs are $\{a,c\}$ and $\{b,d\}$.  The witnessing arc of the first
contains $b$ or $d$, making that endpoint larger than both $a$ and $c$.
The witnessing arc of the second contains $a$ or $c$, making that
endpoint larger than both $b$ and $d$.  These two strict comparisons
contradict one another.  Shared endpoints and ties cause no difficulty.

Replace each cactus cycle by a new vertex joined to all vertices of
that cycle, and retain the bridges. This produces a tree; root it at
the image of the fixed vertex.  Each pair
cuts off a consecutive interval of cycle vertices avoiding the attachment
toward the root, together with all branches attached to that interval.
Nonalternating pairs give nested or disjoint shores within one cycle.
For different cycles, a descendant cycle lies in one branch of an
ancestor cycle, and incomparable branches are disjoint.  The shores
are therefore laminar in the cactus.  Taking preimages under the
contraction map preserves this property.
\end{proof}

The one-respecting fallback is evaluated separately, since it may
cross a candidate cut.

\subsection{Evaluating all candidates together}

\begin{lemma}\label{lem:block3-local-evaluator}
Given $q$ distinct noncrossing cuts, each specified by its two boundary
edges in a rooted tree $T$, their capacities in $G$ can all be computed
in $O(m+n+q)$ time and space.
\end{lemma}

\begin{proof}
Number the tree vertices in preorder.  For a vertex $x$, let $T_x$
denote its rooted subtree.  For a pair of boundary edges, let $x,y$
be their child endpoints.  The shore avoiding the root is the difference
of the larger subtree and the smaller one when the endpoints are
comparable by ancestry, and the union of the two subtrees otherwise.
Consequently each shore is the union of at most two preorder intervals.
Remove empty pieces and merge adjacent pieces.  The intervals left
are exactly the maximal contiguous components of that shore, and its
cardinality is the sum of their lengths.

These component intervals are themselves laminar.  A component of a
shore contained in another shore lies inside one component of the
larger shore; disjoint shores have disjoint components.  Identical
component intervals may occur in distinct nested shores.

Create one interval record for each component, labeled by its shore
and by that shore's total cardinality.  Add a universal interval and
one singleton record per original vertex; give singleton vertex records
cardinality zero.  Sort the records by increasing left endpoint,
decreasing right endpoint, and decreasing full-shore cardinality, in
that order.  Three stable bucket-sort passes suffice because all
keys are integer vertex indices or cardinalities.

A stack of containing intervals assigns every record to its nearest
preceding enclosing record.  For a component of a shore $A$, this
record belongs exactly to the immediate larger shore containing $A$.
To verify this, suppose a component of another shore $B$ contains it.
The shores intersect and hence are nested.  If $B$ is contained in
$A$, maximality of the component of $A$ forces the two component
intervals to coincide, and decreasing cardinality puts the larger
shore first.  Thus preceding enclosing records belong to ancestor
shores.  Conversely, each ancestor shore has a component containing
the interval under consideration.  This proves the parent assertion.
Both components of a two-component shore consequently obtain the same
parent.  The zero-cardinality label places each vertex singleton
below all shore records containing that vertex.

Construct a hierarchy $H$ with one universal root, one node per shore,
and one leaf per original vertex.  The descendant vertex leaves of a
shore node are exactly that shore.  Preprocess least common ancestors
in $H$.  For each original edge, add its capacity at both endpoint
leaves and subtract twice its capacity at their least common ancestor.
A postorder sum then gives each shore's boundary capacity.  An edge
crossing the shore contributes once, whereas an internal edge's two
endpoint contributions cancel at its least common ancestor.  This
uses one scan of $G$ and a linear traversal of $H$.
\end{proof}

\subsection{The complete algorithm}

Algorithm~\ref{alg:fixed-tree-cut} combines the construction.  The only
unbounded random primitive is the exact ratio coin.  Its total work
is capped without conditioning any probability argument on completion.

\begin{algorithm}[ht]
\caption{A minimum cut two-respecting a fixed tree}\label{alg:fixed-tree-cut}
Compute all one-respecting capacities and retain a minimum one-respecting cut. \Return it immediately if its capacity is zero\;
Sample every original edge according to \eqref{eq:block3-local-sampling}. If the ratio coins together exceed $200(m+n)$ iterations, \Return the retained cut\;
Contract sampled components in a separate copy of $T$, retaining tree-edge identities and removing loops\;
Construct the two-edge-cut cactus of each bridgeless component. On each cycle, select the nearest other edge of no larger capacity in both directions, and remove duplicate pairs\;
Evaluate the resulting laminar family by Lemma~\ref{lem:block3-local-evaluator}. \Return its least-capacity cut or the retained one-respecting cut, whichever is smaller\;
\end{algorithm}

\begin{proof}[Proof of Theorem~\ref{thm:block3-local}]
Every selected pair specifies an actual nontrivial cut, and every
returned capacity is evaluated in the original graph.  If a one-respecting
cut is locally optimal, the retained cut is already optimal.  Otherwise,
Lemmas~\ref{lem:block3-local-capture} and
\ref{lem:block3-local-selection} place an optimum in the candidate
family with probability greater than $\tfrac12(1-e^{-1/4})$.

The expected total number of ratio-coin iterations is at most $2m$,
by Lemma~\ref{lem:block3-ratio-coin}.  Markov's inequality therefore
bounds the abort probability by
\[
 \frac{2m}{200(m+n)}\le\frac1{100}.
\]
The inequality uses nonnegativity of the vertex count.  Subtracting
this failure probability from the uncapped capture probability leaves
\[
 \frac12(1-e^{-1/4})-\frac1{100}>\frac1{10}.
\]
This numerical inequality follows, for example, by lower-bounding the
exponential $e^{1/4}$ by the first four terms of its power series.
No independence between capture and abort is required.

All other steps are deterministic and linear-time.  The quotient,
cactus, candidate list, and evaluation hierarchy have linear size.
The cap counts every fair bit used by a ratio coin.  Finally, recover
the winning shore from its at most two preorder intervals in linear
time.  This proves the stated time, space, success, and output guarantees.
\end{proof}

\section{A base distribution of spanning trees}\label{sec:block3-base}

For each original edge $e$, let $c_e$ denote its capacity.
The recursion starts from a distribution with a small expected number
of edges in every prescribed set. A polylogarithmic vertex term is
sufficient here: later recursive calls reduce the number of vertices
before invoking this construction.

\begin{lemma}[Base tree distribution]\label{lem:block3-base-packing}
For every fixed positive error allowance $\delta$, there is a randomized
algorithm returning a valid original-edge spanning tree $T$ and an
event $\mathcal E$ for the analysis, with
\[
 \Prb[\mathcal E^c]\le\delta.
\]
The inequality is the combined error bound for scale selection,
completion of the forest prefix, and the work cap, established below.
For every fixed assignment of labels $z_e$ in the unit interval to
the original edges,
\[
 \E\!\left[\mathbf1_{\mathcal E}\sum_{e\in T}z_e\right]
 \le\frac{101}{50\lambda}\sum_e c_ez_e.
\]
The inequality follows by averaging a completed prefix of forests
and charging its labels to all sampled arrivals. The labels are not
given to the algorithm. Its capped time and space are
$O_\delta(m+n\log^3(2n))$ in the fair-bit model.
\end{lemma}

\subsection{One certificate for all thresholds}

Aggregate parallel endpoint pairs and retain their original-edge lists.
Within this construction, write $c_e$ for the capacity of aggregate
pair $e$, and let $s$ be the number of positive aggregate pairs. Run a maximum
adjacency ordering: each unscanned vertex $u$ has a key equal to its
total capacity to already scanned vertices. When a maximum-key
vertex $v$ is scanned, process its unscanned neighbors. For a processed
pair $e$ with later endpoint $u$, let $\ell_e$ be the key of $u$ just
before the update, and define
\[
 h_e=\ell_e+c_e.
\]
The equality defines the updated key, which is assigned to $u$.
At every later endpoint the intervals from its processed pairs
partition an interval beginning at zero, in their processing order.
Let $R_u$ be the final right endpoint. A Fibonacci heap implements
the ordering in $O(m+n\log(2n))$ time.

For a positive threshold $k$, define clipped capacities by
\[
 c_k(e)=\max\{0,\min\{k,h_e\}-\ell_e\}.
\]
This equality defines the length of the part of the pair's interval
below the threshold. For every cut $S$, the resulting certificate satisfies
\begin{equation}\label{eq:block3-certificate}
 \sum_e c_k(e)\le(n-1)k,
 \qquad
 \min\{k,c(\delta_G(S))\}\le c_k(\delta_G(S))
       \le c(\delta_G(S)).
\end{equation}
The first inequality sums at most $k$ at each later endpoint. The
cut lower bound follows from the layer argument in the next paragraph.
The last inequality holds because clipping decreases no edge's
capacity below zero and increases no edge's original capacity.

Fix a positive real layer $t$ outside all interval endpoints.
The pairs whose intervals contain $t$, oriented toward their later
endpoints, form a forest: the orientation is acyclic and every vertex
has at most one incoming edge. Call it $F_t$. An unscanned vertex
is active when its current key exceeds $t$. All active vertices
belong to one component of the exposed forest. Indeed, the next
scanned vertex is active whenever an active vertex exists, and
each newly active neighbor joins its component. If none exists,
all newly active neighbors join through the vertex just scanned.
It follows that the endpoints of every processed pair are connected
in $F_t$ for every positive layer below its right endpoint, apart
from finitely many endpoints. Below its left endpoint both vertices
are active; inside its interval the pair itself connects them.
Aggregation ensures that the later endpoint's key has not already
been increased by a parallel pair during the same scan.

If clipping decreases some pair of a cut, that pair's right endpoint
exceeds $k$. Every layer below $k$ therefore has a forest edge
crossing this cut. Integrating gives clipped cut capacity at least
$k$. If no crossing pair is decreased, the cut capacity is unchanged.
This proves the lower bound in \eqref{eq:block3-certificate}.
The construction is the continuous-capacity form of the
Nagamochi--Ibaraki certificate
\cite{NI1992,KargerThesis,BK2015,ForsterDeVos2023}.

\subsection{Finding a constant-factor scale}

Build a balanced binary tree on each vertex's ordered intervals.
At a node store the two endpoints $a,b$ of the union interval below
it. At a queried threshold $k$, compute that node's clipped mass
on demand as $\max\{0,\min\{k,b\}-a\}$. Thus a sampled path
requires only constant work at each visited node; changing the
threshold does not scan or rebuild the lists. Pad the total mass at each
later endpoint to $k$ with a dummy outcome. Ratio coins in the
binary tree then return pair $e$ with probability $c_k(e)/k$.
Each choice takes expected logarithmic time, and the trees are
built once in linear time.

Let $L_s$ be a sufficiently large constant multiple of
$\lceil\log_2(2n)\rceil$, rounded up. Make $L_s$ independent choices
at each later endpoint and discard the dummies. Denote the resulting
unweighted multigraph by $H(k)$. It has at most $(n-1)L_s$ edges.
For every cut $S$,
\[
 \E|\delta_{H(k)}(S)|=\frac{L_s}{k}c_k(\delta_G(S)).
\]
The equality adds the probabilities of crossing choices over the
independent draws. Each cut count is a sum of independent Bernoulli
variables. For every real $a$ at least one, the weighted cut-counting
theorem bounds the number of cuts of capacity at most $a$ times the
minimum by $n^{2a}$
\cite{Karger2000}. Applying a Chernoff bound in consecutive
minimum-cut-sized capacity intervals shows that, when the minimum
cut of the certificate is at least $k$, every sampled cut has size
at least $L_s/2$, except with an arbitrarily small inverse-polynomial
probability. In detail, the number of cuts in the $j$th such interval
is at most $n^{2(j+1)}$, whereas each lower-tail probability is
$2\exp(-\Omega(jL_s))$. The product is summable over positive
integers $j$ once the constant in $L_s$ is chosen sufficiently large.

Compute any maximum spanning tree for the aggregate capacities,
and let $w$ be its least edge capacity. Then
\[
 w\le\lambda\le sw.
\]
The first inequality holds because every cut crosses a tree edge.
For the second, the fundamental cut of a least tree edge contains
no heavier pair, by the exchange property, and has at most $s$ pairs.
Construct consecutive dyadic thresholds beginning at $w$ and ending
at the first one at least $32sw$. There are logarithmically many
thresholds, and their construction uses only doubling.

For a tested threshold, apply Matula's linear-time unweighted
three-approximation to $H(k)$ \cite[Section~3.5]{KargerThesis}.
If this graph is disconnected, use its zero cut for this scale test.
Write $\lambda(H(k))$ for its minimum cut value, and let $a_k$ be
the capacity of the approximate cut. The guarantee is
\[
 \lambda(H(k))\le a_k\le3\lambda(H(k)).
\]
The first inequality follows from feasibility; the second is Matula's
approximation guarantee on the unweighted multigraph. Declare the
test true when $a_k$ is smaller than $L_s/2$.
For thresholds at most $\lambda$, the preceding all-cut bound makes
the test false. For thresholds at least $16\lambda$, a fixed original
minimum cut is unchanged by clipping and has expected sample count
at most $L_s/16$. A one-cut Chernoff bound makes that count at most
$L_s/8$, so the approximation makes the test true. Increase the
constant in $L_s$ and union bound over every possible threshold,
including tests not actually queried.

Test the endpoints, then binary-search between an observed false
lower endpoint and a true upper endpoint until they are adjacent.
No monotonicity is needed inside the factor-sixteen transition:
every true test is above $\lambda$, and every false test is below
$16\lambda$. Let $U$ be the final upper threshold. Adjacency gives
\begin{equation}\label{eq:block3-base-scale}
 \lambda<U<32\lambda.
\end{equation}
The first inequality uses that the upper test is true; the second
uses that the lower test is false and the two thresholds differ
by a factor of two. Fresh randomness is used for each test. Exposing
all possible test randomness in advance justifies the adaptive search.
An endpoint failure returns the stored tree. Only doubly logarithmically
many tests are made; their expected total cost is
$O(m+n\log^2(2n)\log\log(4n))$.

\subsection{Completing a prefix of forests}

Condition on a scale history satisfying \eqref{eq:block3-base-scale}.
Use the clipped capacities at $U$ and pad their total to
$(n-1)U$ with a dummy. Let $L$ be a sufficiently large constant
multiple of $\lceil\log_2(2n)\rceil$, rounded up, and make exactly
$(n-1)L$ fresh independent choices from this distribution. Initialize
$L$ empty forests. Insert each real choice into the first forest
whose endpoints are disconnected, discarding it if no such forest
exists. A dummy has no effect. Later forest partitions refine earlier
ones. Hence a choice is accepted among the first $i$ forests exactly
when its endpoints are disconnected in the $i$th forest.

Let $c_i(t)$ be the component count of forest $i$ after $t$ choices.
Define the prefix deficit and its initial boundary value by
\[
 D_i(t)=\sum_{j=1}^{i}(c_j(t)-1),\qquad D_0(t)=0.
\]
The first equality defines the total number of missing forest edges
in the first $i$ indices; the summation starts at the first forest.
The second defines the empty-prefix deficit. Set
\[
 p=\frac{\lambda}{2(n-1)U}.
\]
This equality defines the lower bound per missing component on the
probability that the next choice decreases the prefix deficit.
The certificate has minimum cut value $\lambda$, by
\eqref{eq:block3-certificate} and the scale guarantee. Summing the
cut lower bounds of the current components and dividing by two
therefore gives decrease probability at least $p(c_i(t)-1)$.
Writing $y_i(t)$ for the expected deficit gives
\[
 y_i(t+1)\le(1-p)y_i(t)+p y_{i-1}(t).
\]
The inequality subtracts the conditional expected decrease and uses
the difference of consecutive prefix deficits for the missing edges
of the last forest. The initial expected deficit is $i(n-1)$.
For a real number, a subscript plus denotes its positive part.
A binomial comparison solves this recurrence:
\begin{equation}\label{eq:block3-base-deficit}
 \E D_i(t)\le(n-1)\E[(i-\operatorname{Bin}(t,p))_+].
\end{equation}
The inequality
follows by induction: the right side obeys the corresponding equality
recurrence under one extra independent Bernoulli trial, has the same
initial value, and both recurrence coefficients are nonnegative.
No independence between the forests is assumed.

For analysis only, define
\[
 A=L\lambda/U,\qquad k=\lfloor199A/400\rfloor.
\]
The first equality defines the expected number of arrivals of a
minimum cut, and the second chooses a prefix just shorter than
the binomial mean. The algorithm does not use either quantity.
At the final draw count, that mean is $A/2$. By
\eqref{eq:block3-base-deficit} and a Chernoff lower-tail bound,
\[
 \Prb[D_k((n-1)L)>0]\le(n-1)k e^{-A/160000}.
\]
The event condition asks whether the chosen prefix is incomplete.
The inequality bounds its probability by its expected integer deficit,
then bounds that deficit by $k$ times the probability that the
comparison binomial falls below $k$; its relative deviation is at
least one two-hundredth. Choose the constant in $L$ large enough
that this failure has the prescribed fixed allowance and $A$ is
at least one thousand. Then
\begin{equation}\label{eq:block3-base-ratio}
 A/k\le101/50.
\end{equation}
The inequality follows by replacing the rounded-down denominator
by $199A/400-1$ and using the stated lower bound on $A$.
The algorithm chooses a fresh uniform member of the actually
completed prefix; an empty prefix returns the stored tree.

\subsection{Labels, original representatives, and work}

First place labels on aggregate pairs. On the event that the first
$k$ forests span, the average label cost in the completed prefix
is at most the total label cost of all sampled arrivals divided by
$k$. Conditional on any good scale history, its expectation is at most
\[
 \frac{L}{Uk}\sum_e c_U(e)z_e
 \le\frac{101}{50\lambda}\sum_e c_ez_e.
\]
The inequality uses \eqref{eq:block3-base-ratio} and edgewise
domination by the original capacities. The expression on the left
is obtained from the unconditional mean of the fresh sampled
arrivals, without conditioning their law on successful completion.

For labels on original edges, label each aggregate pair by the
capacity-weighted average of its labels. After selecting a tree,
choose one proportional original representative independently for
each of its pairs. Conditional on that aggregate tree, the
\emph{uncapped} representative choices have exactly its averaged
label cost in expectation. The final event $\mathcal E$ additionally
excludes all cap failures. Dropping the cap indicator only increases
a nonnegative label sum, so the same bound holds with
$\mathbf1_{\mathcal E}$. This is an uncapped coupling argument;
it does not condition representative choices on avoiding the cap.

Represent each forest by a rooted forest of component records. When
merging two components, attach the root of the smaller component below
the other root. A search follows parent links and leaves them unchanged.
Search paths have logarithmic length, because every depth increase
at least doubles the component size. Binary search over the nested
forest indices uses doubly logarithmically many finds per arrival.
The fixed arrival count is $O(n\log(2n))$, so this routing costs
$O(n\log^3(2n))$. The balanced sampler makes each pair choice in
expected $O(\log(2n))$ time, since there are at most quadratically
many aggregate pairs. All representative lists and their at-most-once
choices cost expected $O(m+n)$ time. Scale selection and integer
scalings fit the same overall bound. Allocate the error allowance
among the scale tests, incomplete-prefix event, and one aggregate
work cap. Uniform expected bounds and Markov's inequality give
the stated capped algorithm and complete
Lemma~\ref{lem:block3-base-packing}.

\begin{corollary}[Three base probabilities]\label{cor:block3-base-guarantees}
With error allowance one two-thousandth, the base distribution has
the following properties. It avoids every fixed original-edge set
of capacity at most $49\lambda/100$ with probability at least
$97/10000$; it uses at most one edge of every fixed set of capacity
at most $37\lambda/50$ with probability at least $2521/10000$;
and it crosses every fixed minimum cut at most twice with probability
greater than $326/1000$. All three probabilities exceed $1/200$.
\end{corollary}
\begin{proof}
For a nonnegative integer $r$ and a fixed original-edge set $B$,
Markov's inequality on the event $\mathcal E$ gives
\[
 \Prb[|T\cap B|\le r]
 \ge 1-\frac1{2000}
       -\frac{101c(B)}{50(r+1)\lambda}.
\]
The event condition specifies the permitted edge count. The inequality
first subtracts the exceptional-event probability, then uses the
fact that a count larger than $r$ is at least $r+1$ in the moment
bound of Lemma~\ref{lem:block3-base-packing}. Substitute respectively
the budgets zero, one, and two, with normalized capacities
$49/100$, $37/50$, and one. The stated numerical bounds follow by
rational arithmetic.
\end{proof}

\section{Random priorities and contraction}\label{sec:block3-clock-generator}

The recursive algorithm uses a random spanning tree to introduce
connections gradually. The central estimate bounds the total capacity
that remains between its components, simultaneously over all
thresholds. This avoids classifying every input edge at every scale.

Aggregate parallel endpoint pairs, retaining their original-edge lists.
For a pair $e$, write $c_e$ for its aggregate capacity. Independently
for each pair, let $R_e$ count fair heads before the first tail, and
define its priority by
\[
  \kappa_e=2^{R_e}c_e.
\]
This equality defines the priority by repeated doubling. The expected
number of fair bits and doublings is constant per pair.
Let $T$ be a maximum spanning tree for these priorities, with integer
identifiers breaking ties. For an endpoint pair $e$ with endpoints
$u,v$, let $T[u,v]$ denote its tree path. Define its path minimum
and the analytical sum $Z$ by
\[
  \tau_e=\min_{f\in T[u,v]}\kappa_f,
  \qquad Z=\sum_e\frac{c_e}{\tau_e}.
\]
The first equality defines the path minimum, and the second defines
the sum used to bound the number of generated arrivals. These ratios
are used only in the analysis.

\begin{lemma}[An expected sum of ratios]\label{lem:block3-clock-potential}
The priorities satisfy
\[
  \E Z\le 2(n-1).
\]
The inequality follows by comparing the priority thresholds with
exponential arrivals, as proved below. Moreover, for each positive
threshold $q$, let $W_q$ be the total original capacity between
components formed by the pairs of priority at least $q$. Then,
simultaneously for all thresholds,
\[
  W_q=\sum_{e:\tau_e<q}c_e,
  \qquad \frac{W_q}{q}\le Z.
\]
The equality uses the threshold connectivity property of $T$. The
condition inside the sum selects exactly the separated endpoint pairs.
The inequality follows by replacing the denominator $q$ of every
selected term by its smaller path minimum.
\end{lemma}
\begin{proof}
For the analysis, give each aggregate pair an independent exponential
arrival time of rate $c_e$. Let $t_e^X$ be the time at which its
endpoints first become connected, and let $W(t)$ be the capacity
between current components. Integration of the separation indicators
gives
\[
  \sum_e c_e t_e^X=\int_0^\infty W(t)\,dt.
\]
The equality holds because each pair contributes its capacity for
exactly the time before its endpoints become connected. There are
$n-1$ component-merging events. Conditional on any such event's
history, the next merging event has exponential waiting time with
rate equal to the current crossing capacity. Thus each interval
between consecutive merges contributes one in expectation, and
\[
  \E\sum_e c_e t_e^X=n-1.
\]
The equality sums those conditional expectations over the merges.

Couple the geometric priorities to these arrivals as follows. For
each pair draw an independent uniform random variable $V_e$ in the
open unit interval. In this analytical coupling define
\[
 R_e=\left\lfloor\log_2(1/V_e)\right\rfloor,
 \quad X_e=-\frac{\ln(1-V_e)}{c_e},
 \quad D_e=\frac{2^{-R_e}}{c_e}.
\]
The first equality gives the geometric head-count distribution, the
second gives an exponential variable of rate $c_e$, and the third
defines the reciprocal priority. These definitions give
\[
 D_e<\frac{2V_e}{c_e}\le 2X_e.
\]
The first inequality follows from the defining dyadic interval of
$R_e$; the second follows from the logarithmic inequality bounding
minus the natural logarithm of one minus $V_e$ below by $V_e$.
Connectivity times are monotone in individual arrival times.
Consequently the connectivity time $t_e^D$ under deadlines $D_e$
satisfies
\[
 t_e^D\le 2t_e^X,
 \qquad t_e^D=1/\tau_e.
\]
The inequality follows from the coordinatewise comparison of arrival
times. The equality is the threshold connectivity property of the
maximum spanning tree. Therefore
\[
 \E Z=\E\sum_e c_e t_e^D
       \le 2\E\sum_e c_e t_e^X=2(n-1).
\]
The first equality substitutes the displayed reciprocal identity;
the inequality uses the coupled connectivity-time bound term by term;
the last equality uses the preceding merge-interval calculation.
The threshold statements follow from the same threshold connectivity property.
The algorithm never evaluates the logarithms, reciprocals, or
exponential variables used in this proof.
\end{proof}

\begin{lemma}[Grouping pairs by a tree edge]\label{lem:block3-clock-owners}
In expected $O(m+n)$ work, one can construct $T$ and, for every
aggregate pair $e$, choose a tree edge $f(e)$ attaining the minimum priority on its tree
path. The original edges can then be grouped by their chosen
tree edge in linear additional work.
\end{lemma}
\begin{proof}
Generate the priorities and use the expected-linear comparison
minimum spanning tree algorithm with the order reversed. Given $T$,
apply the linear offline tree-path extremum algorithm
\cite{DRT1992,BuchsbaumEtAl2008}. Ties are broken consistently by tree-edge identifiers,
so each answer includes an attaining edge. Tree-edge queries are
answered directly. If the path-query interface is expressed as
non-tree-edge sensitivity of a supplied minimum spanning tree,
place every query edge in a comparison category above every tree
edge; this makes the supplied tree a minimum spanning tree without
altering its path extrema. This uses comparison categories, not
numeric perturbations. Integer buckets then collect the lists.
\end{proof}

\begin{lemma}[Avoiding a fixed set of original edges]\label{lem:block3-seed-protection}
Choose independently, for each aggregate pair, one original edge
with probability proportional to its capacity. Use this same
representative whenever that pair supplies an initial tree edge.
For a fixed set $B$ of original edges and a positive threshold $a$,
the probability that some representative in $B$ belongs to a pair
of priority at least $a$ is at most $c(B)/a$.
\end{lemma}
\begin{proof}
For each pair $e$, let $c_B(e)$ be the capacity of its original edges
in $B$. The geometric tail and the trivial probability bound give
\[
 \Prb[\kappa_e\ge a]\le\min\{1,c_e/a\}\le c_e/a.
\]
The event condition is the threshold test in the lemma. The first
inequality uses the geometric tail when the threshold exceeds the
capacity and the trivial bound otherwise; the second drops the
minimum with one. The representative and the priority are independent.
The probability that the pair has a marked representative and
reaches the threshold is therefore at most $c_B(e)/a$, by multiplying
the displayed upper bound by the representative's probability.
Summing over the pairs proves the claim. Requiring membership in
$T$ can only restrict the event. Sharing one representative among
all forest indices avoids multiplying this probability by the
number of indices.
\end{proof}

\section{A recursive distribution of spanning trees}
\label{sec:block3-marked-budget-bootstrap}

The recursive step builds a spanning forest with few components and then
completes it on the quotient graph.  An original minimum cut need not
remain a cut of the quotient, because a forest component may meet both
shores.  We therefore preserve two additional guarantees for arbitrary
sets of original edges.  They account for crossings already used by the
forest.

\begin{definition}[Three spanning-tree guarantees]
\label{def:block3-three-properties}
Let \(H\) be a connected undirected multigraph on at least two vertices with positive capacities,
and let \(\lambda_H\) denote its minimum-cut value.  A distribution on
spanning trees has guarantees \((p_0,p_1,p_2)\) if the following hold.
\begin{enumerate}
\item For every fixed set \(B\) of original edges of total capacity at
most \(49\lambda_H/100\), the tree avoids \(B\) with probability at
least \(p_0\).
\item For every fixed set \(B\) of original edges of total capacity at
most \(37\lambda_H/50\), the tree uses at most one edge of \(B\) with
probability at least \(p_1\).
\item For every fixed global minimum cut, the tree crosses that cut at
most twice with probability at least \(p_2\).
\end{enumerate}
The same distribution satisfies all three requirements and receives
neither the marked set nor the cut.
\end{definition}

A recursive call receives a quotient determined by its parent's
randomness and uses fresh randomness of its own.  Thus its guarantees
apply conditionally on the complete parent history.
The base construction of Lemma~\ref{lem:block3-base-packing} has all
three probabilities at least \(1/200\), by
Corollary~\ref{cor:block3-base-guarantees}.  We give one implementation
of the recursive step, charging every fair random bit.

\subsection{Independent Poisson counts}

\begin{lemma}[Independent Poisson counts]
\label{lem:block3-fair-histogram}
Let \(x_1,\ldots,x_p\) be nonnegative opaque masses with positive total
\(W\), let \(d\) be a positive opaque register, and let \(L\) be a
positive integer. Define independent counts by
\[
 N_i\sim\operatorname{Pois}(Lx_i/d).
\]
The displayed relation prescribes the Poisson law of each count;
the counts are independent. They can be generated in expected
\(O(p+L+LW/d)\) work, including every fair random bit. Only addition,
subtraction, and comparison are applied to opaque registers.
\end{lemma}
\begin{proof}
Pad the list to a power-of-two length with zeros.  Repeated deterministic
selection of the upper half partitions the decreasing ranks into groups
of sizes \(1,1,2,4,\ldots\), in linear total work, without sorting within
groups.  For a group \(g\), let \(s_g\) denote its size and \(b_g\) its
maximum.  Then
\[
 \sum_gs_gb_g\le2W.
\]
The inequality charges the first two singleton groups to twice the
largest mass and every subsequent group's envelope to twice its
preceding group's actual mass.  These preceding groups are disjoint.

Skip groups with zero maximum.  Form \(Lb_g\) once per remaining group
by doubling and addition.  For every member \(i\), independently
generate a Poisson count of mean \(Lb_g/d\), using
Lemma~\ref{lem:block3-ratio-poisson}, and retain each occurrence with
an exact ratio coin of probability \(x_i/b_g\).  Independent retention multiplies each Poisson mean by the retention
probability, giving the required independent counts. Generating the
counts and retaining occurrences have expected cost \(O(p+LW/d)\), by the preceding envelope bound.  There are
\(O(1+\log(p+1))\) groups; their register scalings cost
\(O((1+\log(p+1))(1+\log(L+1)))\), which is \(O(p+L)\).
This last bound follows by bounding logarithms by constant multiples
of square roots and applying the arithmetic--geometric mean inequality.
Both exact primitives charge all fair bits and their integer arithmetic.
\end{proof}

\subsection{Generating rounded occurrences}
\label{sec:block3-rounded}
\label{sec:block3-fair-bits}

Consider a connected recursive input with \(n\) vertices, \(m\) original
positive-capacity edges, and minimum-cut value \(\lambda\).
Choose a shrink parameter \(R\) at least one and an integer \(K\) at
least \(1000\).  The lower bound on \(K\) is used in the estimates below.
Define the component threshold by
\[
 q_{\rm cap}=\max\{2,\lceil n/R\rceil\}.
\]
This definition leaves a constant-size quotient when rounding would
otherwise give fewer than two vertices.  Set
\begin{equation}\label{eq:block3-rounded-parameters}
 L=1024K,\qquad D=1024K,\qquad \delta=\frac1{D+1}.
\end{equation}
The first equality fixes the lifetime multiplier, the second fixes the
number of subdivisions in each dyadic interval, and the third defines
the relative error used in the analysis.

Use the construction of Lemma~\ref{lem:block3-clock-potential}.
Each aggregated pair \(e\) has priority \(\kappa_e\), and \(T\) is a
maximum spanning tree for these priorities.  For every pair \(e\),
choose a tree edge \(f(e)\) attaining the path minimum \(\tau_e\).
For a tree edge \(f\), let \(E_f\) contain the original edges whose
aggregated pair is assigned to \(f\), and let \(W_f\) be their total
capacity.  These groups are constructed in linear time by
Lemma~\ref{lem:block3-clock-owners}.
Choose one original-edge representative for every pair in \(T\),
independently with probability proportional to capacity.  The same
representative is used at all indices.  These choices are independent
of the priorities and ordinary occurrences.  We refer to the priorities
and these representatives together as the initial-tree history.

Introduce the representative of each tree edge \(f\) into each of
\(K\) connectivity partitions at analytical time \(L/\kappa_f\).
For each group \(E_f\), apply Lemma~\ref{lem:block3-fair-histogram}
with denominator \(\kappa_f\).  Thus an original edge \(a\) of capacity
\(c_a\) in that group has count distributed as
\(\operatorname{Pois}(Lc_a/\kappa_f)\).  Counts are independent
conditional on the priorities.  For each occurrence, independently
choose a nonnegative integer \(r\) and an integer \(k\) with the following
laws, specified for every nonnegative integer \(j\):
\[
 \Prb[r=j]=2^{-j-1},\qquad
 k\text{ uniform in }\{D+1,\ldots,2D\}.
\]
The event in the probability fixes the dyadic-interval index to \(j\);
the outer equality specifies its geometric probability law.
Assign the occurrence physical time
\begin{equation}\label{eq:block3-rounded-token-time}
 t=\frac{Lk}{2D\,2^r\kappa_f}.
\end{equation}
This equality defines its rounded time; the quotient is analytical
and is not evaluated.  At a common physical time, insert all
initial-tree representatives first, then process ordinary occurrences
by original-edge identifier and finally copy identifier.

\begin{lemma}[Cell distribution and total count]
\label{lem:block3-rounded-cell-law}
Conditional on priorities, consider an original edge whose aggregated
pair is \(e\).  It has independent Poisson counts in consecutive cells ending at
\[
 \frac{L}{\tau_e}\frac{k}{2D\,2^r},
 \qquad r\text{ a nonnegative integer},\quad
 k\in\{D+1,\ldots,2D\},
\]
Each cell mean is its length
times the original-edge capacity.  Consecutive endpoints have ratio
at most \(1+1/D\).  If \(N\) is the total occurrence count, then
\begin{equation}\label{eq:core-occurrence-mean}
 \E N\le2L(n-1).
\end{equation}
The inequality is the clock-potential estimate applied to the total
conditional mean.
\end{lemma}
\begin{proof}
Independently assigning Poisson occurrences to cells gives independent
Poisson cell counts, with each mean multiplied by its assignment
probability. Indeed, substituting the Poisson mass formula into the
probabilities for independently assigning occurrences factors the
joint probability into the cellwise Poisson probabilities. For an
original edge \(a\), the cell mean is
\[
 \frac{Lc_a}{\tau_e}\frac{2^{-r-1}}D
   =c_a\frac{L}{2D\,2^r\tau_e}.
\]
The equality collects the denominator factors.  The rightmost factor
is the cell length, also at the boundary between dyadic intervals.
Counts for distinct cells and original edges are independent.  Thus the
construction rounds Poisson arrival times upward by a factor at most
\(1+1/D\).  Summing conditional means gives
\[
 \E N=L\,\E\sum_e\frac{c_e}{\tau_e}\le2L(n-1).
\]
The equality aggregates original capacities within each pair; the
inequality is Lemma~\ref{lem:block3-clock-potential}.
\end{proof}

Generating the independent counts takes expected work \(O(m+nK)\),
including the additive \(L\) term for each group \(E_f\).  Sampling the
dyadic interval uses constant expected fair-bit work, and sampling
\(k\) by binary rejection uses \(O(\log(K+1))\) expected bits.
Choosing the initial-tree representatives takes expected work
\(O(m+n)\): a proportional draw from \(p\) parallel edges uses
\(O(1+\log(p+1))\) work, and the original-edge lists for distinct
pairs in \(T\) are disjoint.

\subsection{Recovering the forests without sorting}
\label{subsec:core-routing}

An ordinary occurrence stores a doubled priority register \(A\).
Its key representation is
\[
 A=2^r\kappa_f,\qquad t=\frac{L}{2D}\frac{k}{A}.
\]
The first equality defines the register, formed by \(r\) doublings;
the second rewrites the time definition.  An initial-tree edge uses the pair
\((\kappa_f,2D)\).  For each register \(A\), precompute the multiples
\(uA\) for integers \(u\) from \(D+1\) through \(2D\), using \(O(D)\)
additions.  Two physical times are compared by comparing the stored
registers \(k_1A_2\) and \(k_2A_1\).  Initial-tree edges precede ordinary
occurrences at the same time, and the identifiers resolve the remaining
ties.  Thus all total keys are distinct.

The conceptual process maintains \(K\) nested partitions, indexed from
one through \(K\).  An initial-tree representative is inserted into every partition.  An ordinary occurrence joins only the
first partition separating its endpoints, and is discarded if no such
partition exists.  Each partition only coarsens, and each later
partition refines every earlier one.

\begin{lemma}[Ordered MST implementation]
\label{lem:block3-rounded-routing}
The complete assignment of occurrences to partitions and the first completed physical
time at which all partitions have at most \(q_{\rm cap}\) components
can be recovered in expected \(O(K(N+n))\) work, conditional on the
generated event list.  Original-edge lists are not rescanned in these
passes.
\end{lemma}
\begin{proof}
Compute an MST of the event graph under its distinct total keys and
call it \(B_1\).  Kruskal's characterization shows that its ordinary
edges are precisely the first partition's accepted occurrences.
Its selected initial-tree edges have the same effect on components as
inserting all initial-tree edges, since a rejected edge is already internal.

Delete only the selected ordinary occurrences, retaining every initial-tree edge,
and compute the next ordered MST \(B_2\).  Repeat through all indices.
The ordinary input to each pass consists exactly of the occurrences
rejected at earlier indices.  This holds at every prefix, because
Kruskal's acceptance depends only on earlier keys.  Hence the prefix of
\(B_i\) has the same components as the conceptual partition at index
\(i\).  Every pass is connected because the entire initial spanning
tree is retained.

Use the exact expected-linear MST algorithm of
Karger, Klein, and Tarjan~\cite{KKT1995}, with fresh private randomness
at each pass.  Its expected work and list filtering are \(O(N+n)\).
Uniqueness of the ordered MST means these private bits affect work
but do not change any forest or its probability law.

If the initial partition meets the threshold, select the empty prefix.
Otherwise, use linear selection to find the \((n-q_{\rm cap})\)-th
key in \(B_K\) and take its complete physical time.  Each edge in a
tree prefix reduces its component count by one.  Thus this is the
first complete boundary meeting the threshold in the last partition,
and nestedness gives the same guarantee for all preceding partitions.
\end{proof}

Choose an index \(I\) uniformly and independently after finding this
boundary.  Let \(F\) be the prefix of \(B_I\) through the complete
selected physical time.  This prefix is already a spanning forest of
the desired components.  Construct its quotient \(Q\) by one
original-edge scan, invoke the recursive procedure once with fresh
randomness, and lift the resulting tree.  Write \(\lambda_Q\) for the
quotient's minimum-cut value when it has multiple vertices.  Then
\begin{equation}\label{eq:block3-mb-quotient-mincut}
 \lambda_Q\ge\lambda.
\end{equation}
The inequality holds because every quotient cut pulls back to an
original cut.  A one-vertex quotient needs no completion.

\begin{lemma}[Selected-prefix reconstruction]
\label{lem:core-prefix-reconstruction}
The procedure returns a valid original-edge spanning tree.  If the
initial-tree representatives avoid a marked set through the selected time, the
marked-edge count of \(F\) is at most the historical marked-acceptance
count at index \(I\).
\end{lemma}
\begin{proof}
An MST prefix is a forest with the maintained components, by
Lemma~\ref{lem:block3-rounded-routing}.  Its ordinary edges were
accepted at the chosen index, and its initial-tree edges avoid the marked
set under the hypothesis.  A quotient tree completes it to a
spanning tree, and lifting preserves original-edge identities.
\end{proof}

\begin{algorithm}[ht]
\caption{Recursive tree construction}\label{alg:core-recursion}
\KwIn{A connected positive-capacity multigraph; the shrink parameter \(R\); forest count \(K\); and a recursive tree procedure.}
Aggregate endpoint pairs, retaining original-edge lists. Generate geometric priorities and an initial maximum spanning tree. For each pair, choose a tree edge attaining its path minimum. Choose independent proportional original representatives for the tree edges\;
Generate independent Poisson counts for the original edges. Assign rounded keys and deterministic tie order, and construct the integer-multiple comparison tables\;
Compute the successive ordered MSTs, deleting selected ordinary occurrences after each pass and retaining every initial-tree edge\;
Select the first complete physical time meeting the component bound in the last MST. Choose an index uniformly and take its MST prefix as \(F\)\;
Contract \(F\), make one fresh recursive call, and \Return \(F\) together with the lifted quotient tree\;
\end{algorithm}
\noindent A one-vertex quotient returns the empty completion. A surrounding work cap, when used, returns a previously stored valid output.

The unconditional expected outer work is \(O(m+nK^2)\).
The tables and MST passes have expected cost \(O(K(N+n))\);
the expected count is \(O(nK)\); and sampling the integers \(k\) costs
\(O(nK\log(K+1))\).  Forest extraction and quotient formation add
\(O(m+n)\).  These estimates include all fair random bits.

\section{Analysis of the recursive step}
\label{sec:core-closure}

We first show that the process reaches a small quotient.  We then show
that a constant fraction of its indices retain an appropriate guarantee
for recursive completion.  The minimum-cut value is used only in the
analysis.  Define two analytical horizons by
\begin{equation}\label{eq:block3-rounded-horizons}
 T_*=\frac{2.002K}{\lambda},\qquad
 T_{\max}=\frac{2.004K}{\lambda}.
\end{equation}
The first equality defines the time used to bound the component
deficit; the second defines the later horizon for all probability
estimates.  Neither horizon is computed.

Fix a marked set \(B\) of total capacity at most \(\lambda\).
An initial-tree edge introduced by \(T_{\max}\) has priority at least
\(L/T_{\max}\), so Lemma~\ref{lem:block3-seed-protection} gives
\begin{equation}\label{eq:block3-rounded-protection}
 \Prb[\text{a marked initial-tree representative appears by }T_{\max}]
 \le\frac{c(B)T_{\max}}L
 \le\frac{2.004}{1024}<.002.
\end{equation}
The first inequality applies the bound for an initial-tree representative
belonging to the marked set; the second uses the capacity bound on \(B\)
and the definitions of the horizon and \(L\); the last is a numerical
comparison.  This event depends only on the priorities and initial-tree
representatives.  Condition below on any fixed choice of these quantities
for which no marked representative appears through \(T_{\max}\).
The ordinary counts retain their independent conditional laws.

\subsection{Integrated rates for rounded arrivals}

For analysis, extend each geometric grid beyond its lifetime and
restore the omitted Poisson cells.  Such occurrences are already
internal to a component of the shared initial-tree edges when processed,
so all partitions reject them.  Restoring them changes no accepted edge
or component.  If the marked set is a genuine cut and no initial-tree
representative crosses it, no edge of the cut has endpoints connected
by initial-tree edges through the analytical horizon.

Expand each physical grid point into consecutive blocks in the
algorithm's original-edge order.  A block has length equal to its cell
mean and contains an independent rate-one Poisson process.  Insert
initial-tree representatives before these blocks.  Copies of one edge have identical endpoints,
so this representation has exactly the implemented edge order.
Use physical time plus cumulative block length as expanded time.
Let \(E\) denote the original edge set of this recursive input.
The length inserted through physical time \(t\) is at most \(c(E)t\);
thus the potential grid points accumulating at zero cause no problem.
Let \(t(s)\) denote physical time at expanded time \(s\).  The revealed
history at an expanded time consists of all occurrences already exposed;
future occurrences in the precomputed lists remain hidden in the analysis.
For an original edge \(a\), let \(\mu_a(s)\) be its accumulated Poisson
mean through expanded time \(s\), namely the total length already exposed
in blocks belonging to \(a\). Thus \(\mu_a\) is the integral of its
arrival rate. It is deterministic after the initial-tree history
is fixed.  For a count that accepts occurrences according to the preceding
history, integrate each edge's arrival rate only while the rule would
accept its next occurrence. We use this integral for accepted counts
as well. It can be random because the acceptance decisions depend
on earlier occurrences.

Whenever a state or count process has a physical-time argument, it
denotes the state after all initial-tree insertions and ordinary occurrences
at that time.  An expanded-time argument denotes the intermediate state
within a batch.  The two conventions agree away from batch times.

\begin{lemma}[Bounds on cumulative arrival means]
\label{lem:block3-rounded-envelope}
Each function \(\mu_a\) is continuous and satisfies
\begin{equation}\label{eq:block3-rounded-envelope}
 (1-\delta)c_at(s)\le\mu_a(s)\le c_at(s).
\end{equation}
The first inequality uses the preceding grid endpoint and the relative
mesh error; the second uses the current physical endpoint.
Conditional on the fixed initial-tree history, the original-edge processes are
independent Poisson processes with these deterministic cumulative means.
\end{lemma}
\begin{proof}
Before its block, an edge's cumulative arrival mean equals its capacity times the
preceding grid endpoint.  During the block it advances continuously
to its capacity times the current endpoint and afterward remains there.  Consecutive endpoints have
ratio at most \(1+1/D\), and
\[
 \frac1{1+1/D}=1-\delta.
\]
The equality substitutes the definition of \(\delta\).
The same bounds apply to edges without a block at the current time.
Independence follows from the independent Poisson cell counts and the
independent processes in the blocks.
\end{proof}

Every acceptance decision is determined by the revealed history just
before the occurrence.  Lemma~\ref{lem:block3-rounded-routing} identifies
the offline computation with this ordered process at every prefix, and
therefore also identifies its selected stopping boundary.  In particular,
the analysis never uses advance knowledge of the precomputed lists.

\begin{lemma}[Weighted integral estimate]
\label{lem:block3-rounded-weighted-compensator}
Let \(g\) be a finite-valued, nonnegative, nonincreasing function,
with its value at each time determined by the revealed history.  Let \(\alpha\) be nonnegative, and define the exponential
weight by \(w(t)=e^{\alpha t}\).  This equality specifies the weight.
For a physical-time argument, \(g(t)\) denotes its end-of-batch
value; values at the isolated batch times do not affect a
physical-time integral.  At an expanded-time endpoint of physical
coordinate \(T\),
\begin{equation}\label{eq:block3-rounded-weighted-compensator}
 \int w(t(s))g(s-)\,d\mu_a(s)
 \ge c_a\int_0^T w(t)[1-\delta(1+\alpha t)]g(t)\,dt.
\end{equation}
The inequality follows by integration by parts, using the bounds of
Lemma~\ref{lem:block3-rounded-envelope} and the fact that \(g\) is nonincreasing; the proof below
bounds the resulting endpoint and variation terms.
\end{lemma}
\begin{proof}
Define the error by \(\nu_a(s)=\mu_a(s)-c_at(s)\).
This equality defines the difference between the expanded process's
cumulative rate and the unrounded cumulative rate.
The error is continuous, nonpositive, and at least
\(-\delta c_at(s)\).  Let \(\sigma\) denote the final expanded-time
position, with physical coordinate \(T\).  In the next calculation,
integrals without explicit arguments are in expanded time.
Integration by parts gives
\[
 \int wg\,d\mu_a
 =c_a\int wg\,dt+w(T)g(\sigma)\nu_a(\sigma)-\int\nu_a\,d(wg).
\]
The equality separates the physical-clock term and integrates its
error by parts, using the zero initial error.  The contribution from
differentiating \(w\) in the last integral is nonnegative.
The endpoint and decreasing-\(g\) contributions are bounded below by
\[
 -\delta c_a\left(Tw(T)g(\sigma)+\int tw(t)(-dg)\right).
\]
This lower bound uses the error envelope and the nonpositivity of
increments of \(g\).  A second integration by parts gives
\[
 Tw(T)g(\sigma)+\int tw(t)(-dg)
 =\int_0^T w(t)(1+\alpha t)g(t)\,dt.
\]
The equality differentiates \(tw(t)\).  Substituting proves the
claim.  Continuity of the error in expanded time makes the jump
correction terms in integration by parts vanish.
\end{proof}

\subsection{Reaching the component threshold}

Let \(c_i(s)\) be the component count at index \(i\).
For an original edge \(a\), let \(j_a(s)\) be its first separated
index, with infinite value if all indices connect its endpoints.
These indices are nondecreasing because partitions only coarsen.

\begin{lemma}[Stopping estimate]
\label{lem:block3-rounded-deficit}
Conditional on every fixed initial-tree history,
\[
 \E(c_K(T_*)-1)\le1001(n-1)e^{-K/3000000}.
\]
The inequality follows by geometrically weighting the nested component
deficits and integrating their decay, as proved below.
Consequently the probability of stopping later than \(T_{\max}\)
is at most
\begin{equation}\label{eq:block3-rounded-stop-error}
 2002R e^{-K/3000000}.
\end{equation}
\end{lemma}
\begin{proof}
Define coefficients and a deficit potential by
\[
 \rho=\frac{1000}{1001},\qquad
 a_i=\rho^{i-1},\qquad
 \Psi(s)=\sum_{i\in\{1,\ldots,K\}}a_i(c_i(s)-1).
\]
The first equality chooses the geometric coefficient; the second
defines the index weights; the third defines their weighted deficit.
An ordinary occurrence accepted at index \(j\) decreases the potential
by \(a_j\); initial-tree insertions can only decrease it.
Let \(g_a(s)\) be \(a_{j_a(s)}\), or zero at the infinite index.
It is nonincreasing.

Let \(W_i\) be the capacity between distinct components at index \(i\).
Summing their cut lower bounds gives
\[
 W_i\ge\frac{\lambda}{2}(c_i-1).
\]
The inequality counts each crossing edge twice when there are multiple
components; with one component both sides vanish.  Hence
\[
 \sum_a c_ag_a
 =a_KW_K+\sum_{j\in\{1,\ldots,K-1\}}(a_j-a_{j+1})W_j
 \ge\frac{\lambda(1-\rho)}2\Psi.
\]
The equality groups edges by first separated index and telescopes.
The inequality uses the preceding capacity bound and the differences
of geometric coefficients.  Define
\[
 \alpha_0=\frac{\lambda}{2002},\qquad
 \alpha=(1-10^{-5})\alpha_0.
\]
The first equality names the resulting rate coefficient; the second
leaves slack for rounding.  Through \(T_{\max}\),
\[
 \delta(1+\alpha t)
 \le\frac{1+(2.004/2002)K}{1024K}
 <2\cdot10^{-6}<10^{-5}.
\]
The first inequality substitutes the parameter and horizon bounds;
the second uses \(K\) at least \(1000\); the last is a numerical
comparison.

Apply Lemma~\ref{lem:block3-rounded-weighted-compensator} to every
\(g_a\) and take expectations in the evolution of the exponentially
weighted potential.  The rate lower bound cancels the exponential
weight's derivative; initial-tree insertions contribute nonpositive terms.  Thus
\[
 \E[e^{\alpha t}\Psi(t)]\le\Psi(0).
\]
The inequality is the integrated drift bound; it asserts no
pointwise physical-time rate for a rounded batch.
Initially the geometric weights sum to at most \(1001\).
Dividing by the final coefficient gives
\[
 \E(c_K(T_*)-1)
 \le1001(n-1)
 \exp\left(K\log\frac{1001}{1000}
            -(1-10^{-5})\frac K{1000}\right)
 \le1001(n-1)e^{-K/3000000}.
\]
The first inequality substitutes the initial bound and \(T_*\),
also replacing \(K-1\) by \(K\) in the positive exponent.
The second applies the alternating cubic upper bound for
the natural logarithm of one plus its argument, evaluated at one
one-thousandth; its decay coefficient exceeds
\(4.89\cdot10^{-7}\).

Finally,
\[
 q_{\rm cap}-1\ge\frac n{2R}.
\]
The inequality follows from the ceiling when \(n/R\) is at least two
and from the constant threshold otherwise.  Markov's inequality gives
the stated stopping error.  If the component threshold holds at
\(T_*\), an eligible complete boundary has already occurred because
partitions are constant between boundaries.
\end{proof}

\subsection{Uniform counting estimates}

\begin{lemma}[Uniform count deviations]
\label{lem:block3-rounded-count-martingale}
Let \(N\) be a counting process whose jumps have size one, and let
\(\mathcal A\) be the continuous integral of its conditional arrival rate,
as defined above. Both start from zero, either at a fixed expanded
time or at a random expanded time whose occurrence can be recognized
from the history already revealed. Suppose \(\mathcal A\) remains at most \(v\) through the
remaining horizon.  For a positive \(u\) at most \(2v\),
\[
 \Prb[\exists s:N(s)-\mathcal A(s)\ge u]\le e^{-u^2/(4v)},\qquad
 \Prb[\exists s:N(s)-\mathcal A(s)\le-u]\le e^{-u^2/(4v)}.
\]
Inside the first probability, the inequality defines the
upper-deviation event; the outer inequality follows by stopping the
upper exponential process. Inside the second probability, the
inequality defines the lower-deviation event; the outer inequality
follows by stopping the lower exponential process. Each process is
nonnegative and has conditional expected future value at most its
present value, as proved next. The same reasoning holds conditionally
after a restart at such a random time.
\end{lemma}
\begin{proof}
Set \(\theta=u/(2v)\), defining an exponential parameter at most one.
Consider the nonnegative exponential processes
\[
 \exp\{\theta N-(e^\theta-1)\mathcal A\},\qquad
 \exp\{-\theta N-(e^{-\theta}-1)\mathcal A\}.
\]
Their conditional expected future values are at most their current
values: a unit arrival multiplies the first process
by \(e^\theta\) and the second by \(e^{-\theta}\), while the integrated-rate
terms cancel the corresponding conditional rates of change.  This
calculation uses the acceptance rule immediately before each arrival;
it does not require accepted arrivals to have independent increments.
On the interval from zero to one, each of
\(e^\theta-1-\theta\) and \(e^{-\theta}-1+\theta\) is at most
\(\theta^2\), by its Taylor expansion.  On the corresponding
deviation event the relevant process is at least
\[
 \exp(\theta u-\theta^2v)=\exp(u^2/(4v)).
\]
The equality substitutes the chosen \(\theta\).
Both processes start at one.  Stop the relevant process when it first
reaches the displayed threshold, and also at a bounded horizon.  Its
stopped expectation is at most one, by its conditional-expectation
property.  Nonnegativity therefore bounds the probability of reaching
the threshold by its reciprocal.  One may first bound the number of
exposed arrivals and the horizon, then let both bounds increase; the
crossing events increase to the desired event.  The same argument,
conditional on the revealed history, applies after a restart at such a
random time.
\end{proof}

Every marked count used below has integrated conditional arrival rate
at most \(3K\) through
\(T_{\max}\), since its total original capacity is at most \(\lambda\).
We always apply the lemma with this bound.

\subsection{Arbitrary marked sets}

Fix a marked set \(B\).  Let \(x_i\) be its historical accepted count
at index \(i\), let \(S_i\) be the sum of these counts through index
\(i\), let \(b_i\) be its currently external capacity at that index,
and let \(Z_B\) count all its conceptual occurrences, including restored
ones.

First suppose the capacity of \(B\) is at most \(.49\lambda\).
At the fixed horizon, \(Z_B\) is Poisson with mean at most
\(.98196K\).  A Poisson upper-tail bound therefore makes it at most
\(.99K\), except with error exponentially small in \(K\).
At least \(K/100\) indices have no marked acceptance.  At any such
index, the selected forest avoids \(B\), and its quotient's residual
marked capacity is at most \(.49\lambda_Q\), by
\eqref{eq:block3-mb-quotient-mincut}.  The first recursive guarantee
completes this forest without a marked edge.

Now suppose the capacity of \(B\) is at most \(.74\lambda\).
Set \(\eta=10^{-4}\), fixing the slack in the next estimates.
Let \(\mathcal A_i\) be the integral of the conditional arrival rate
of \(S_i\).
An edge still external at index \(i\) was external at every earlier
time, since partitions only coarsen.  Thus
\begin{equation}\label{eq:block3-rounded-weak-compensator}
 \mathcal A_i(s)\ge
 \sum_{a\text{ external at }s}\mu_a(s)
 \ge(1-\delta)b_i(s)t(s).
\end{equation}
The first inequality retains only occurrences from edges external
at the current time, all of which would have been accepted by this
index or an earlier one.  The second applies Lemma~\ref{lem:block3-rounded-envelope}.

Lemma~\ref{lem:block3-rounded-count-martingale}, with deviation
\(\eta K/8\), and a union over indices give
\[
 S_i(s)\ge(1-\delta)b_i(s)t(s)-\eta K/8
\]
simultaneously at all times, except with probability
\(K e^{-\eta^2K/768}\).  The inequality combines the lower
count deviation with \eqref{eq:block3-rounded-weak-compensator}.
At physical times at least \(K/(4\lambda)\), this implies
\begin{equation}\label{eq:block3-rounded-weak-hazard}
 b_i(t)\le\lambda\left(\frac{S_i(t)}{\lambda t}+\eta\right).
\end{equation}
The inequality absorbs the rounding and additive errors into
\(\eta\lambda t\): the rounding error is below \(.002\),
whereas \(\eta K/8\) is at least \(.0125\), and
\(\eta\lambda t\) is at least \(\eta K/4\).
The upper deviation estimate with deviation \(\eta K/4\) gives
\begin{equation}\label{eq:block3-rounded-mark-count}
 Z_B(t)\le(.74+\eta)\lambda t
\end{equation}
at the same times, except with probability
\(e^{-\eta^2K/192}\).  The inequality uses the upper bound on the integrated conditional
rate \(.74\lambda t\) and absorbs the deviation by the lower bound
on \(t\).  At earlier times, Poisson domination and monotonicity make
\(Z_B\) at most \(K/2\), except with error exponentially small in
\(K\).  These estimates are uniform through the horizon; they are
not conditioned on successful stopping.

At the selected boundary \(t\), if it is in the later range, put \(A=\lambda t\), defining its
normalized physical time, and put \(u=(.49-\eta)A\), defining a
count threshold.  Call an index useful if it has no marked acceptance,
or if it has exactly one and \(S_i\) is at most \(u\).
The former uses the second recursive guarantee; the latter uses the
first, by \eqref{eq:block3-rounded-weak-hazard}.

Charge cumulative marked acceptances at rate \(1/2\) below \(u\)
and rate one above \(u\).  A nonuseful index below the threshold
uses at least two acceptances, and one above it uses at least one.
Only the single threshold-crossing step can require an extra unit.
Consequently
\[
 \#\{\text{nonuseful indices}\}
 \le\tfrac12\min\{Z_B,u\}+\max\{0,Z_B-u\}+1
 \le(.495+1.5\eta)A+1
 \le .9922806K+1.
\]
The first inequality is this charging argument.  The second uses
monotonicity in \(Z_B\), substitutes
\eqref{eq:block3-rounded-mark-count}, and evaluates the resulting
piecewise expression.  The third substitutes the horizon and the
value of \(\eta\).  For the specified lower bound on \(K\), at least
\(K/200\) indices are useful.  If the selected boundary is in the earlier range,
at least half the indices have no marked acceptance, which is
stronger.  This argument never asserts a prefix property for an
arbitrary marked set.

\subsection{A fixed minimum cut}

Fix an original global minimum cut \(C\).  Use \(x_i,S_i,b_i\)
for this cut and \(Z\) for all its processed occurrences.
Condition on an initial-tree history in which no representative crosses
\(C\) through the analytical horizon.

\begin{lemma}[First target occurrences]
\label{lem:block3-mb-forced-prefix}
Indices with a positive target-acceptance count form an initial
segment.  For each index \(i\), the first \(\min\{i,Z\}\) target
occurrences are all accepted among the first \(i\) indices.
Before index \(i\) has a target acceptance, every edge of \(C\)
is external to that index.
\end{lemma}
\begin{proof}
A partition with no target acceptance has every component contained
in one shore, because all its defining edges are nontarget edges
or initial-tree representatives that do not cross \(C\).  Conversely, a component meeting both shores
requires a target acceptance.  Nestedness implies that a mixed
component at an index is mixed at every earlier index.  Hence positive
target counts form a prefix.  Before the \(j\)-th target occurrence,
index \(j\) cannot be mixed: doing so would require at least \(j\)
preceding target acceptances.  The occurrence is therefore accepted
no later than that index.  A pure partition separates every target
edge, proving the final assertion.
\end{proof}

The lemma is special to a cut.  Unmarked paths may connect the
endpoints of an arbitrary marked edge, so the same assertion would
be false for a general marked set.

\begin{lemma}[Residual target capacity]
\label{lem:block3-rounded-strong-hazard}
Put \(h=K/1000\), defining a count slack.
Except with probability \(2K e^{-K/192000000}\), the following
holds for both thresholds \(\beta\) in \(\{.49,.74\}\), every index
\(i\), and every expanded time through \(T_{\max}\) at which \(Z\)
has reached \(i\):
\begin{equation}\label{eq:block3-rounded-strong-hazard}
 S_i-i+h\le\beta(Z-i)
 \quad\Longrightarrow\quad b_i\le\beta\lambda.
\end{equation}
The first inequality is the assumed bound on accepted target
occurrences. The second follows by restarting the count-deviation
estimates at the \(i\)-th target occurrence and applying
Lemma~\ref{lem:block3-rounded-envelope} to the edges still external
afterward, as proved below.
\end{lemma}
\begin{proof}
Let \(\sigma_i\) be the expanded time of the \(i\)-th target occurrence,
when it exists, and define \(t_i=t(\sigma_i)\).
This equality names its physical time.  Immediately afterward,
\[
 S_i=Z=i.
\]
The first equality holds because Lemma~\ref{lem:block3-mb-forced-prefix}
ensures that all target occurrences so far were accepted among the
first \(i\) indices. The second holds because the restart occurs
immediately after the \(i\)-th target occurrence.
For a later expanded time \(\sigma\), define \(t=t(\sigma)\).
This equality names its physical coordinate.  Abbreviate all counts
at this current expanded time, and define the tail counts by
\[
 N_i=S_i-i,\qquad Z_i^\circ=Z-i.
\]
The first equality removes the forced acceptances and the second
removes the first \(i\) target occurrences.
Let \(\mathcal A_i^\circ\) and \(M_i^\circ\) be the integrals of their conditional arrival
rates from \(\sigma_i\) through \(\sigma\).  Both remain at most \(3K\).
Apply Lemma~\ref{lem:block3-rounded-count-martingale} after the restart,
with deviation \(h/4\).  Outside the stated total error,
\begin{equation}\label{eq:block3-rounded-tail-deviations}
 N_i\ge\mathcal A_i^\circ-h/4,\qquad
 Z_i^\circ\le M_i^\circ+h/4
\end{equation}
simultaneously for all indices and subsequent times.  The first
inequality is the lower count deviation; the second is the upper
target-count deviation.  A union bound over two tails and \(K\)
indices gives the stated error.

Freeze the target edges external at expanded time \(\sigma\).  Each was external
throughout the interval after \(\sigma_i\).  Therefore
\begin{equation}\label{eq:block3-rounded-residual-tail}
 \mathcal A_i^\circ
 \ge\sum_{a\text{ external at }\sigma}
       \bigl(\mu_a(\sigma)-\mu_a(\sigma_i)\bigr)
 \ge b_i(\sigma)\max\{0,(1-\delta)t-t_i\}.
\end{equation}
The first inequality counts only these surviving external edges;
the second applies the lower envelope at the final time, the upper
envelope at the restart, and nonnegativity of increments.
The integrated target arrival rate satisfies
\[
 M_i^\circ\le\lambda\bigl(t-(1-\delta)t_i\bigr).
\]
The inequality uses the upper envelope at the final time and the
lower envelope at the restart.

Suppose the conclusion of
\eqref{eq:block3-rounded-strong-hazard} fails.  Combining the last
two displays gives
\[
 \mathcal A_i^\circ\ge\beta M_i^\circ-2\beta\delta\lambda t
 >\beta M_i^\circ-.004.
\]
The first inequality uses that \(t_i\) is at most \(t\) and that the
surviving capacity exceeds \(\beta\lambda\), retaining validity if
the intermediate time difference is negative.  The strict inequality
substitutes the mesh, horizon, and largest threshold.
Together with \eqref{eq:block3-rounded-tail-deviations}, it implies
\[
 S_i-i\ge\beta(Z-i)-(1+\beta)h/4-.004.
\]
The inequality substitutes the lower tail bound and the upper target
tail bound into the preceding comparison.
Because \(h\) is at least one, the total subtraction is strictly
smaller than \(h\).  This contradicts the antecedent.
The same two deviation events work for both thresholds.
\end{proof}

At the fixed horizon, \(Z\) is Poisson with mean at most \(2.004K\).
It exceeds \(2.01K\) only with exponentially small probability.
Together with Lemma~\ref{lem:block3-rounded-strong-hazard}, this yields
a constant useful-index fraction as follows.

Define two residual-capacity thresholds by
\[
 \beta_0=.49,\qquad \beta_1=.74.
\]
The first equality specifies the threshold for using no additional
target edge; the second specifies the threshold for using at most one.  Call an index useful in any of three cases:
it has no target acceptance; it has one and satisfies the antecedent
of \eqref{eq:block3-rounded-strong-hazard} for \(\beta_1\); or it
has two and satisfies that antecedent for \(\beta_0\).
At a zero-target index, every component is pure, so \(C\) descends
to a nontrivial quotient cut of unchanged value.  It is a quotient
minimum cut by \eqref{eq:block3-mb-quotient-mincut}, and the third
recursive guarantee applies.  At one or two target acceptances,
the residual-capacity lemma permits the second or first guarantee,
respectively.  Prefix reconstruction can only reduce the historical
target count.  Thus each useful index succeeds with probability at
least the smallest predecessor probability.

\begin{lemma}[Useful-index count]
\label{lem:block3-mb-potential}
On these concentration events, at least \(K/100\) indices are useful.
\end{lemma}
\begin{proof}
Let \(\ell\) be the positive-prefix length.  Set \(S_0=0\),
because the cumulative count over no indices is zero.
For indices from zero through \(\ell\), define
\[
 D_i=Z-i,\qquad Y_i=S_i-i,\qquad Y'_i=Y_i+h.
\]
The first equality defines the remaining count, the second removes
the forced prefix, and the third adds the concentration slack.
Along the positive prefix, \(Y_i\) is nondecreasing, \(D_i\) is
nonnegative, and the endpoint has \(Y_\ell\) at most \(D_\ell\).

Define constants
\[
 p=\frac{11026}{21663},\qquad
 a=1-p=\frac{10637}{21663},\qquad
 b=\frac{37}{87}.
\]
The first equality chooses the coefficient \(p\); the second defines
\(a\) as its complement; the third evaluates that complement; the
fourth chooses \(b\).
For nonnegative real arguments \(D\) and \(Y'\), define the convex
potential
\begin{equation}\label{eq:block3-mb-potential}
 \Phi(D,Y')=\max\{pD+(p/2)Y',\ b(D+Y'),\ Y'\}.
\end{equation}
The equality defines the maximum of three affine functions.
For positive \(D\), the active expression changes when the ratio
\(Y'/D\) passes \(\beta_0\) and \(\beta_1\), in that order.
This follows by equating the adjacent affine expressions and
substituting the constants.  The ratio is nondecreasing along the
positive prefix.  At zero \(D\), use the displayed maximum directly.

A step with acceptance count \(x\) decreases \(D\) by one and increases
\(Y'\) by \(x-1\).  If \(x\) is at least three, no affine expression
decreases: every ratio of its \(D\) coefficient to its \(Y'\)
coefficient is at most two.  A nonuseful step with two acceptances
ends beyond the first threshold.  If it starts there, an active
second or third expression is nondecreasing.  Its unique possible
threshold-crossing step loses less than one.  A nonuseful step
with one acceptance similarly ends beyond the second threshold;
after crossing it, the active last expression is constant.
Its unique crossing step also loses less than one.
Every useful positive step decreases the maximum by at most one.

Let \(g\) count useful indices within the positive prefix.
Summing these changes gives
\[
 \Phi(D_\ell,Y'_\ell)\ge\Phi(D_0,Y'_0)-g-2\ge pZ-g-2.
\]
The first inequality charges at most one per useful step and one per
possible threshold crossing.  The second retains the first affine
expression at the initial point and discards its nonnegative slack.
At the endpoint, each affine expression is at most \(D_\ell+h\):
its \(D\)-coefficient after substituting
\(Y'_\ell\) at most \(D_\ell+h\) is at most one, as is its
\(h\)-coefficient.  Consequently
\[
 \#\{\text{nonuseful indices}\}
   =\ell-g\le(1-p)Z+h+2.
\]
The equality holds because all indices after the prefix are useful;
the inequality rearranges the preceding potential estimate and uses
the definition of \(D_\ell\).
The numerical constants satisfy
\[
 (1-p)\frac{201}{100}+\frac1{1000}
 =\frac{7134011}{7221000}<.988.
\]
The equality substitutes \(p\) and combines the fractions; the strict
inequality compares the resulting rational number with \(.988\).
Since the target count is at most \(2.01K\), at least
\(.012K-2\) indices are useful.  The lower bound of \(1000\) on \(K\)
makes this at least \(K/100\).
\end{proof}

\begin{figure}[ht]
\centering
\begin{tikzpicture}[x=8cm,y=4.4cm, every node/.style={font=\small}]
\draw[-{Stealth[length=2mm]}] (0,0)--(1.12,0) node[right] {$Y'/D$};
\draw[-{Stealth[length=2mm]}] (0,0)--(0,1.12)
 node[above] {$\Phi(D,Y')/D$};
\draw[thick] (0,0.50897844343)--(.49,.63367816092)--(.74,.74)--(1.06,1.06);
\draw[dashed,gray] (.49,0)--(.49,.63367816092);
\draw[dashed,gray] (.74,0)--(.74,.74);
\node[below] at (.49,0) {$49/100$};
\node[below] at (.74,0) {$37/50$};
\node[left] at (0,.50897844343) {$p$};
\node[above] at (.20,.63) {first piece};
\node[above] at (.60,.73) {second piece};
\node[right] at (.88,.78) {third piece};
\end{tikzpicture}
\caption{The potential of Lemma~\ref{lem:block3-mb-potential},
normalized by a positive $D$. Its two breakpoints are exactly the
surviving-capacity thresholds. The ratio on the horizontal axis is
nondecreasing along the positive prefix, so only two steps can
cross into a new affine piece.}
\label{fig:potential}
\end{figure}
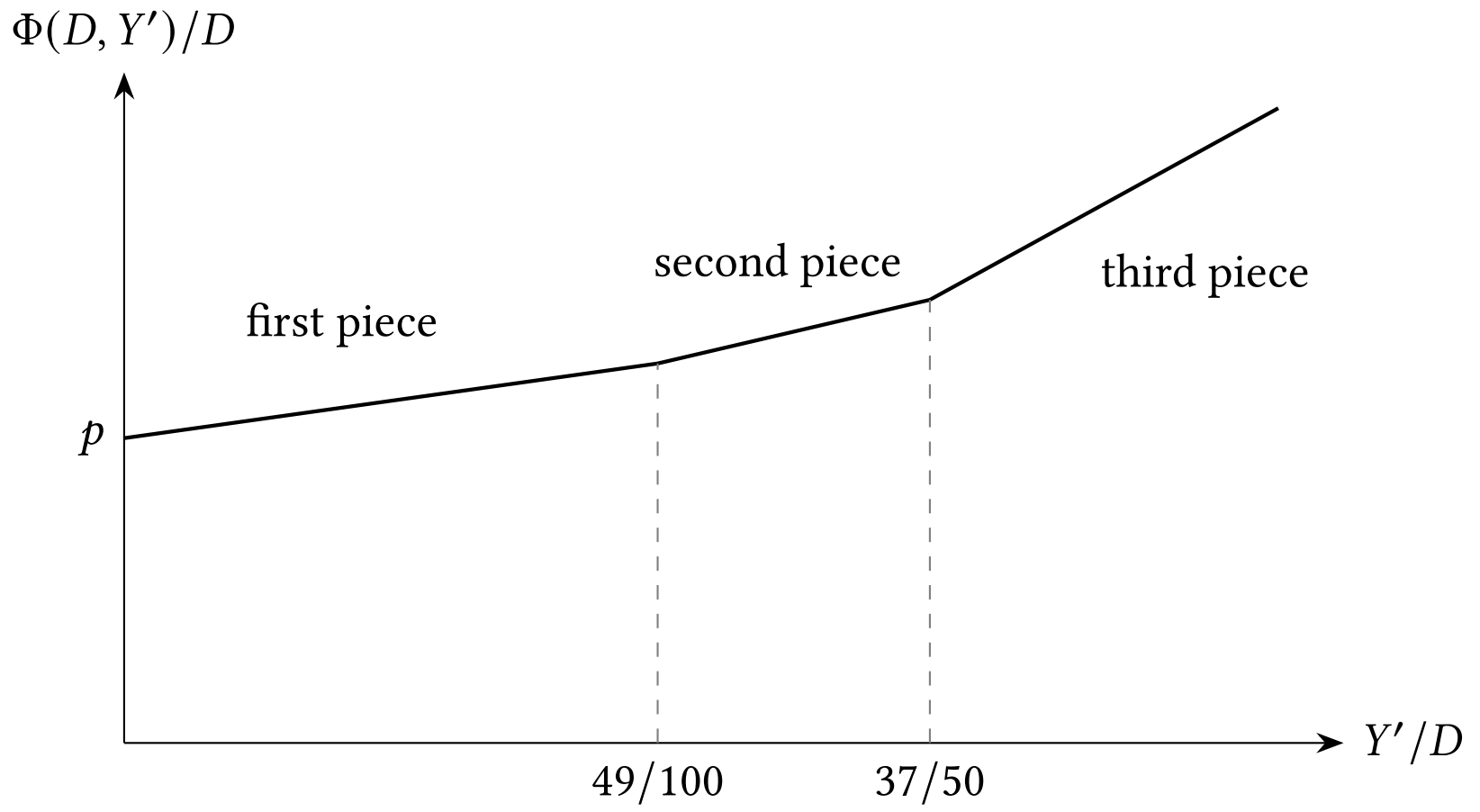

Figure~\ref{fig:potential} illustrates the two changes of affine piece used in the preceding charge.

\subsection{The recursive-step bound}
\label{subsec:core-wrapper-bound}

\begin{lemma}[Uniform estimates for one recursive step]
\label{lem:block3-uniform-wrapper-estimates}
There are positive absolute constants \(A_0,a_0,b_0,C_0\) with the
following properties.  For a connected input with \(v\) vertices and
\(M\) original edges, shrink parameter \(R\) at least one, and
integer \(K\) at least \(1000\), fix a target edge set or minimum cut.
Conditional on any initial-tree history in which no introduced
representative belongs to that target through \(T_{\max}\), the total
failure probability needed for the corresponding one of the
three guarantees is at most
\begin{equation}\label{eq:block3-uniform-error}
 A_0(R+1)K^{b_0}e^{-a_0K}.
\end{equation}
The conditioning event has probability greater than \(.998\), and the useful
index fraction is at least \(1/200\).  Before its single recursive
call, the procedure has unconditional expected work at most
\begin{equation}\label{eq:block3-uniform-outer-cost}
 C_0(M+vK^2).
\end{equation}
These constants are independent of the predecessor's success
probabilities.  The returned forest has at most
\(\max\{2,\lceil v/R\rceil\}\) components on every execution.
\end{lemma}
\begin{proof}
The stopping-error bound is \eqref{eq:block3-rounded-stop-error}.
All counting failures proved above are a fixed finite collection of
polynomials in \(K\) times exponentials with negative absolute
linear coefficients.  Enlarging \(A_0,b_0\) and decreasing \(a_0\)
bounds their sum by \eqref{eq:block3-uniform-error}.
The initial-tree avoidance bound and useful fractions were proved separately
for each target type.  The work estimate was established after the
algorithm, including the expected work of all exact MST calls.
Every realization eventually meets the component threshold because
all initial-tree edges are retained, so they eventually form a spanning
tree at every index.
\end{proof}

\begin{theorem}[Closure under one recursive step]
\label{thm:block3-rounded-wrapper}
Suppose the predecessor always returns a valid tree and satisfies the
three guarantees with a positive common probability \(p\).
For every shrink ratio \(R\) at least one, a choice of
\(K\) of order \(1+\log(R+2)\), with constants depending only on \(p\),
preserves the three guarantees with common probability at least
\(p/1000\).  The recursive call has at most
\(\max\{2,\lceil n/R\rceil\}\) vertices and at most \(m\) original
edges.  Let \(q\) and \(M\) denote its quotient's vertex and
original-edge counts, and let \(R_{\rm prev}\) denote the predecessor's
capped running-time bound.  With fixed work caps, total running time is
\begin{equation}\label{eq:block3-rounded-recurrence}
 O\bigl(m+nK^2+R_{\rm prev}(q,M)\bigr).
\end{equation}
\end{theorem}
\begin{proof}
First store a DFS spanning tree for use on abort, in linear time.
Choose the coefficient in \(K\) sufficiently large that
\eqref{eq:block3-uniform-error} is at most \(p/10000\).
Lemma~\ref{lem:block3-uniform-wrapper-estimates} permits fixed outer
work caps whose total unconditional failure probability is at most
\(p/10000\), by Markov's inequality.  Conditional on an outer history
and a useful index, the fresh predecessor succeeds with probability
at least \(p\).  The common success probability is thus at least
\[
 .998\left(\frac p{200}-\frac p{10000}\right)
       -\frac p{10000}>\frac p{1000}.
\]
The expression multiplies the initial-tree avoidance probability by useful-index success,
subtracts the conditional counting failures, and then subtracts
the unconditional cap failures; the strict inequality is numerical.
The quotient bound is enforced by the stopping rule, and the work
bound follows from the outer estimate and the single predecessor
call.  Caps are coupled to the uncapped laws and never used to
condition Poisson or independence assertions.  Every fallback is
a valid tree.
\end{proof}

\section{Choice of parameters and the two running-time bounds}
\label{sec:core-parameters}

\subsection{Any fixed number of iterated logarithms}

Recall the truncated iterated logarithms \(L_j(\cdot)\) defined in
Section~\ref{sec:block3-primitives}.

\begin{theorem}[Fixed-depth distribution]
\label{thm:block3-rounded-fixed-depth}
\label{thm:block3-fair-fixed-depth}
For every fixed positive integer \(s\), there is a distribution
satisfying the three guarantees with positive probabilities depending
only on \(s\), in capped time and space
\[
 O_s\bigl(m+nL_{s+1}(n)^2\bigr).
\]
Combining it with the linear fixed-tree cut oracle gives a global
minimum cut with any prescribed fixed success probability below one
in the same asymptotic time.
\end{theorem}
\begin{proof}
The base has work \(O(m+nL_1(n)^3)\) and common success probability
at least \(1/200\).  Construct the distributions successively.
At the \(r\)-th step, use shrink ratio \(L_r(n)^4\) and choose
\(K\) of order \(L_{r+1}(n)\), with constants permitted to depend
on the fixed depth.  Each recursive call uses its own vertex count
in these size functions.

For the first step the quotient's cubic logarithmic vertex term is
\(O(n)\): its size is at most the maximum of two and the ceiling of
\(n/L_1(n)^4\), and the logarithm is monotone.
At a later step the predecessor's vertex term is
\(qL_r(q)^2\), which is again \(O(n)\) by the fourth-power shrink
and monotonicity.  Bounded quotient sizes contribute only a constant.
The recurrence
\eqref{eq:block3-rounded-recurrence} therefore gives the stated
vertex term.  A fixed chain makes only a fixed number of scans
of lists containing at most \(m\) original edges.  Closure preserves
positive probabilities through this fixed number of calls.

For any fixed original optimum, the final tree crosses it at most
twice with a fixed positive probability.  Conditional on such a tree,
a fresh invocation of Theorem~\ref{thm:block3-local} finds the optimum
with probability greater than \(1/10\).  A fixed number of independent
complete trials, retaining the least exact cut value, gives any
specified fixed success probability.  Sequential trials reuse space.
\end{proof}

In particular, choosing depth \(h\) for any fixed positive integer
\(h\) yields the claimed \(O_h(m+nL_h(n))\) bound, because
\(L_{h+1}(n)^2\) is \(O(L_h(n))\).  The constant in this comparison
is uniform in the argument for each fixed \(h\).  The constants of
the algorithm may depend on its fixed depth.  The next analysis
accounts explicitly for a growing depth.

\subsection{A uniform depth choice}
\label{sec:block3-uniform-depth}

For the original input define
\begin{equation}\label{eq:block3-uniform-depth}
 s=s(n)=\min\{j\text{ a positive integer}:L_j(n)=2\}.
\end{equation}
The first equality names the depth function; the second defines its
first stabilized iterate, and the equality inside the set is its
stopping condition.  This depth differs from \(\log^*n\) by an
absolute additive constant.  Every constant below is absolute.

Use the uncapped recursive construction; only its terminal base
retains its fixed cap.  A call has an integer parameter \(r\), the
number of remaining recursive steps, from zero through \(s\).
Use the base when \(r\) is zero or the current graph has at most
two vertices.  Fix a sufficiently large absolute positive integer
\(C\).  On a current input with \(v\) vertices that does not meet a
base case, set
\begin{equation}\label{eq:block3-uniform-parameters}
 R=L_r(v)^4,\qquad
 K=C\bigl(s+1+L_{r+1}(v)\bigr).
\end{equation}
The first equality defines the shrink ratio; the second defines the
forest count.  Make one child call with remaining parameter \(r-1\).
The value \(s\) is computed once for the original input.

Write \(p_0=1/200\), defining the base probability lower bound, and
define the target probability at depth \(r\) by
\begin{equation}\label{eq:block3-uniform-p-r}
 p_r=p_0\,1000^{-r}.
\end{equation}
This equality specifies the loss allowed at each recursive step.
One absolute choice of \(C\) ensures
\begin{equation}\label{eq:block3-uniform-error-budget}
 A_0(R+1)K^{b_0}e^{-a_0K}
 \le\frac{p_0\,1000^{-s}}{10000}
 \le\frac{p_{r-1}}{10000}.
\end{equation}
The first inequality is obtained by the coefficient choice justified
next; the second follows because \(r-1\) is at most \(s\).
To justify a common \(C\), first choose an absolute threshold after
which \(b_0\log K\) is at most \(a_0K/4\).
The logarithm of \(R+1\) is at most \(5L_{r+1}(v)\), since \(R\)
is the fourth power of that iterate's predecessor and all iterates
are at least two.  A sufficiently large \(C\) absorbs this term into
another \(a_0K/4\), exceeds the fixed threshold and \(1000\), and
makes the remaining \(a_0K/2\) dominate
\(s\log1000+\log(10000A_0/p_0)\).  This proves the first inequality
uniformly over all calls.  Only integer size parameters are computed.

Induction now establishes all three guarantees with probability at
least \(p_r\).  The base is already covered.  In a recursive step,
the fresh child has common probability \(p_{r-1}\), and the outer
probability is at least
\[
 .998\left(\frac{p_{r-1}}{200}
                   -\frac{p_{r-1}}{10000}\right)
 >\frac{p_{r-1}}{1000}=p_r.
\]
The expression applies initial-tree avoidance, the useful fraction, and
the error budget; the strict inequality is numerical; the equality
is the definition of \(p_r\).  There is no cap subtraction at these
outer calls.  Consequently one complete depth-\(s\) tree call followed
by a fresh local oracle finds the optimum with probability at least
\(p_s/10\).

\subsection{Charging the work of a recursive chain}

Let \(v_0,v_1,\ldots\) be successive vertex counts along the unique
recursive chain, starting with the original graph.
A call on \(v\) vertices with remaining parameter \(r\) has child
size at most
\[
 q=\max\{2,\lceil v/L_r(v)^4\rceil\}.
\]
The equality defines a deterministic upper bound for that child.
For inputs with more than two vertices this is at most \(2v/3\),
because every iterate is at least two.  Thus the sum of vertex
counts is at most \(3n\).  There are at most \(s\) outer calls and
one base, each retaining at most \(m\) original edges.  Their edge
work is therefore \(O((s+1)m)\).

The square of the local forest count is
\(O(s^2+L_{r+1}(v)^2)\).  The \(s^2\) vertex terms sum to
\(O(ns^2)\).  Charge each nonroot logarithmic vertex term to its
parent.  Put \(x=L_r(v)\), defining the parent's shrink iterate.
If \(v/x^4\) is at least one, monotonicity and ceiling rounding give
\[
 qL_r(q)^2\le\frac{2v}{x^4}x^2\le\frac v2.
\]
The first inequality bounds the child size and its iterate by the
parent quantities; the second uses \(x\) at least two.
Otherwise the quotient-size bound is two and the logarithmic
vertex term is eight.  The root's iterate has already stabilized,
so its own logarithmic vertex term is \(4n\).
Every nonroot term is charged once, and their total is \(O(n)\).

The terminal cubic base term has a separate parent charge.
At a parent with remaining parameter one, if \(v/L_1(v)^4\) is at
least one, then
\[
 qL_1(q)^3\le\frac{2v}{L_1(v)}\le v.
\]
The first inequality uses the fourth-power shrink and monotonicity;
the second uses the iterate's lower bound of two.
Otherwise the quotient bound is two and the cubic vertex term is
sixteen.  An earlier termination at at most two vertices also has
constant vertex work.  Inputs with at most two vertices can be
solved directly.

Consequently, there is an absolute constant \(C_1\) such that one
complete tree-and-oracle trial has expected work
\begin{equation}\label{eq:block3-uniform-one-trial}
 \E[\text{trial work}]
 \le C_1\bigl((s+1)m+n(s+1)^2\bigr)
\end{equation}
The inequality sums the parent charges,
the original-edge scans, and the final local oracle.
All expected local bounds hold uniformly conditional on every
quotient history, so no independence of work across levels is needed.

The size iterates require no operation on a real capacity.  A table
of ceiling binary logarithms for all integers through \(n\) can be
built in linear time by doubling interval endpoints.  At most
\(O(s^2)\) table accesses compute all iterates in a chain.
This work fits the displayed bound and the table is reused by
all trials.

\subsection{Amplification and a single work cap}

Run the following number of independent complete trials:
\begin{equation}\label{eq:block3-uniform-trials}
 B=4000\cdot1000^s.
\end{equation}
The equality defines the trial count.  Keep the smallest returned
exact cut value.  The probability that every trial misses the optimum
is at most
\[
 (1-p_s/10)^B\le e^{-Bp_s/10}=e^{-2}<1/6.
\]
The inequality uses the exponential upper bound on one minus a
probability; the equality substitutes \(B,p_s,p_0\); the final
strict inequality is numerical.

First compute a nontrivial original cut and its exact value.
Place one step counter around the entire randomized computation,
with budget six times a known absolute upper bound on its total
expected work.  If that budget is reached, return the best stored
cut.  Markov's inequality bounds this abort probability by \(1/6\).
The union of abort and missed-optimum events has probability below
\(1/3\).  They need not be independent.

\begin{theorem}[Uniform iterated-logarithm bound]
\label{thm:block3-uniform-logstar}
There is a uniform Monte Carlo algorithm for exact undirected global
minimum cut with arbitrary nonnegative real capacities and parallel
original edges.  It returns an optimum with probability at least
\(2/3\), in capped time and space
\begin{equation}\label{eq:block3-uniform-final-bound}
 O\left(1000^{s(n)}
       \bigl((s(n)+1)m+n(s(n)+1)^2\bigr)\right)
 \subseteq O\bigl((m+n)2^{O(\log^*n)}\bigr).
\end{equation}
The inclusion uses the additive-constant relation between \(s(n)\)
and \(\log^*n\), absorbing its polynomial factors into the
exponential.  Every output, including a cap fallback, is a
nontrivial original cut with its exact capacity.
\end{theorem}
\begin{proof}
The preceding amplification and cap give the probability and running
time.  Allocations are charged to the same counter, so the time bound
also bounds space.  Disconnected positive-capacity support is handled
directly by its zero cut.  All sampling and exact MST work has already
been charged, including its fair random bits.
\end{proof}

For completeness, the cap also enforces the word convention.
The depth is \(O(\log n)\), so the trial count and cap are bounded by
a fixed polynomial in input size.  Every primitive checks the counter
before exceeding the budget.  Stored counts, occurrence identifiers,
and exponent counters therefore fit the stated index words.
The Poisson routine's rare larger rational computations retain their
charged multiword implementation.  The uncapped process defines only
the coupling and expected-work majorant; no unbounded identifier is
assigned unit cost on the implemented run.

The fixed-depth and uniform algorithms serve different density
regimes.  The former retains a linear edge term.  On graphs whose
edge and vertex counts are of the same order, the uniform bound
is asymptotically smaller than every fixed iterated-logarithm overhead.
Neither argument establishes a uniform linear-time algorithm.

\section*{AI Acknowledgement}
OpenAI's GPT Astra 6.2 and GPT Astra 6 assisted with the development of some proofs and with drafting and exposition. Gemini Neon and GPT Sol also assisted with the writing. Responsibility for the results' correctness and the paper's final content rests with the authors. Lean files for the single-source shortest paths results are released on \url{https://github.com/igholami/sssp-lean}, and Lean files for the minimum-cut formalization are released on \url{https://github.com/igholami/mincut-lean}.

\printbibliography

@article{DBLP:journals/nm/Dijkstra59,
  author       = {Edsger W. Dijkstra},
  title        = {A note on two problems in connexion with graphs},
  journal      = {Numerische Mathematik},
  volume       = {1},
  pages        = {269--271},
  year         = {1959},
  doi          = {10.1007/BF01386390},
  bibsource    = {dblp computer science bibliography, https://dblp.org}
}

@article{DBLP:journals/jacm/Thorup99,
  author       = {Mikkel Thorup},
  title        = {Undirected Single-Source Shortest Paths with Positive Integer Weights
                  in Linear Time},
  journal      = {J. {ACM}},
  volume       = {46},
  number       = {3},
  pages        = {362--394},
  year         = {1999},
  doi          = {10.1145/316542.316548},
  bibsource    = {dblp computer science bibliography, https://dblp.org}
}

@article{DBLP:journals/jacm/FredmanT87,
  author       = {Michael L. Fredman and
                  Robert Endre Tarjan},
  title        = {Fibonacci heaps and their uses in improved network optimization algorithms},
  journal      = {J. {ACM}},
  volume       = {34},
  number       = {3},
  pages        = {596--615},
  year         = {1987},
  doi          = {10.1145/28869.28874},
  bibsource    = {dblp computer science bibliography, https://dblp.org}
}

@article{DBLP:journals/siamcomp/PettieR05,
  author       = {Seth Pettie and
                  Vijaya Ramachandran},
  title        = {A Shortest Path Algorithm for Real-Weighted Undirected Graphs},
  journal      = {{SIAM} J. Comput.},
  volume       = {34},
  number       = {6},
  pages        = {1398--1431},
  year         = {2005},
  doi          = {10.1137/S0097539702419650},
  bibsource    = {dblp computer science bibliography, https://dblp.org}
}

@inproceedings{DBLP:conf/stoc/DuanMMSY25,
  author       = {Ran Duan and
                  Jiayi Mao and
                  Xiao Mao and
                  Xinkai Shu and
                  Longhui Yin},
  editor       = {Michal Kouck{\'{y}} and
                  Nikhil Bansal},
  title        = {Breaking the Sorting Barrier for Directed Single-Source Shortest Paths},
  booktitle    = {Proceedings of the 57th Annual {ACM} Symposium on Theory of Computing,
                  {STOC} 2025, Prague, Czechia, June 23-27, 2025},
  pages        = {36--44},
  publisher    = {{ACM}},
  address      = {New York, NY, USA},
  year         = {2025},
  doi          = {10.1145/3717823.3718179},
  bibsource    = {dblp computer science bibliography, https://dblp.org}
}

@inproceedings{DBLP:conf/focs/DuanMSY23,
  author       = {Ran Duan and
                  Jiayi Mao and
                  Xinkai Shu and
                  Longhui Yin},
  title        = {A Randomized Algorithm for Single-Source Shortest Path on Undirected
                  Real-Weighted Graphs},
  booktitle    = {64th {IEEE} Annual Symposium on Foundations of Computer Science, {FOCS}
                  2023, Santa Cruz, CA, USA, November 6-9, 2023},
  pages        = {484--492},
  publisher    = {{IEEE} Computer Society},
  address      = {Los Alamitos, CA, USA},
  year         = {2023},
  doi          = {10.1109/FOCS57990.2023.00035},
  bibsource    = {dblp computer science bibliography, https://dblp.org}
}

@misc{DBLP:journals/corr/abs-2609-15247,
  author       = {Avi Kadria and
                  Liam Roditty},
  title        = {A Faster Undirected Single-Source Shortest Path Algorithm},
  year         = {2026},
  archiveprefix = {arXiv},
  eprint       = {2609.15247},
  bibsource    = {dblp computer science bibliography, https://dblp.org}
}

@inproceedings{DBLP:conf/icalp/DuanMSY26,
  author       = {Ran Duan and
                  Xiao Mao and
                  Xinkai Shu and
                  Longhui Yin},
  editor       = {Sayan Bhattacharya and
                  Danupon Nanongkai and
                  Michael Benedikt and
                  Gabriele Puppis},
  title        = {A Faster Directed Single-Source Shortest Path Algorithm},
  booktitle    = {53rd International Colloquium on Automata, Languages, and Programming,
                  {ICALP} 2026, Royal Holloway, University of London, Egham, United
                  Kingdom, July 7-10, 2026},
  series       = {LIPIcs},
  volume       = {374},
  pages        = {81:1--81:23},
  publisher    = {Schloss Dagstuhl - Leibniz-Zentrum f{\"{u}}r Informatik},
  address      = {Dagstuhl, Germany},
  year         = {2026},
  doi          = {10.4230/LIPIcs.ICALP.2026.81},
  bibsource    = {dblp computer science bibliography, https://dblp.org}
}

@article{DBLP:journals/jcss/FakcharoenpholRT04,
  author       = {Jittat Fakcharoenphol and
                  Satish Rao and
                  Kunal Talwar},
  title        = {A tight bound on approximating arbitrary metrics by tree metrics},
  journal      = {J. Comput. Syst. Sci.},
  volume       = {69},
  number       = {3},
  pages        = {485--497},
  year         = {2004},
  doi          = {10.1016/J.JCSS.2004.04.011},
  bibsource    = {dblp computer science bibliography, https://dblp.org}
}

@article{DBLP:journals/jacm/Tarjan75,
  author       = {Robert Endre Tarjan},
  title        = {Efficiency of a Good But Not Linear Set Union Algorithm},
  journal      = {J. {ACM}},
  volume       = {22},
  number       = {2},
  pages        = {215--225},
  year         = {1975},
  doi          = {10.1145/321879.321884},
  bibsource    = {dblp computer science bibliography, https://dblp.org}
}

@article{DBLP:journals/tcs/BenderF04,
  author       = {Michael A. Bender and
                  Martin Farach{-}Colton},
  title        = {The Level Ancestor Problem simplified},
  journal      = {Theor. Comput. Sci.},
  volume       = {321},
  number       = {1},
  pages        = {5--12},
  year         = {2004},
  doi          = {10.1016/J.TCS.2003.05.002},
  bibsource    = {dblp computer science bibliography, https://dblp.org}
}

@article{DBLP:journals/siamcomp/HarelT84,
  author       = {Dov Harel and
                  Robert Endre Tarjan},
  title        = {Fast Algorithms for Finding Nearest Common Ancestors},
  journal      = {{SIAM} J. Comput.},
  volume       = {13},
  number       = {2},
  pages        = {338--355},
  year         = {1984},
  doi          = {10.1137/0213024},
  bibsource    = {dblp computer science bibliography, https://dblp.org}
}

@article{DBLP:journals/jcss/BlumFPRT73,
  author       = {Manuel Blum and
                  Robert W. Floyd and
                  Vaughan R. Pratt and
                  Ronald L. Rivest and
                  Robert Endre Tarjan},
  title        = {Time Bounds for Selection},
  journal      = {J. Comput. Syst. Sci.},
  volume       = {7},
  number       = {4},
  pages        = {448--461},
  year         = {1973},
  doi          = {10.1016/S0022-0000(73)80033-9},
  bibsource    = {dblp computer science bibliography, https://dblp.org}
}

@inproceedings{Yan2026,
  author    = {Shuyi Yan},
  title     = {Lossless Derandomization for Undirected Single-Source Shortest Paths and Approximate Distance Oracles},
  booktitle = {2026 SIAM Symposium on Simplicity in Algorithms (SOSA)},
  pages     = {404--412},
  publisher = {SIAM},
  address   = {Philadelphia, PA, USA},
  year      = {2026},
  doi       = {10.1137/1.9781611978964.32}
}

@misc{GrishinZabokritskiy2026,
  author  = {Grishin, Valery (Binyamin) and Zabokritskiy (Yohananov), Aryeh Lev},
  title   = {Improved Bounds for Directed {SSSP} in Sparse and Intermediate-Density Graphs},
  year    = {2026},
  archiveprefix = {arXiv},
  eprint  = {2611.26453v2}
}

@article{Karger2000,
  author = {David R. Karger},
  title = {Minimum Cuts in Near-Linear Time},
  journal = {Journal of the ACM},
  volume = {47}, number = {1}, pages = {46--76}, year = {2000},
  doi = {10.1145/331605.331608}
}

@article{KKT1995,
  author = {David R. Karger and Philip N. Klein and Robert E. Tarjan},
  title = {A Randomized Linear-Time Algorithm to Find Minimum Spanning Trees},
  journal = {Journal of the ACM},
  volume = {42}, number = {2}, pages = {321--328}, year = {1995},
  doi = {10.1145/201019.201022}
}

@article{DRT1992,
  author = {Brandon Dixon and Monika Rauch and Robert E. Tarjan},
  title = {Verification and Sensitivity Analysis of Minimum Spanning Trees in Linear Time},
  journal = {SIAM Journal on Computing},
  volume = {21}, number = {6}, pages = {1184--1192}, year = {1992},
  doi = {10.1137/0221070}
}

@article{BuchsbaumEtAl2008,
  author = {Adam L. Buchsbaum and Loukas Georgiadis and Haim Kaplan and
            Anne Rogers and Robert E. Tarjan and Jeffery R. Westbrook},
  title = {Linear-Time Algorithms for Dominators and Other Path-Evaluation Problems},
  journal = {SIAM Journal on Computing},
  volume = {38}, number = {4}, pages = {1533--1573}, year = {2008},
  doi = {10.1137/070693217},
  note = {Author manuscript arXiv:cs/0207061v2, Section 6.3 and Theorem 6.1}
}

@article{NI1992,
  author = {Hiroshi Nagamochi and Toshihide Ibaraki},
  title = {A Linear-Time Algorithm for Finding a Sparse {$k$}-Connected
           Spanning Subgraph of a {$k$}-Connected Graph},
  journal = {Algorithmica}, volume = {7}, pages = {583--596}, year = {1992}
}

@phdthesis{KargerThesis,
  author = {David R. Karger},
  title = {Random Sampling in Graph Optimization Problems},
  school = {Stanford University}, year = {1995}
}

@article{ForsterDeVos2023,
  author = {Sebastian Forster and Tijn {de Vos}},
  title = {Faster Cut Sparsification of Weighted Graphs},
  journal = {Algorithmica}, volume = {85}, pages = {929--964}, year = {2023},
  doi = {10.1007/s00453-022-01053-4}
}

@article{GMW2024,
  author = {Pawe{\l} Gawrychowski and Shay Mozes and Oren Weimann},
  title = {Minimum Cut in {$O(m\log^2 n)$} Time},
  journal = {Theory of Computing Systems}, year = {2024},
  doi = {10.1007/s00224-024-10179-7},
  note = {Conference version in ICALP 2020}
}

@inproceedings{GMW2021,
  author = {Pawe{\l} Gawrychowski and Shay Mozes and Oren Weimann},
  title = {A Note on a Recent Algorithm for Minimum Cut},
  booktitle = {Symposium on Simplicity in Algorithms (SOSA)},
  pages = {74--79}, year = {2021}
}

@article{MNS2017,
  author = {Kurt Mehlhorn and Adrian Neumann and Jens M. Schmidt},
  title = {Certifying 3-Edge-Connectivity},
  journal = {Algorithmica}, volume = {77}, number = {2},
  pages = {309--335}, year = {2017}
}

@misc{KPS2026,
  author = {Sanjeev Khanna and Aaron Putterman and Junkai Song},
  title = {A Deterministic Linear Time Algorithm for Minimum Spanning Tree},
  year = {2026}, eprint = {2611.31187}, archivePrefix = {arXiv},
  note = {arXiv:2611.31187, version 1, November 23, 2026},
  url = {https://arxiv.org/abs/2611.31187}
}

@article{BK2015,
  author = {Andr{\'a}s A. Bencz{\'u}r and David R. Karger},
  title = {Randomized Approximation Schemes for Cuts and Flows in Capacitated Graphs},
  journal = {SIAM Journal on Computing},
  volume = {44}, number = {2}, pages = {290--319}, year = {2015},
  doi = {10.1137/070705970}
}
\end{document}